\documentclass[aps,prd,preprintnumbers,showpacs,showkeys,nofootinbib,
superscriptaddress,fleqn,floatfix,tightenlines,10pt]{revtex4-2}
\usepackage{amsmath,amsfonts,amssymb,amscd,amsxtra,amsthm}
\usepackage{graphicx}  
\usepackage{epstopdf}
\usepackage{dcolumn}  
\usepackage{bm}          
\usepackage{slashed}
\usepackage{cancel}
\usepackage{float}
\usepackage{mathtools}
\usepackage{amsbsy}
\usepackage{amstext}

\usepackage[utf8]{inputenc} 
\usepackage{booktabs} 
\usepackage[normalem]{ulem} 
\usepackage[dvipsnames]{xcolor} 
\usepackage{tabularx}
\usepackage{enumitem}  
\usepackage{array} 
\usepackage{slashed}
\usepackage{tikz}
\usepackage{float}
\usepackage{multirow}
\usepackage{cancel}
\usepackage{physics}

\renewcommand{\sout}{\bgroup \color{red} \ULdepth=-.5ex \ULset}

\makeatletter

\begin{document}
\preprint{INHA-NTG-11/2022}
\title{Gravitational form factors of the baryon octet with
   flavor SU(3) symmetry breaking} 
\author{Ho-Yeon Won}
\email[E-mail: ]{hoywon@inha.edu}
\affiliation{Department of Physics, Inha University, Incheon 402-751,
  South Korea} 

\author{June-Young Kim}
\email[E-mail: ]{Jun-Young.Kim@ruhr-uni-bochum.de}
\affiliation{Department of Physics, Inha University, Incheon 402-751,
  South Korea} 
\affiliation{Theory Center, Jefferson Lab, Newport News, VA 23606, USA} 

\author{Hyun-Chul Kim}
\email[E-mail: ]{hchkim@inha.ac.kr}
\affiliation{Department of Physics, Inha University, Incheon 402-751,
  South Korea} 
\affiliation{School of Physics, Korea Institute for Advanced Study
  (KIAS), Seoul 02455, South Korea}

\date {\today}
\begin{abstract}
We investigate the gravitational form factors of the baryon
octet within the framework of the SU(3) chiral quark-soliton model, 
considering the effects of flavor SU(3) symmetry breaking, and the
corresponding energy-momentum tensor distributions. We examine the
effects of flavor SU(3) symmetry breaking to the mass, angular
momentum, pressure, and shear force distributions of the baryon
octet. We first find that a heavier baryon is energetically
more compact than a lighter one. For the spin distributions of the
baryon octet, they are properly normalized to their spins and are
decomposed into the flavor-singlet axial charge and the orbital
angular momentum even when the flavor SU(3) symmetry is broken. While 
the effects of the flavor SU(3) symmetry breaking differently
contribute to the angular momentum distributions for the octet
baryons, they are found to be rather small. The spin and orbital
angular momentum almost equally contribute to the angular momentum
distributions for the octet baryons. We also estimate the effects of
the flavor SU(3) symmetry breaking to the pressure and shear force 
distributions. Interestingly, even if we include the effects of the
SU(3) flavor symmetry breaking, the shear force distributions are kept
to be positive over $r$. It indicates that the Polyakov \& Schweitzer
local stability condition is kept to be intact with the flavor SU(3)
symmetry broken. Lastly, we discuss how much the gravitational form
factors vary with the effects of flavor SU(3) symmetry 
breaking considered.  
\end{abstract}
\pacs{}
\keywords{}
\maketitle

\section{Introduction}
It is of great importance to understand the mechanical
structure of a baryon as much as the electromagnetic~(EM) one,
since it reveals how the baryon is mechanically shaped by its
partons. The gravitational form factors (GFFs) of a baryon provide
information on its mechanical properties such as the mass, spin,
pressure, and shear force. At an early stage, the GFFs were
considered as a purely academic subject~\cite{Kobzarev:1962wt, 
  Pagels:1966zza} due to the difficulty in having access to them
experimentally. However, the generalized parton
distributions~(GPDs) have paved way for extracting the GFFs
experimentally, since the EM form factors and GFFs are defined
respectively as the first and second Mellin moments of the GPDs that 
can be measured by the hard exclusive process such as deeply virtual 
Compton scattering~(DVCS) or hard exclusive meson
production. Recently, the first measurement of the nucleon $D$-term 
form factors from DVCS was reported~\cite{Burkert:2018bqq,
  Kumericki:2019ddg, Burkert:2021ith}. The transition GPDs will soon
be extracted from the experimental data on the hard exclusive
meson production $p\to\Delta^{++}\pi^{-}$ at Jefferson
Lab~(JLab)~\cite{Proceedings:2020fyd, Joo}. This measurement 
will lead to the $N\to \Delta$ transition GFFs~\cite{Kim:2022bwn}. 
Moreover, the upcoming Electric-Ion Collider~(EIC) project will unveil
the fractions of the mass and spin of the nucleon, which are taken up
by quarks and gluons inside it. It is well known that the
quark content of the nucleon spin is small~(see a recent
review~\cite{Aidala:2012mv}) and the strange quark is polarized 
negatively ($\Delta s \sim -0.10 $~\cite{Bass:2004xa}). This implies
that the gluon spin and the orbital motion of the quarks and
gluon should considerably contribute to the nucleon spin. The future
EIC project will provide a clue to the spin structure of the nucleon.

The GFFs for spin-1/2 particles parametrize the matrix element of the
energy-momentum tensor (EMT) current~\cite{Kobzarev:1962wt,
  Pagels:1966zza, Kobsarev:1970qm, Ng:1993vh}. It was
recently generalized to higher-spin
particles~\cite{Cotogno:2019vjb} in a systematic way. Based on this  
parametrization, the GFFs of the nucleon have been intensively
investigated in various approaches~\cite{Polyakov:2002wz, Ji:1997gm,
  Schweitzer:2002nm, Jung:2013bya, Hagler:2003jd, Gockeler:2003jfa,
  Pasquini:2007xz,   Hwang:2007tb, Abidin:2008hn, Brodsky:2008pf,
  Pasquini:2014vua,   Chakrabarti:2015lba, Lorce:2018egm,
  Teryaev:2016edw,   Shanahan:2018nnv, Shanahan:2018pib,
  Neubelt:2019sou, Anikin:2019kwi,   Alharazin:2020yjv,
  Gegelia:2021wnj, Varma:2020crx, Fujita:2022jus, Mamo:2022eui,
  Pefkou:2021fni, Azizi:2019ytx, Polyakov:2018exb, Freese:2021mzg,
  Freese:2021qtb, Freese:2021czn}. The parity flip 
transition~\cite{Polyakov:2020rzq, Azizi:2020jog, Ozdem:2019pkg} and
$N \to \Delta$ transition~\cite{Kim:2022bwn} matrix elements of the
EMT current were also parametrized.  For a spin-1 particle, the
model-independent formalism for the GFFs and  distributions
were studied in Refs.~\cite{Polyakov:2019lbq, Cosyn:2019aio,
  Kim:2022wkc, Freese:2022yur, Freese:2022ibw} and the GFFs
were obtained by many theoretical works~\cite{Freese:2019bhb,
  Sun:2020wfo, Epelbaum:2021ahi, Pefkou:2021fni}. The GFFs
for a spin-3/2 particle were also examined~\cite{Pefkou:2021fni,
  Fu:2022rkn, Alharazin:2022wjj, Panteleeva:2020ejw}. On the other
hand, the GFFs of the baryon octet were much less
studied~\cite{Ozdem:2020ieh}. To compute them, we need to
consider the flavor SU(3) symmetry and its breakdown. Since the
effects of the flavor SU(3) symmetry breaking on the GFFs and related
distributions have never been examined, it is worthwhile to
investigate them. In particular, it is critical to check whether the
local and global stability conditions are satisfied with the flavor
SU(3) symmetry broken. 

While the three-dimensional (3D) EMT distributions, which
show how partons are spatially distributed inside a baryon in the
Breit frame (BF)~\cite{Polyakov:2002yz}, were obtained by the 3D
Fourier transform of the corresponding GFFs, there have been 
serious criticisms of the 3D distributions of the
nucleon~\cite{RevModPhys.29.144, Burkardt:2000za, Burkardt:2002hr,
  Miller:2007uy, Jaffe:2020ebz}. The 3D distributions depend on the
shape of the wave packet of a baryon, and this wave packet cannot be
localized below the Compton wavelength. It brings about ambiguous
relativistic effects of which the contribution is approximately 
by around $20~\%$ for the nucleon. Thus, they cannot be neglected
anymore. To circumvent these ambiguous relativistic effects, the
two-dimensional (2D) spatial EMT distributions have been considered in
the infinite momentum frame~(IMF) or on the light-front~(LF). The 
ambiguous relativistic corrections are kinematically suppressed
then~\cite{Lorce:2018egm, Freese:2021czn, Freese:2021qtb}. However,
we have to pay the price that we lose information in the longitudinal 
direction. 

There is yet another way of understanding the 3D distributions by
defining them using the Wigner phase-space distribution. 
While it does not furnish the 3D distributions with the probabilistic
meaning, it allows us to treat their relativistic effects. Moreover,
it shows that the 3D BF and 2D IMF distributions can naturally be 
interpolated in the Wigner sense. Thus, we can trace down the
origin of the relativistic corrections to the 2D IMF distributions. At
the same time, a direct connection between the 3D BF and 2D IMF
distributions was found to be the IMF Abel
transform~\cite{Panteleeva:2021iip, Kim:2021jjf}. Note that, very 
recently, a novel concept of the 3D strict probabilistic distribution
was introduced to remove ambiguous relativistic
corrections~\cite{Epelbaum:2022fjc, Panteleeva:2022khw}. In this work,
we first define the 3D BF distributions in the Wigner sense and then
map out the 2D IMF ones by using the IMF Abel transform. 

In the current work, we will scrutinize the GFFs of the baryon octet
and pertinent three-dimensional distributions within the framework of
a pion mean-field approach or the chiral quark-soliton
model~($\chi$QSM)~\cite{Diakonov:1987ty, Wakamatsu:1990ud,
  Christov:1995vm}. E. Witten in his seminal paper~\cite{Witten:1979kh, 
  Witten:1983tx} inspired the idea of the meson mean-field 
approach. In the limit of a large number of colors ($N_{c}$), the
quantum fluctuations are of order $1/N_c$, so that it can be
ignored. Thus, a baryon can be viewed as $N_{c}$ valence quarks
bound by a pion mean field that arises from a classical solution of
the equation of motion. To put more explicitly, the presence of the
$N_{c}$ valence quarks polarizes the vacuum, which produces the pion
mean field. Then the $N_{c}$ valence quarks are also influenced by the
pion mean field in a self-consistent way. As a result, a classical
baryon appears as a chiral soliton with a hedgehog symmetry, which is
composed of the $N_{c}$ valence quarks. 
While we ignore the $1/N_c$ mesonic quantum fluctuations, we have to
consider the fluctuations of the pion field along the
zero-mode direction. The translational and rotational zero modes are
related to the symmetries of the baryon. Integrating over the zero
modes completely, we can restore the correct quantum numbers of the
baryon~\cite{Diakonov:1987ty, Christov:1995vm}. The $\chi$QSM
successfully described various properties of the baryon octet and
decuplet such as the EM properties~\cite{Kim:1995mr, Kim:1995ha,
  Wakamatsu:1996xm, Kim:1997ip, Silva:2013laa, 
  Kim:2020lgp, Kim:2019gka}, axial-vector
structures~\cite{Silva:2005fa, Jun:2020lfx, Suh:2022guw}, tensor
charges~\cite{Kim:1995bq, Kim:1996vk},  GFFs~\cite{Goeke:2007fp,
Goeke:2007fq,Wakamatsu:2007uc, Kim:2020nug, Kim:2021jjf}, and
partonic structures~\cite{Pobylitsa:1996rs, Schweitzer:2001sr,
Diakonov:1996sr, Diakonov:1997vc, Wakamatsu:1997en, Son:2019ghf,
Son:2022qro, Kim:2021zbz, Son:2022aue}. It has also been extended to
singly heavy baryons~\cite{Yang:2016qdz, Kim:2018cxv, Kim:2018xlc, 
  Yang:2018uoj, Kim:2018nqf, Kim:2019rcx, Yang:2019tst, Yang:2020klp, 
  Kim:2021xpp}. The GFFs of the singly heavy baryons were 
also studied within the $\chi$QSM~\cite{Kim:2020nug}.  
The $\chi$QSM can also be associated with quantum chromodyanmics (QCD)
via the instanton vacuum~\cite{Diakonov:1985eg, Diakonov:2002fq}. 
The low-energy QCD effective partition function can be derived from
the instanton vacuum. The dynamical quark mass, which is obtained from
the Fourier transform of the fermionic zero mode, is originally
momentum-dependent. In the present work, we turn off the momentum
dependence and introduce a regularization scheme to tame the
divergence coming from the quark loops. 

The present work is organized as follows: In Section~\ref{sec:2}, we
define the GFFs of a spin-1/2 baryon from the matrix elements of the
EMT current. In Section~\ref{sec:3}, we explain the general formalism
for the EMT distributions in both 3D and 2D cases.  
In Section~\ref{sec:4}, we show how the GFFs and the EMT distributions
can be computed within a framework of the SU(3) $\chi$QSM, considering
the effects of the flavor SU(3) symmetry breaking. In Sec~\ref{sec:5},
the numerical results for the GFFs and the EMT distributions of the
octet baryons are presented and discussed. The last Section devotes to
the summary of the present work and draw conclusions.

\section{Gravitational form factors of a spin-1/2 particle \label{sec:2}}
The symmetric EMT current in QCD can be derived by varying the QCD
action under the Poincar{\'e} transformation according to N\"{o}ther's
theorem with the symmetrization imposed for a particle with nonzero
spin~\cite{BELINFANTE1939887,Pauli:1940dq,Callan:1970ze}. More
directly, one can derive the symmetric EMT current by taking a
functional derivative of the QCD
action~\cite{Kobzarev:1962wt,Parker:2009uva} with respect to the
metric tensor of a curved background field. The symmetric total EMT 
operator consists of the quark $(q)$ and gluon $(g)$ parts, which 
are respectively expressed as 
\begin{align}
\hat{T}_{q}^{\mu\nu}&=\frac{i}{4}\bigg[
 \bar{\psi}_{q}\gamma^{\mu}\overrightarrow{\mathcal{D}}^{\nu}\psi_{q}
+\bar{\psi}_{q}\gamma^{\nu}\overrightarrow{\mathcal{D}}^{\mu}\psi_{q}
-\bar{\psi}_{q}\gamma^{\mu}\overleftarrow{\mathcal{D}}^{\nu}\psi_{q}
-\bar{\psi}_{q}\gamma^{\nu}\overleftarrow{\mathcal{D}}^{\mu}\psi_{q}
\bigg]
-g^{\mu\nu}\bar{\psi}_{q}\left(
  \frac{i}{2}\overrightarrow{\slashed{\mathcal{D}}}
- \frac{i}{2}\overleftarrow{\slashed{\mathcal{D}}}
- \hat{m}_{q}
  \right)\psi_{q},\cr
\hat{T}_{g}^{\mu\nu}&=
F^{a,\mu\eta}F_{\;\;\;\eta}^{a,\;\nu}
+\frac{1}{4}g^{\mu\nu}F^{a,\kappa\eta}F_{\;\;\;\kappa\eta}^{a,}.
\end{align}
Here, the covariant derivatives are defined as
$\overleftrightarrow{\mathcal{D}}_{\mu}
=\overleftrightarrow{\partial}_{\mu}+igt^{a}A_{\mu}^{a}$.  
$t^{a}$ represent the SU(3) color group generators that satisfy
the commutation relations $[t^{a},t^{b}]=if^{abc}t^{c}$ and are
normalized to be $\mathrm{tr}(t^{a}t^{b})=\frac{1}{2}\delta^{ab}$.
$\psi_{q}$ denotes the quark field with flavor $q$ and $\hat{m}_{q}$
designates the corresponding current quark mass. $F^{a,\mu\eta}$
stands for the gluon field strength expressed as $F_{\mu\nu}^{a}
= \partial_{\mu}A_{\nu}^{a} - \partial_{\nu}A_{\mu}^{a} -
gf^{abc}A_{\mu}^{b}A_{\nu}^{c}$. The total EMT operator is conserved
as follows: 
\begin{align} 
\partial^{\mu}\hat{T}_{\mu\nu}=0,  \quad
  \hat{T}^{\mu\nu}=\sum_{q}\hat{T}_{q}^{\mu\nu}+\hat{T}_{g}^{\mu\nu}, 
\end{align}

For the lowest-lying octet baryon, the matrix element of the EMT
current can be parametrized in terms of the three
GFFs~\cite{Pagels:1966zza, Kobzarev:1962wt, Polyakov:2018zvc}:  
\begin{align}
&\mel{B,p',J_{3}'}{\hat{T}_{\mu\nu}(0)}{B,p,J_{3}}\cr
&=\bar{u}(p',J_{3}')\Bigg[
  A^{B}(t)\frac{P_{\mu}P_{\nu}}{m_{B}}
+ J^{B}(t) \frac{i(P_{\mu}\sigma_{\nu\rho} + P_{\nu}\sigma_{\mu\rho})
  \Delta^{\rho}}{2m_{B}} + D^{B}(t) \frac{\Delta_{\mu} \Delta_{\nu} -
  g_{\mu\nu} \Delta^{2}}{4m_{B}} \Bigg]u(p,J_{3}),
\end{align}
which depends on the spin polarizations $J_{3}$ and $J'_{3}$, the
average momentum $P= (p+p')/2$ of the initial and final states, and
the four-momentum transfer $\Delta= p'-p$. The squared momentum 
transfer is denoted by $t=\Delta^{2}$. The on-shell conditions of the
final and initial four momenta are given by $p'^{2}=p^{2}=m^{2}_{B}$
where $m_{B}$ denotes the mass of the octet baryon. In the BF, these 
GFFs $A^{B}(t)$, $J^{B}(t)$, and $D^{B}(t)$ are traditionally 
understood as the mass, angular momentum, and $D$-term form factors,
respectively. Here, one should keep in mind that in the level of the
quark and gluon degrees of freedom we have one additional form factor
$\bar{c}$, which is constrained to satisfy the relation
$\sum_{a=q,g}\bar{c}^{a}(t)=0$. It can be dropped because of the  
conservation of the total EMT current. 

\section{Energy-momentum tensor distributions}
In the BF, a 3D distribution is traditionally defined as a Fourier
transformation of the corresponding form factor. Since, however, the
baryon cannot be localized below the Compton wavelength, it causes
ambiguous relativistic corrections. These corrections are up to
20~\% for the nucleon. In the non-relativistic picture, they are often
neglected. In the large $N_{c}$ limit, the frame dependence of the
distribution was carefully examined in
Ref.~\cite{Lorce:2022cle}. These 3D distributions in the BF    
can be understood quasi-probabilistic distributions in phase space or
the Wigner distributions~\cite{Lorce:2018zpf, Lorce:2018egm,
  Lorce:2020onh, Lorce:2022jyi}. To obtain the quantum-mechanical  
probabilistic distributions, one should take the IMF or the LF
frame such that the relativistic corrections are kinematically
suppressed and the nucleon is described as a transversely localized
state. This yields 2D transverse densities in the IMF or on the LF.  

The matrix element of the EMT current for a physical
state $|\psi \rangle$ can be expressed in terms of the Wigner
distribution as~\cite{Lorce:2020onh}   
\begin{align}
\langle \hat{T}^{\mu \nu}(\bm{r}) \rangle= \int
  \frac{d^{3} \bm{P}}{(2\pi)^{3}}\int d^{3} \bm{R} \, W(\bm{R},\bm{P})
  \langle \hat{T}^{\mu \nu}(\bm{r}) \rangle_{\bm{R},\bm{P}}, 
\label{eq:20}
\end{align}
where $W(\bm{R},\bm{P})$ represents the Wigner distribution 
given by 
\begin{align}
W(\bm{R},\bm{P}) &=\int \frac{d^{3} \bm{\Delta}}{(2\pi)^{3}}
   e^{-i\bm{\Delta}\cdot \bm{R}}   \tilde{\psi}^{*}\left(\bm{P} +
   \frac{\bm{\Delta}}{2}\right)   \tilde{\psi}\left(\bm{P} -
   \frac{\bm{\Delta}}{2}\right) \cr 
&=\int d^{3}\bm{z} \,e^{-i\bm{z}\cdot \bm{P}} {\psi}^{*}\left(\bm{R} -
  \frac{\bm{z}}{2}\right) {\psi}\left(\bm{R} +
  \frac{\bm{z}}{2}\right). 
\label{eq:Wig}
\end{align}
The average position $\bm{R}$ and momentum $\bm{P}$ are defined as
$\bm{R}=(\bm{r}'+\bm{r})/2$ and $\bm{P}=(\bm{p}'+\bm{p})/2$, respectively. 
$\bm{\Delta}=\bm{p}'-\bm{p}$ denotes the three-momentum transfer,
which enables us to get access to the internal structure of a
particle. The variable $\bm{z}=\bm{r}'-\bm{r}$ stands for the position
separation between the initial and final particles. 
The Wigner distribution contains information on the wave packet of a 
particle 
\begin{align}
 \psi(\bm{r}) = \langle \bm{r} |\psi \rangle  = \int \frac{d^3
  \bm{p}}{(2\pi^3)} e^{i\bm{p}\cdot \bm{r}} 
  \tilde{\psi}(\bm{p}), \ \ \  \tilde{\psi}(\bm{p}) =
  \frac{1}{\sqrt{2p^{0}}}\langle p| \psi \rangle, 
\end{align}
where the plane-wave states $|p\rangle$ and
  $|\bm{r}\rangle$ are respectively normalized as  
 $\langle p'|p\rangle = 2 p^{0}(2\pi)^3
\delta^{(3)}(\bm{p}'-\bm{p})$ and $\langle \bm{r}'|\bm{r}\rangle
=  \delta^{(3)}(\bm{r}'-\bm{r})$. The position state
$|\bm{r}\rangle$ localized at $\bm{r}$ at time $t=0$ is defined
as a Fourier transform of the momentum eigenstate $|p\rangle$
\begin{align}
|\bm{r} \rangle = \int \frac{d^{3}\bm{p}}{(2\pi)^{3}\sqrt{2p^{0}}}
       e^{-i\bm{p}\cdot \bm{r}} | p \rangle. 
\end{align}
If we integrate
over the average position and momentum, then the probabilistic density
in either position or momentum space is recovered to be 
\begin{align}
  \int   \frac{d^{3}\bm{P}}{(2\pi)^{3}}\,W_{N}(\bm{R},\bm{P}) =|
  \psi_N\left(\bm{R} \right) |^{2},\;\;\;
\int d^{3}\bm{R}\,W_{N}(\bm{R},\bm{P}) =| \tilde{\psi}_N\left(\bm{P} \right)
  |^{2}.
\end{align}

Given $\bm{P}$ and $\bm{R}$, the matrix element $\langle
\hat{T}^{\mu \nu} (\bm{r})\rangle_{\bm{R},\bm{P}}$ conveys information on
the internal structure of the particle localized around the average
position $\bm{R}$ and average momentum $\bm{P}$. 
This can be expressed as the 3D Fourier transform of the matrix
element $\langle  B, p', J_{3}' | \hat{T}^{\mu \nu}(0) | B,p, J_{3}
\rangle$:   
\begin{align}
\langle \hat{T}^{\mu \nu}(\bm{r}) \rangle_{\bm{R},\bm{P}}= \langle
  \hat{T}^{\mu \nu}(0) \rangle_{-\bm{x},\bm{P}} = \int
  \frac{d^{3}\bm{\Delta}}{(2\pi)^{3}} e^{-i\bm{x} \cdot \bm{\Delta} }
  \frac{1}{\sqrt{2p^{0}}\sqrt{2p'^{0}}}\langle  p', J_{3}' |
  \hat{T}^{\mu \nu}(0) | p, J_{3} \rangle, 
  \label{eq:25}
\end{align} 
with the shifted position vector $\bm{x}=\bm{r}-\bm{R}$. Note that,
very recently, a novel concept of the 3D strict probabilistic
distribution was introduced to remove ambiguous relativistic
corrections~\cite{Epelbaum:2022fjc, Panteleeva:2022khw}. 

\subsection{Three-dimensional energy-momentum tensor distributions in
  the Breit frame \label{sec:3}}
Having integrated over $\bm{P}$ of Eq.~\eqref{eq:20}, we find that the
part of the wave packet can be factorized. Thus, the target in the BF
is understood as a localized state around $\bm{R}$ from the Wigner
perspective. In this frame, Eq.~\eqref{eq:25} is reduced to   
\begin{align}
T^{\mu\nu}_{\mathrm{BF},B}(\bm{r},J_{3}',J_{3})
=\int \frac{d^{3}\Delta}{(2\pi)^{3} 2P_{0}}e^{-i\bm{\Delta}\cdot\bm{r}}
\mel{B, p',J_{3}'}{\hat{T}^{\mu\nu}(0)}{B, p,J_{3}}.
\label{eq:26}
\end{align} 
From now on we use $\bm{r}$ instead of $\bm{x}$, i.e.,
$\bm{x}=\bm{r}-\bm{R} \to \bm{r}$. In the Wigner sense, the temporal
component of the EMT current yields mass distribution: 
\begin{align}
 T^{00}_{\mathrm{BF},B}(\bm{r},J'_{3},J_{3}) = \varepsilon^{B}(r)
  \delta_{J'_{3} J_{3}} =  m_{B} \int 
  \frac{d^{3}\Delta}{(2\pi)^{3}} e^{-i\bm{\Delta} \cdot \bm{r}} \left[
  A^{B}(t)  - \frac{t}{4m_{B}^{2}}\left(
  A^{B}(t) - 2J^{B}(t) + D^{B}(t) \right) \right]
  \delta_{J'_{3} J_{3}}.
  \label{eq:mass}
\end{align}

By Integrating $T^{00}_{\mathrm{BF},B}$ over 3D space, one obviously
gets the mass of a baryon in the rest frame 
\begin{align}
 \int d^{3}r T^{00}_{\mathrm{BF},B}(\bm{r},J'_{3},J_{3}) = m_{B} A^{B}(0)=m_{B},
\end{align}
with the normalization $A^{B}(0)=1$. Note that, for a higher-spin
particle $(J\geq 1)$, a quadrupole distribution of the 
energy inside the particle appears~\cite{Polyakov:2018rew,
  Polyakov:2019lbq, Cosyn:2019aio, Panteleeva:2020ejw, Kim:2020lrs,
  Freese:2022yur, Kim:2022wkc}. The size of the mass distribution can
be quantified by the mass radius. It is given by either integral of
the mass distribution or derivative of the mass form factor $A^{B}(t)$
with respect to the momentum squared, 
\begin{align}
\langle r^{2}_{\varepsilon} \rangle_{B} = \frac{\int d^{3}r \, r^{2}
  \varepsilon^{B}(r)}{ \int d^{3}r \, \varepsilon^{B}(r)} =
  \frac{6}{A^{B}(0)} \frac{d A^{B}(t)}{dt} \bigg{|}_{t=0}. 
  \label{eq:mass_rad}
\end{align}

The $0k$-component of the EMT current is related to the spatial
distribution of the spin carried by the partons inside a baryon: 
 \begin{align}
J^{i}_{B}(\bm{r},J_{3}',J_{3})& =\epsilon^{ijk} r^{j}
    T^{0k}_{\mathrm{BF},B}(\bm{r},J_{3}',J_{3})    \cr  
  &= 2S^{j}_{J_{3}'J_{3}}\int \frac{d^{3}\Delta}{(2\pi)^{3}}
  e^{-i\bm{\Delta} \cdot \bm{r}} \left[ \left(J^{B}(t)+ \frac{2}{3}t
  \frac{J^{B}(t)}{dt}\right)\delta^{ij} + \left(\Delta^{i}\Delta^{j}-
  \frac{1}{3}\bm{\Delta}^{2} \delta^{ij}\right) \frac{J^{B}(t)}{dt}
  \right].
  \label{eq:angmo}
\end{align}
In principle, both the monopole and quadrupole distributions should be
considered when we deal with the spin distribution. However, we drop
the quadrupole contribution for simplicity, which does not affect the
normalization of the spin form factor $J^{B}(0)$. The quadrupole
structure of the spin distribution was intensively discussed and
related to the monopole distribution in Refs.~\cite{Lorce:2017wkb,
  Schweitzer:2019kkd}. The monopole contribution to the spin
distribution, which is the first term in Eq.~\eqref{eq:angmo}, is
defined as  
\begin{align}
\rho^{B}_{J}(r) := \int \frac{d^{3}\Delta}{(2\pi)^{3}}
  e^{-i\bm{\Delta} \cdot \bm{r}} \left[ \left(J^{B}(t)+ \frac{2}{3}t
  \frac{J^{B}(t)}{dt}\right)
  \right].
  \label{eq:angmodis}
\end{align}
Integrating $J^{i}_{B}(\bm{r},J_{3}',J_{3})$ over space gives the
spin of the baryon as follows
\begin{align}
\int d^{3}r  J^{i}_{B}(\bm{r},J_{3}',J_{3})=
  2\hat{S}^{i}_{J_{3}'J_{3}} \int d^{3}r \,  \rho^{B}_{J}(r) = 
  2\hat{S}^{i}_{J_{3}'J_{3}} J^{B}(0) = \hat{S}^{i}_{J_{3}'J_{3}},
\end{align}
which is just the spin operator of a baryon. The quadrupole
contribution, the second term in Eq.~\eqref{eq:angmo},  obviously
vanishes after the integration over the 3D space. 

The spatial components EMT current $T^{ij}_{\mathrm{BF},B}$ provides
information on the mechanical properties of a baryon. It can be
decomposed into isotropic and anisotropic contributions. This
anisotropic contribution plays a significant role in the mechanical
structure of a baryon~\cite{Lorce:2018egm, Polyakov:2018zvc}. They are
respectively referred to as the pressure $p^{B}(r)$ and shear force
$s^{B}(r)$ and expressed as~\cite{Polyakov:2002yz, Polyakov:2018zvc} 
\begin{align}
T^{ij}_{\mathrm{BF},B}(\bm{r},J_{3}',J_{3}) =  p^{B}(r) \delta^{ij} \delta
_{J_{3}'J_{3}}+ s^{B}(r)\bigg{(} \frac{r^{i}r^{j}}{r^{2}} -
  \frac{1}{3} \delta^{ij}\bigg{)} \delta_{J_{3}'J_{3}},
\end{align}
where the pressure and shear force distributions are respectively
defined as 
\begin{align}
  p^{B}(r)=\frac{1}{6m_{B}}\frac{1}{r^{2}}\frac{1}{dr}r^{2}\frac{d}{dr}
  \tilde{D}^{B}(r), \ \ 
  s^{B}(r)=-\frac{1}{4m_{B}}r\frac{d}{dr}\frac{1}{r}\frac{d}{dr}
  \tilde{D}^{B}(r), 
\ \ \text{with} \ \
\tilde{D}^{B}(r) = \int \frac{d^{3}\Delta}{(2\pi)^{3}}
  e^{-i\bm{\Delta}\cdot \bm{r}} D^{B}(t).
 \label{eq:mech_dis}
\end{align}
From Eq.~\eqref{eq:mech_dis}, it is easy to see that the 3D von Laue
stability condition for the pressure is automatically satisfied:
\begin{align}
\int d^{3}r \, p^{B}(r)=0.
\label{eq:von_Laue}
\end{align}
It indicates that the pressure should have at least one nodal
point. In addition, the pressure and shear force distributions
automatically comply with the differential equation derived from the
total EMT conservation: 
\begin{align} 
\partial^{i}T^{ij}_{\mathrm{BF},B}(\bm{r},J_{3}',J_{3}) =
  \frac{r^{j}}{r}\left[ \frac{2}{3}\frac{ 
\partial s^{B}(r)}{ \partial r}+ \frac{2s^{B}(r)}{r} + \frac{ \partial
  p^{B}(r)}{ \partial r}\right] \delta_{J_{3}'J_{3}}= 0.
  \label{eq:diffEq}
\end{align}
It gives a number of the integral relations
between pressure and shear force. One of them is the 2D von Laue
stability condition that is derived as 
\begin{align}
  \int^{\infty}_{0} dr \; r \; \left(-\frac{1}{3}s^{B}(r) +
  p^{B}(r)\right) =0. 
  \label{eq:stability2D}
\end{align}
The combination $-\frac{1}{3}s^{B}(r) + p^{B}(r)$ carries the meaning
of the tangential force distribution. It is an eigenvalue of the
stress tensor and it must at least have one nodal point such that it 
complies with the 2D von Laue condition~\eqref{eq:stability2D}.  

Moreover, in Refs.~\cite{Perevalova:2016dln, Lorce:2018egm,
  Polyakov:2018zvc}, the local stability conditions were conjectured:
\begin{align}
\frac{2}{3}s^{B}(r) + p^{B}(r) > 0, \quad s^{B}(r)>0.
  \label{eq:mecstab1}
\end{align}
The combination $\frac{2}{3}s^{B}(r) + p^{B}(r)$ bears the meaning of
the normal force distribution and is again identified as an eigenvalue
of the stress tensor. Equation~\eqref{eq:mecstab1} implies that at any
distance $r$ the normal force should be directed outwards. This
Polyakov-Schweitzer local stability condition was examined in various 
contexts~\cite{Perevalova:2016dln, Lorce:2018egm, Polyakov:2018zvc,
  Kim:2020nug}. The value of the $D$-term form factor at zero momentum
transfer is obtained by integrating the pressure or shear-force
distributions over 3D space as   
\begin{align}
D^{B}(0) &= - \frac{4m_{B}}{15} \int d^{3}r \, r^{2} s^{B}(r)=   m_{B}
           \int d^{3}r \, r^{2} p^{B}(r), 
\label{eq:dterm}
\end{align}
and the positive shear force for any value of $r$ implies the negative
$D$-term. In addition, the positivity of the normal
forces~\eqref{eq:mecstab1} enables us to define the mechanical radius:  
\begin{align}
\langle r^{2}_{\mathrm{mech}} \rangle_{B} = \frac{\int d^{3} r~r^{2}
 \bigg{(}\frac{2}{3}s^{B}(r) + p^{B}(r)\bigg{)} }{\int d^{3} r~
  \bigg{(}\frac{2}{3}s^{B}(r) + p^{B}(r)\bigg{)}} = 
  \frac{6D^{B}(0)}{\int^{0}_{-\infty}  D^{B}(t) dt}.
  \label{eq:mecradius}
\end{align}

\subsection{Two-dimensional energy-momentum tensor distributions in
  the infinite momentum frame} 
In Refs.~\cite{Lorce:2018egm, Lorce:2020onh}, the elastic frame~(EF) 
was introduced. This frame naturally interpolates between the 2D BF
and 2D IMF for both the nucleon~\cite{Lorce:2020onh, Chen:2022smg} and the
deuteron~\cite{Lorce:2022jyi}. In the EF, the average momentum and
momentum transfer of the initial and final states are respectively
given by $P= (P_{0}, \bm{0}_{\perp}, P_{z})$ and $\Delta= (0,
\bm{\Delta}_{\perp}, 0)$. Accordingly, the EF distributions depend on
the impact parameter $x_{\perp}$ ($\bm{r}=(\bm{x}_{\perp}, \, x_{z})$)
and momentum $\bm{P}=(\bm{0},P_{z})$, where an octet baryon moves along
the $z$-direction without loss of generality. In this frame,
Eq.~\eqref{eq:25} is reduced to   
\begin{align}
T^{\mu \nu}_{\mathrm{EF},B}(\bm{x}_{\perp},P_{z},J_{3}', J_{3})=
  \int
  \frac{d^{2}\bm{\Delta}_{\perp}}{2P_{0}(2\pi)^{2}}
  e^{-i\bm{x}_{\perp} \cdot \bm{\Delta}_{\perp} } 
  \langle B, p', J_{3}' |
  \hat{T}^{\mu \nu}(0) | B, p, J_{3} \rangle \bigg{|}_{\Delta_{z}=0}.
\label{eq:6}
\end{align} 
To proceed to the IMF from the EF, we set $P_{z} \to \infty$ in
Eq.~\eqref{eq:6}. As explored in Refs.~\cite{Lorce:2018egm, 
  Freese:2021czn, Lorce:2017wkb, Schweitzer:2019kkd, Lorce:2018egm,
  Panteleeva:2021iip, Freese:2021czn}, we obtain the 2D IMF or 2D LF
EMT distributions by taking the limit $P_{z} \to \infty$. Note that we
only consider the longitudinally polarized octet baryons instead of
the transversely polarized ones. As a result, the corresponding 2D
distributions for the mass $\mathcal{E}^{B}(x_{\perp})$, angular
momentum $\rho_{J}^{(2D),B}(x_{\perp})$, pressure
$\mathcal{P}^{B}(x_{\perp})$, and shear force
$\mathcal{S}^{B}(x_{\perp})$ are obtained by the 2D inverse Fourier 
transform 
\begin{align}
& \mathcal{E}^{B}(x_{\perp}) =  m_{B} \tilde{A}^{B}(x_{\perp}), \ \ \ 
\rho^{(2D),B}_{J}(x_{\perp})= -\frac{1}{2}x_{\perp} \frac{d}{dx_{\perp}}
  \tilde{J}^{B}(x_{\perp}),  \cr 
 & \mathcal{S}^{B}(x_{\perp}) = -\frac{1}{8m_{B}}
   x_{\perp}\frac{1}{dx_{\perp}} \frac{1}{x_{\perp}}
   \frac{d}{dx_{\perp}}  \tilde{D}^{B}(x_{\perp}), \ \ \
   \mathcal{P}^{B}(x_{\perp})=
   \frac{1}{16m_{B}}\frac{1}{x_{\perp}}
   \frac{d}{dx_{\perp}}x_{\perp}\frac{d}{dx_{\perp}}\tilde{D}^{B}(x_{\perp}),   
 \label{eq:2DFF}
\end{align}
where the 2D Fourier transform of the corresponding GFFs are defined
as follows  
\begin{align}
\tilde{F}^{B}(x_{\perp}) = \int \frac{d^{2}
  \bm{\Delta}_{\perp}}{(2\pi)^{2}}e^{-i\bm{\Delta}_{\perp} \cdot
  \bm{x_{\perp}}} F^{B}(-\bm{\Delta}^{2}_{\perp}). 
\end{align}
with $F^{B}=A^{B},J^{B},D^{B}$. In the IMF, we divide the mass and
mechanical densities respectively by the Lorentz factors $P_{0}/m_{B}$
and $2m_{B}/P_{0}$ to remove the kinematical divergence and
suppression in these densities~\cite{Panteleeva:2021iip,
  Kim:2021jjf}. On the other hand, since the longitudinal boost does
not mix the longitudinal component of the angular momentum, its
distribution does not need to have an additional Lorentz
factor~\cite{Lorce:2017wkb}. In addition, since the $\mathcal{E}^{B}$
is normalized to the mass of an octet baryon, we refer to it as ``mass
distribution'' instead of the ``momentum distribution''. It is
different from the higher-twist mass distribution that arises from the
``bad'' component of the EMT current.  

These distributions~\eqref{eq:2DFF} defined in the IMF can be related
to those in the 3D BF~\eqref{eq:mass},~\eqref{eq:angmodis},
and~\eqref{eq:mech_dis} through the IMF Abel
transform~\cite{Panteleeva:2021iip, Kim:2021jjf} as follows:  
\begin{align}
\left(1 -
  \frac{\partial^{2}_{(2D)}}{4m^{2}_{B}}\right)\mathcal{E}^{B}(x_{\perp}) &=
 2  \int^{\infty}_{x_{\perp}}
   \frac{rdr}{\sqrt{r^{2}-x^{2}_{\perp}}}
   \left[
   \varepsilon^{B}(r)
   +
   \frac{3}{2}p^{B}(r)
   +
   \frac{3}{2m_{B}}\frac{1}{r^{2}}\frac{d}{dr}r
   \rho^{B}_{J}(r)\right],
   \cr 
  \rho^{(2D),B}_{J}(x_{\perp})&= 3 \int^{\infty}_{x_{\perp}}
  \frac{\rho^{B}_{J}(r)}{r}
  \frac{x^{2}_{\perp}dr}{\sqrt{r^{2}-x^{2}_{\perp}}}, \cr
\mathcal{S}^{B}(x_{\perp}) &=\int^{\infty}_{x_{\perp}} \frac{s^{B}(r)}{r}
  \frac{ x^{2}_{\perp}dr}{\sqrt{r^{2}-x^{2}_{\perp}}}, \cr
  \frac{1}{2}\mathcal{S}^{B}(x_{\perp})+\mathcal{P}^{B}(x_{\perp})
  &=\frac{1}{2}\int^{\infty}_{x_{\perp}}
  \left(\frac{2}{3}s^{B}(r)+p^{B}(r)\right) \frac{ r
  dr}{\sqrt{r^{2}-x^{2}_{\perp}}}. 
\label{eq:Abel_PS}
\end{align}
They convey the same physical meaning from the 3D
distributions to the 2D ones. Integrating those distributions over the
transversal plane $\bm{x}_{\perp}$, we obtain 
\begin{align}
\int d^{2}x_{\perp}\mathcal{E}^{B}(x_{\perp})=m_{B}A^{B}(0), \ \ \ \int
  d^{2}x_{\perp}\rho_{J}^{(2D)}(x_{\perp})=J^{B}(0),  
\label{eq:normal}
\end{align}
with the normalized form factor $A^{B}(0)=1$ and
$J^{B}(0)=1/2$, respectively. We can define the 2D mass radius in the same
manner as the 3D one, which are related each other as follows    
\begin{align}
&\langle  x^{2}_{\perp\mathcal{E}} \rangle_{B} =\frac{1}{m_{B}}\int
  d^{2}x_{\perp} x^{2}_{\perp} \mathcal{E}^{B}(x_{\perp})
  =\frac{2}{3}\langle r^{2}_{\varepsilon} \rangle_{B}
  +\frac{D^{B}(0)}{m^{2}_{B}}.
\label{eq:radius}  
\end{align} 
Interestingly, because of the Lorentz boost effects, the 2D mass
radius is associated with to the mechanical properties, i.e.,
$D$-term.  

The conservation of the EMT current also furnishes the 2D stability
condition of the nucleon. We can easily derive the 2D equilibrium
equation from the conservation of the EMT current
\begin{align}
\mathcal{P}^{B\prime}(x_{\perp})+
  \frac{\mathcal{S}^{B}(x_{\perp})}{x_{\perp}} + 
  \frac{1}{2}\mathcal{S}^{B\prime}(x_{\perp})=0, 
\label{eq:diff_eq}
\end{align}
which is similar to the 3D case. One can clearly see that
Eq.~\eqref{eq:2DFF} satisfies the equilibrium
equation~\eqref{eq:diff_eq}. In addition, the 2D pressure distribution
complies with the 2D von Laue condition and that of its lower
dimension subsystem in the octet baryon as     
\begin{align}
\int d^{2}{x}_{\perp} \mathcal{P}^{B}(x_{\perp})=0, \quad
  \int^{\infty}_{0} d{x}_{\perp} \left[\mathcal{P}^{B}(x_{\perp})-  
  \frac{1}{2}\mathcal{S}^{B}(x_{\perp})\right]=0. 
\label{eq:stab}
\end{align}
Furthermore, the Polyakov-Schweitzer local stability condition for
the 3D~\cite{Perevalova:2016dln} and 2D ~\cite{Freese:2021czn}
pressure and shear force distributions can be considered. The 3D
normal force directed outward implies that the 2D normal force 
should also be directed outward
\begin{align}
\frac{1}{2}\mathcal{S}(x_{\perp})+\mathcal{P}(x_{\perp}) >0 
\label{eq:loc_stab}
\end{align}
as shown in Eq.~\eqref{eq:Abel_PS}. 

Note that the combination $\frac{1}{2}\mathcal{S}^{B}(x_{\perp}) +
\mathcal{P}^{B}(x_{\perp})$ and
$-\frac{1}{2}\mathcal{S}^{B}(x_{\perp}) + \mathcal{P}^{B}(x_{\perp})$
mean the normal and tangential force distributions in
the 2D IMF, respectively. The positivity of the 2D normal force is
guaranteed by the fact that the Abel image of a positive function is
also positive and vice versa. This implies that the positivity of the
3D local stability condition is equivalent to the 2D ones. Thus, it 
enables us directly to relate the 3D mechanical radius to the 2D
one  
\begin{align}
\langle x^{2}_{\perp\mathrm{mech}} \rangle_{B} = \frac{\int
  d^{2}x_{\perp} x^{2}_{\perp} \left(\frac{1}{2}\mathcal{S}^{B}(x_{\perp})
  +\mathcal{P}^{B}(x_{\perp})\right)}{\int d^{2}x_{\perp}
  \left(\frac{1}{2}\mathcal{S}^{B}(x_{\perp})
  +\mathcal{P}^{B}(x_{\perp})\right)} = \frac{4D^{B}(0)}{\int^{0}_{-\infty}dt
  D^{B}(t)} = \frac{2}{3}\langle r^{2}_{\mathrm{mech}} \rangle_{B}. 
\label{eq:mech}
\end{align}

\section{Gravitational form factors of the baryon octet in the SU(3) 
chiral quark-soliton model  \label{sec:4}}
We start from the low-energy effective partition function in  
Euclidean space 
\begin{align}
\mathcal{Z}_{\chi\mathrm{QSM}}
&=\int\mathcal{D}\psi\mathcal{D}\psi^{\dagger}\mathcal{D}U
  \exp\left[\int d^{4}x\psi^{\dagger}D(U)\psi\right]\cr
&=\int\mathcal{D}U\exp\left[-S_{\mathrm{eff}}(U)\right],
\label{partition}
\end{align}
where $S_{\mathrm{eff}}$ is the effective chiral action
\begin{align}
  S_{\mathrm{eff}}(U)=-N_{c}\mathrm{Tr}\;\mathrm{ln}D(U).
\end{align}
The Dirac operator $D(U)$ is defined by 
\begin{align}
D(U)=i\slashed{\partial}+i\hat{m}+iMU^{\gamma_{5}},
\label{DiracOp.}
\end{align}
where $\hat{m}$ represents the diagonal matrix of the current quark
masses, i.e., $\hat{m}=\mathrm{diag}(m_{u},m_{d},m_{s})$, in the SU(3)
flavor space. Assuming isospin symmetry, we set the current quark
masses of  $u$- and $d$- quarks to be equal, i.e. $m_{u}=m_{d}$. So,
the matrix of the current quark masses is written as  
\begin{align}
\hat{m}=m_{0}\bm{1}+m_{8}\lambda^{8},
\end{align}
where $m_{0}$ and $m_{8}$ respectively stand for the singlet and octet
components of the current quark mass matrix. They are written as  
\begin{align}
m_{0}&=\frac{2\bar{m}+m_{s}}{3},\;\;\;
m_{8}=\frac{\bar{m}-m_{s}}{\sqrt{3}}.
\end{align}
By introducing the average current quark mass
$\bar{m}=(m_{u}+m_{d})/2$, we can rewrite the matrix of the current
quark masses in terms of $\delta m$ that will be treated
perturbatively: 
\begin{align}
\delta m=\hat{m}-\bar{m}=(m_{0}\bm{1}-\bar{m})
+ m_{8}\lambda^{8} =M_{1}\bm{1}+M_{8}\lambda^{8},
\end{align}
with $M_{1}=m_{0}-\bar{m}$ and $M_{8}=m_{8}$. $M$ stands for the
dynamical quark mass in Eq.~\eqref{DiracOp.}. Note that the original
dynamical quark mass depends on the quark momentum $k$. It is derived
from the zero-mode quark solution in the QCD instanton
vacuum~\cite{Diakonov:1985eg, Diakonov:2002fq} and plays a role of
the natural regulator for a quark loop. Since we turn off the
momentum dependence for simplicity, it is necessary to introduce an
explicit regularization scheme. Here, we use the proper-time
regularization. 

$U^{\gamma_{5}}$ denotes the SU(3) chiral field, which is defined by
\begin{align}
U^{\gamma_{5}}
&=\frac{1+\gamma_{5}}{2}U
+\frac{1-\gamma_{5}}{2}U^{\dagger}
\end{align}
with
\begin{align}
  U&=\exp[i\pi^{a}\lambda^{a}],
\end{align}
where $\pi^{a}$ represents the pseudo-Nambu-Goldstone (pNG) fields
and $\lambda^{a}$ are the Gell-Mann matrices.

Introducing the hedgehog symmetry in flavor SU(2), we regard each pion
field with $a=1,\;2,\;3$ as being aligned along the corresponding 3D
space 
\begin{align}
\pi^{a}(\bm{r})=\hat{n}^{a}P(r),
\end{align}
where $\hat{n}^{a}=x^{a}/r$ with $r=\abs{\bm{r}}$. $P(r)$ is the
profile function of the chiral soliton. It will be determined by
solving the classical equation of motion self-consistently. The SU(2)
chiral field is expressed as 
\begin{align}
U_{\mathrm{SU(2)}}^{\gamma_{5}}
&=\exp[i\gamma_{5}\bm{\hat{n}}\cdot\bm{\tau}P(r)].
\end{align}
To construct the chiral soliton in flavor SU(3), we embed the SU(2)
soliton into SU(3) one~\cite{Witten:1983tx}: 
\begin{align}
  U^{\gamma_{5}}=
\begin{pmatrix}
  U_{\mathrm{SU(2)}}^{\gamma_{5}} & 0 \\
  0 & 1
\end{pmatrix}
.
\end{align}
The Dirac Hamiltonian $h(U)$ is defined by
\begin{align}
h(U)=\gamma_{4}\gamma_{i}\partial_{i}
+\gamma_{4}MU^{\gamma_{5}}
+\gamma_{4}\bar{m}\bm{1}.
\end{align}
The corresponding eigenenergies and eigenfunctions are obtained by
diagonalizing the one-body Dirac Hamiltonian  
\begin{align}
h(U)\psi_{n}(\bm{r})= E_{n}\psi_{n}(\bm{r}),
\end{align}
where $E_{n}$ and $\psi_{n}(\bm{r})$ denote the eigenenergies and
eigenfunctions of the Hamiltonian $h(U)$, respectively. 

Since we employ the saddle-point approximation in the large $N_{c}$
limit, we can easily perform the integration over $U$ in
Eq.~\eqref{partition}. The result is simply given by the value of the
integrand at the stationary mesonic configuration, which can be found
by solving the saddle point equation $\delta S_{\mathrm{eff}}/\delta
P(r)=0$. By minimizing self-consistently the energy around the saddle
point of the pion mean field, we obtain the classical soliton mass
(see review~\cite{Christov:1995vm} in detail), which is expressed as 
\begin{align}
M_{\mathrm{sol}} = N_{c} E_{\mathrm{val}} + E_{\mathrm{sea}}.
\end{align}
Here, $E_{\mathrm{val}}$ is the energy of the discrete bound level, 
and $E_{\mathrm{sea}}$ is the sum of the Dirac-continuum energies. 

In the $\chi$QSM, the symmetrized EMT current is derived as
\begin{align}
\hat{T}^{\mu\nu}_{\mathrm{eff}}(x) =& -\frac{i}{4}\psi^\dagger (x)\left(
   i\gamma^{\mu} \overrightarrow{\partial}^{\nu}
   +i\gamma^{\nu} \overrightarrow{\partial}^{\mu} -
   i\gamma^{\mu} \overleftarrow{\partial}^{\nu} -
   i\gamma^{\nu} \overleftarrow{\partial}^{\mu} \right){\psi}(x), 
\label{eq:EMT_current}
\end{align}
and the matrix element of this EMT current can be computed as follows:
\begin{align}
&\mel{B, p',J_{3}'}{\hat{T}_{\mathrm{eff}}^{\mu\nu}(0)} {B,
  p,J_{3}}\cr 
&=N^{*}(\bm{p}')N(\bm{p}) \lim_{T\to\infty} \exp \left(-i(p'+p)
  \frac{T}{2} \right)
\int d^{3}\bm{x}d^{3}\bm{y} \exp(-i\bm{p}' \cdot \bm{y} + i\bm{p}
  \cdot \bm{x})\cr
&\quad\times \int \mathcal{D} \psi \mathcal{D} \psi^{\dagger}
  \mathcal{D}U J_{B}(\bm{y},T/2) 
T_{\mathrm{eff}}^{\mu\nu}(0)J_{B}^{\dagger}(\bm{x},-T/2)
\exp\left[\int d^{4}z\psi^{\dagger}(z)D(U)\psi(z)\right],
\end{align}
where the baryon states $\ket{B(p,J_{3})}$ and $\bra{B(p',J_{3}')}$
are respectively defined as 
\begin{align}
\ket{B,p,J_{3}}&=\lim_{x_{4}\to-\infty}\exp(ip_{4}x_{4})N(\bm{p})
\int d^{3}\bm{x} \exp(i\bm{p} \cdot \bm{x}) J_{B}^{\dagger}(\bm{x},
   x_{4}) \ket{0},\cr
\bra{B,p',J_{3}'}&=\lim_{y_{4}\to\infty}\exp(-ip_{4}'y_{4})N^{*}(\bm{p}') 
\int d^{3}\bm{y}\exp(-i\bm{p}'\cdot\bm{y})\bra{0}J_{B}(\bm{y},y_{4}).
\end{align}
$J_{B}$ represents the Ioffe-type current consisting of the $N_{c}$ 
valence quarks~\cite{Ioffe:1981kw}
\begin{align}
J_{B}(y)&=\frac{1}{N_{c}!}\epsilon_{\alpha_{1}\cdots\alpha_{N_{c}}}
\Gamma_{(TT_{3}Y)(JJ_{3}'Y_{R})}^{f_{1}\cdots f_{N_{c}}}
\psi_{f_{1}\alpha_{1}}(y)\cdots\psi_{f_{N_{c}}\alpha_{N_{c}}}(y),\cr
J_{B}^{\dagger}(x)&=\frac{1}{N_{c}!}\epsilon_{\beta{1}\dots\beta_{N_{c}}}
\Gamma_{(TT_{3}Y)(JJ_{3}Y_{R})}^{g_{1}\cdots g_{N_{c}}*}
(-i\psi^{\dagger}(x)\gamma_{4})_{g_{1}\beta{1}}\cdots
(-i\psi^{\dagger}(x)\gamma_{4})_{g_{N_{c}}\beta_{N_{c}}},
\end{align}
where the Greek and Latin indices respectively denote the color and 
spin-isospin ones. The matrices $\Gamma_{(TT_{3}Y)(JJ_{3}Y_{R})}$
carry the spin and flavor quantum  
numbers of the corresponding baryon. The right hypercharge
$Y_{R}=N_{c}/3$ with $N_{c}=3$ selects the lowest-lying
representations of the SU(3) baryons such as the baryon 
octet $(\bm{8})$ and decuplet $(\bm{10})$.

Having performed the zero-mode quantization, we obtain the
collective Hamiltonian 
\begin{align}
H_{\mathrm{coll}}=H_{\mathrm{sym}}+H_{\mathrm{sb}},
\label{eq:collect_H}
\end{align}
where
\begin{align}
H_{\mathrm{sym}}
&=M_{\mathrm{sol}}
 +\frac{1}{2I_{1}}\sum_{i=1}^{3}\hat{J}_{i}^{2}
 +\frac{1}{2I_{2}}\sum_{p=4}^{7}\hat{J}_{p}^{2}, \quad H_{\mathrm{sb}}
=\alpha D_{88}^{(8)}
 +\beta \hat{Y}
 +\frac{\gamma}{\sqrt{3}}\sum_{i=1}^{3}D_{8i}^{(8)}\hat{J}_{i}.
\end{align}
$I_{1}$ and $I_{2}$ stand for the moments of inertia. $D^{(8)}_{ab}$
denotes the SU(3) Wigner $D$ function. Three dynamical parameters
$\alpha$, $\beta$, and $\gamma$ are related to the flavor SU(3)
symmetry breaking and given as follows: 
\begin{align}
\alpha
&=\left(
  \frac{1}{\sqrt{3}}\frac{\Sigma_{\pi N}}{\bar{m}}
- \sqrt{3}\frac{K_{2}}{I_{2}}Y_{R}
 \right)M_{8},
\quad
\beta
 =\sqrt{3}\frac{K_{2}}{I_{2}}M_{8}, \quad\gamma
=-2\sqrt{3}\left(
  \frac{K_{1}}{I_{1}}
- \frac{K_{2}}{I_{2}}
\right)M_{8},
\end{align}
where $K_{1}$ and $K_{2}$ are anomalous moments of inertia and
$\Sigma_{\pi N}$ is the pion-nucleon $\Sigma$ term. The collective
wavefunction of a baryon with flavor $(YTT_{3})$ and  
spin $(Y_{R}JJ_{3})$ in the SU(3) representation $\mu$ is derived as
\begin{align}
\psi_{(YTT_{3})(Y_{R}JJ_{3})}^{(\mu)}(A)
=\sqrt{\mathrm{dim{(\mu)}}}(-1)^{J_{3}-Y_{R}/2}
D_{(YTT_{3})(Y_{R}J-J_{3})}^{(\mu)*}(A),
\end{align}
where $\mathrm{dim}(\mu)$ denotes the dimension of the representation
$\mu$. In the presence of the flavor SU(3) symmetry breaking term
$H_{\mathrm{sb}}$, the collective wavefunctions of the baryon octet
should be mixed with those in higher representations. Thus, those for
the baryon octet are derived as  
\begin{align}
\ket{B_{\bm{8}_{1/2}}}
&=\ket{\bm{8}_{1/2},B}
 +c_{\bm{\overline{10}}}^{B}\ket{\bm{\overline{10}}_{1/2},B}
 +c_{\bm{27}}^{B}\ket{\bm{27}_{1/2},B},
\end{align}
with the mixing parameters
\begin{align}
  c_{\bm{\overline{10}}}^{B}=c_{\bm{\overline{10}}}
  \mqty[
    \sqrt{5} \\
    0        \\
    \sqrt{5} \\
    0         ],\qquad
    c_{\bm{27}}^{B}=c_{\bm{27}}
    \mqty[
      \sqrt{6} \\
      3        \\
      2        \\
      \sqrt{6}  ],
\end{align}
in the basis $[N,\Lambda,\Sigma,\Xi]$. 
The coefficients $c_{\bm{\overline{10}}}$ and $c_{\bm{27}}$ are expressed 
in terms of $\alpha$ and $\gamma$
\begin{align}
  c_{\bm{\overline{10}}}=-\frac{I_{2}}{15} \left(\alpha + \frac{1}{2}
  \gamma \right),\qquad 
  c_{\bm{27}}=-\frac{I_{2}}{25} \left(\alpha - \frac{1}{6} \gamma
  \right). 
\end{align}

The matrix elements of the various components of the EMT current 
are written as in 
the large $N_{c}$ limit,
\begin{align}
\mel{B, p',J_{3}'}{\hat{T}_{\mathrm{eff}}^{00}}{B,p,J_{3}}
&=2m_{B}^{2}\left(A^{B}(t) - \frac{t}{4m_{B}^{2}} D^{B}(t) \right)
  \delta_{J_{3}'J_{3}},\cr 
\mel{B,p',J_{3}'}{\hat{T}_{\mathrm{eff}}^{ik}}{B,p,J_{3}}
&=\frac{1}{2}\left(\Delta^{i}\Delta^{k} - \delta^{ik} \bm{\Delta}^{2}
  \right) D^{B}(t)\delta_{J_{3}'J_{3}},\cr
\mel{B,p',J_{3}'}{\hat{T}_{\mathrm{eff}}^{0k}}{B,p,J_{3}}
&=-2i m_{B}\varepsilon^{klm}\Delta^{l}\hat{S}^{m}_{J_{3}'J_{3}}J^{B}(t).
\end{align}
Having considered the rotational $1/N_{c}$ and linear $m_{s}$
corrections, we obtain the final expressions of the GFFs for a octet
baryon $B$ as follows: 
\begin{align}
A^{B}(t)-\frac{t}{4m_{B}^{2}}D^{B}(t)& = \frac{1}{m_{B}} \int d^{3}r
  j_{0}(r\sqrt{-t})\varepsilon^{B}(r),\cr 
D^{B}(t)&=4m_{B}\int d^{3}r \frac{j_{2}(r\sqrt{-t})}{t}s^{B}(r),\cr 
J^{B}(t)&=3\int d^{3}r \frac{j_{1}(r\sqrt{-t})}{r\sqrt{-t}} 
\rho_{J}^{B}(r).
\end{align}
where the corresponding 3D densities $\varepsilon^{B}$, $s^{B}$, and
$\rho_{J}^{B}$ are given by 
\begin{align}
\varepsilon^{B}(r)&=\mathcal{E}(r) + \left(M_{1} + \frac{1}{\sqrt{3}}
                    M_{8}\expval{D_{88}}_{B}\right) 
  \left(\mathcal{S}(r)-2\mathcal{C}(r)\right),\cr
s^{B}(r)&=\mathcal{N}_{1}(\bm{r})
-2\left(M_{1}+\frac{1}{\sqrt{3}}M_{8}\expval{D_{88}}_{B}\right)
\mathcal{N}_{2}(\bm{r}),\cr
\rho_{J}^{B}(r)&=-\frac{1}{2I_{1}}\mathcal{I}_{1}(\bm{r})
+2M_{8}\expval{D_{83}}_{B}\left(
  \frac{K_{1}}{I_{1}}\mathcal{I}_{1}(\bm{r})-\mathcal{K}_{1}(\bm{r})
\right).
\end{align}
Here, $\langle ... \rangle_B$ denotes the matrix element of the
collective operator for a baryon state $B$. The explicit expressions
for the densities $\mathcal{E}, \ \mathcal{S}, \ \mathcal{C}, \
\mathcal{N}_{1}, \ \mathcal{N}_{2}, \ \mathcal{I}_{1}, \ \mathrm{and}
\ \mathcal{K}_{1}$ are listed in Appendix~\ref{app:A}. 

\section{Results and discussion \label{sec:5}}
Before we discuss the numerical results for the EMT distributions and
the GFFs of the baryon octet, we first describe how the parameters in
the $\chi$QSM are fixed. Since the quark loops cause the
divergences, we need to regularize them by
introducing a cutoff mass $\Lambda$. It is fixed by reproducing the
experimental data on the pion decay constant $f_{\pi}=93$ MeV. All
other quark loops such as the pion mass are tamed by this fixed value
of $\Lambda$. The current $u$- and $d$-quark masses are determined to
be $m_{u}=m_{d}= 17.6~\mathrm{MeV}$ by reproducing the pion mass
$m_{\pi}=140$ MeV~(see Ref.~\cite{Christov:1995vm, Goeke:2005fs} for
detail). In principle the only free parameter of the $\chi$QSM is the  
dynamical quark mass $M$. It is fixed by computing the various nucleon
form factors and mass splitting between light
baryons~\cite{Christov:1995vm}. The most preferable value is found  
to be $420$~MeV. Since we are interested in the effect of
flavor SU(3) symmetry breaking on the GFFs and EMT distributions, we
need to fix the value of the strange current quark mass. 
We employ it as $m_{s}=180~\mathrm{MeV}$, which 
describes the mass splitting of both the light and singly heavy 
baryons~\cite{Christov:1995vm, Kim:2018xlc} very well. Note that we
consider the linear $m_{\mathrm{s}}$ corrections.  

The rotational and translational zero modes yield the $1/N_{c}$
corrections. While the translational corrections give an overall shift
of the mass spectra of the baryons, the rotational corrections 
make the nucleon and $\Delta$ baryon states split. In the octet and
decuplet representations, the linear $m_{s}$ corrections take charge
of splitting hyperon states. In this work, while we neglect the $1/N_{c}$
corrections\footnote{Since the $N_{c}$ leading contribution to the
  spin distribution arises from the $1/N_{c}$ rotational corrections,
  we take into account these corrections only for the spin
  distribution.}, we aim at scrutinizing the effects of the $m_{s}$
corrections on the GFFs and the EMT distributions of the baryon
octet. In the following subsection, we will exhibit the $m_{s}$
corrections to the EMT distributions of the baryon octet.

\subsection{Energy density}
We start with examining the energy density $\varepsilon^{B}(r)$. It
comes from the temporal component of the EMT density
$T^{00}_{\mathrm{BF},B}$ defined as the Fourier transform of the mass
form factor~\eqref{eq:mass}. If we integrate $\varepsilon^{B}(r)$ for
a octet baryon over the 3D space, then we obtain the mass of the
corresponding baryon 
 \begin{align}
\int d^{3}r \, \varepsilon^B(r)&= \int d^{3}r \, \left(
  \mathcal{E}(r) + \left[M_{1} + \frac{1}{\sqrt{3}} M_{8}
  \expval{D_{88}}_B \right]   \left[\mathcal{S}(r) - 2\mathcal{C}(r)
                                 \right] \right)  \cr
  &= M_{\mathrm{sol}} + \left[M_{1} + \frac{1}{\sqrt{3}} M_{8}
    \expval{D_{88}}_B \right]  \frac{\Sigma_{\pi N}}{\bar{m}}
    :=m_{B} 
  \label{eq:mass_nor}
\end{align}
with
\begin{align}
\int d^{3} r \, \mathcal{E}(r) = M_{\mathrm{sol}}, \quad \int d^{3} r
  \, \mathcal{S}(r) = \frac{\Sigma_{\pi N}}{\bar{m}}, \quad \int d^{3}
  r \, \mathcal{C}(r) =0. 
\end{align}
Equation~\eqref{eq:mass_nor} coincides with the expression for the
collective Hamiltonian~\eqref{eq:collect_H}. The form
factor $A^{B}(t)$ is naturally normalized as  
\begin{align}
  A^{B}(0)=\frac{1}{m_{B}}\int d^{3}r \, \varepsilon^{B}(r) = 1.
\end{align}
We list the values of the octet baryon masses in Table~\ref{tab:1}. 
We find an obvious fact that the mass of the baryon becomes larger as
the number of the strange quark in the valence level increases. 
One may wonder why the nucleon and hyperon masses are deviated from
the experimental data. Firstly, the classical soliton mass is
typically overestimated at around $\sim 300~\mathrm{MeV}$. Its origin
may be understood as translational zero-mode
corrections~\cite{Pobylitsa:1992bk} and mesonic $1/N_c$
corrections. Secondly, the mass splittings of the hyperons are not
described well without the $m_{s}$ contributions 
mixed with the rotational ones, i.e., $\mathcal{O}(N^{0}_{c},
m_{s})$. Introducing this mixed contribution~\cite{Blotz:1992pw,
  Christov:1995vm}, one can describe the mass splitting of the SU(3)
baryons very well. In the current work, hoswever, since we consider
the baryon masses by using the mass form factors, it is
technically very complicated to take into account such corrections. 
One may encounter triple sums of the quark states. Thus, we
restrict ourselves to examine how the explicit flavor SU(3) symmetry
breaking affects the EMT distribution and stability conditions. 
\begin{table}[htb] 
\caption{Masses of the baryon octet} 
\begin{center}
  \renewcommand{\arraystretch}{1.5}
\scalebox{1.05}{%
\begin{tabular}{c | c c | c} 
  \hline
  \hline
   Baryon & $M_{\mathrm{sol}}$ [MeV]  &   $m_{s}$ correction [MeV]  &
      $m_{B}$  [MeV]  \\ 
  \hline
  $N$         &  $1256$  &  $95 $    &  $1351$ \\
  $\Lambda$   &  $1256$  &  $122$    &  $1378$ \\      
  $\Sigma$    &  $1256$  &  $149$    &  $1405$ \\     
  $\Xi$       &  $1256$  &  $162$    &  $1418$ \\      
  \hline
  \hline
\end{tabular}}
\end{center}
\label{tab:1}
\end{table}

\begin{figure}[htp]
  \centering
  \includegraphics[scale=0.17]{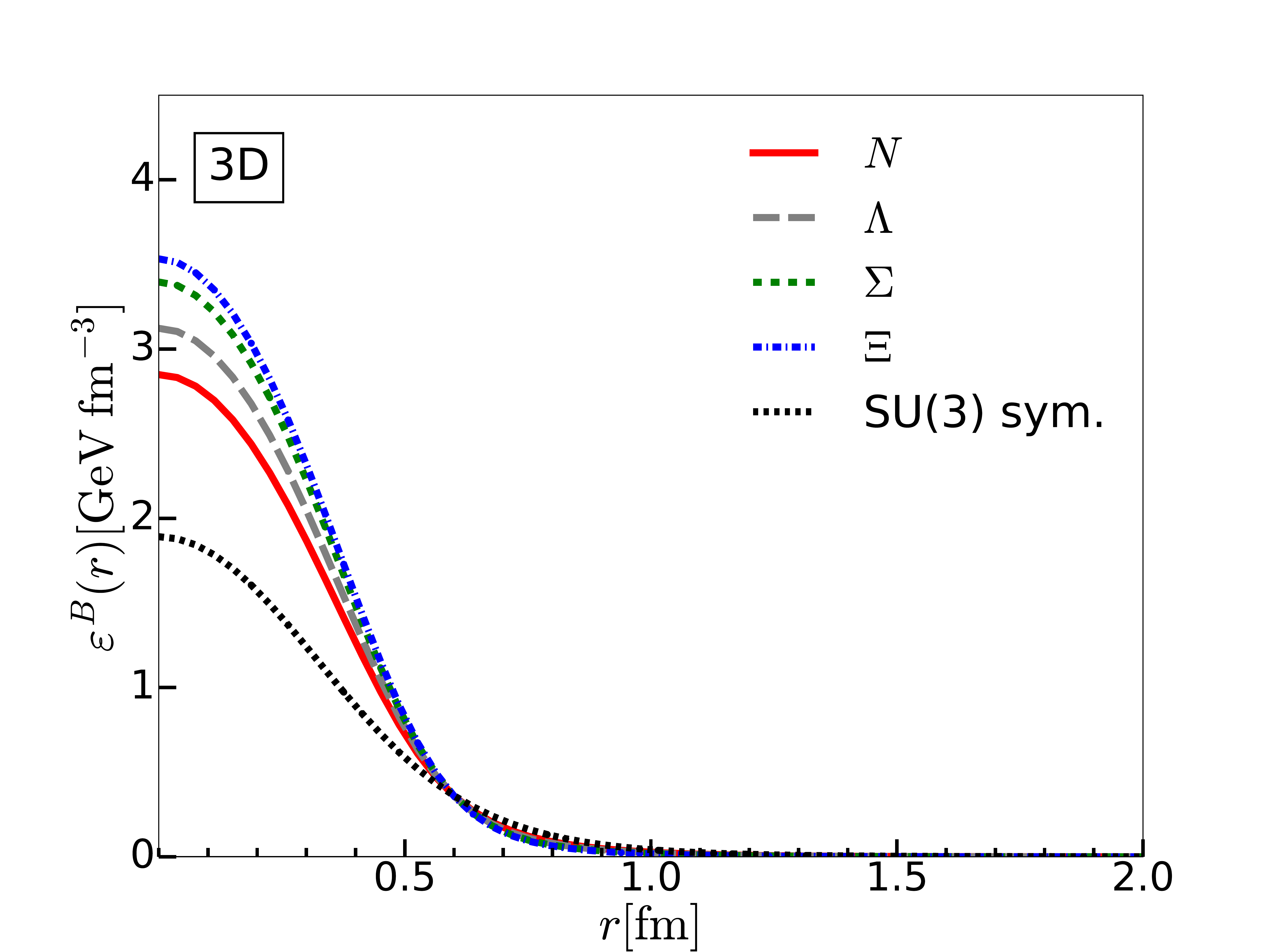}
  \includegraphics[scale=0.17]{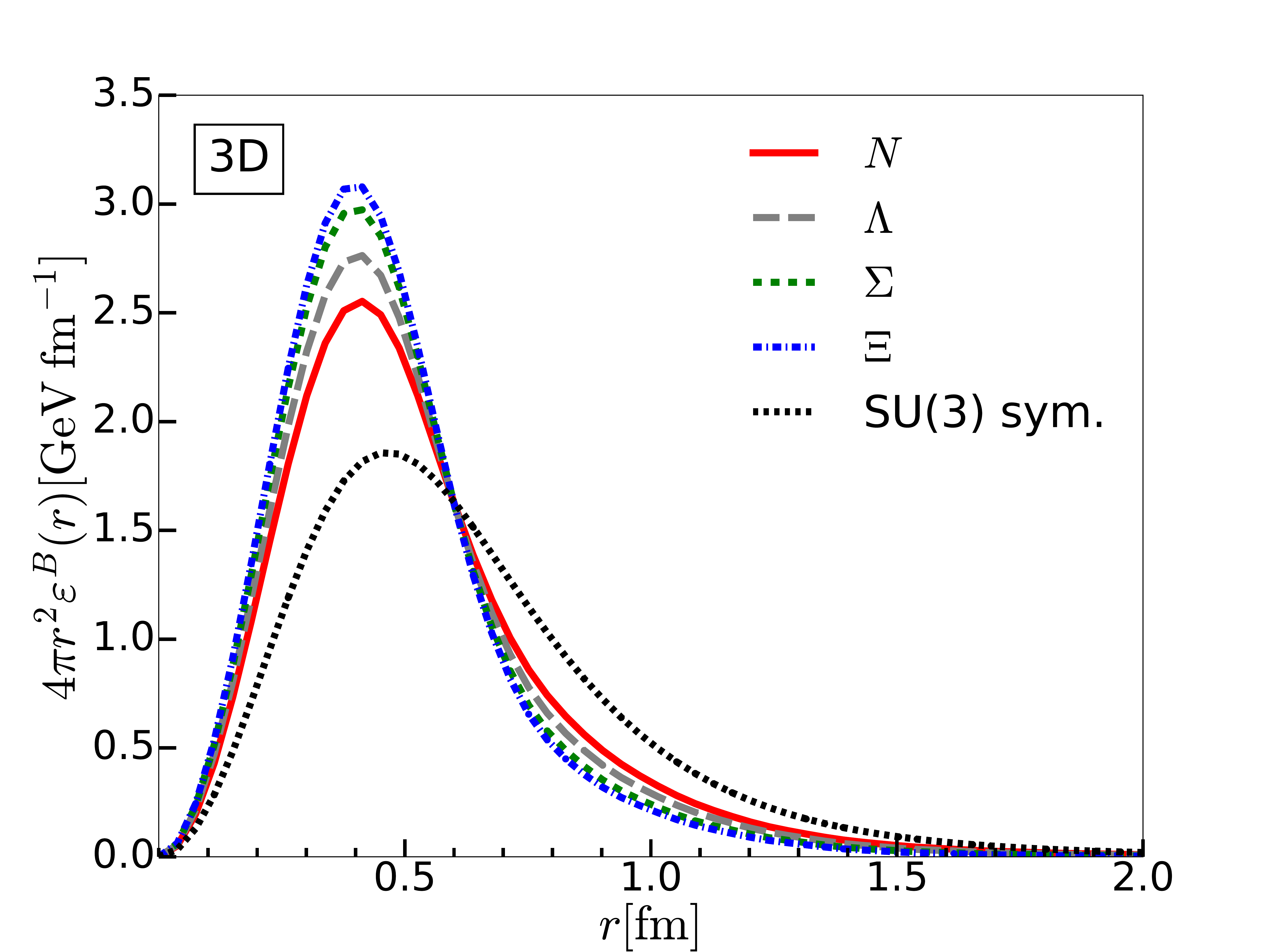}
  \includegraphics[scale=0.17]{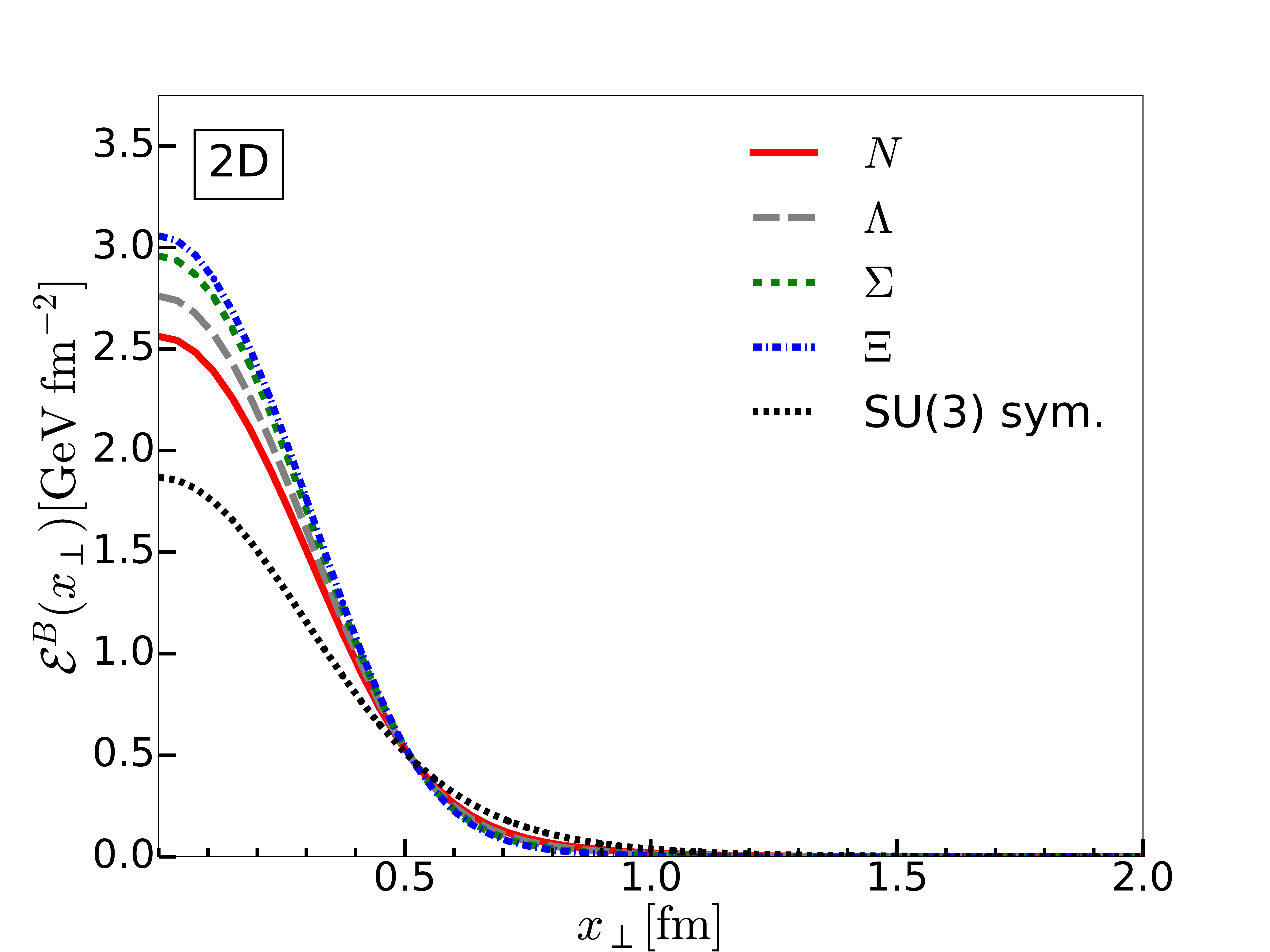}
  \includegraphics[scale=0.17]{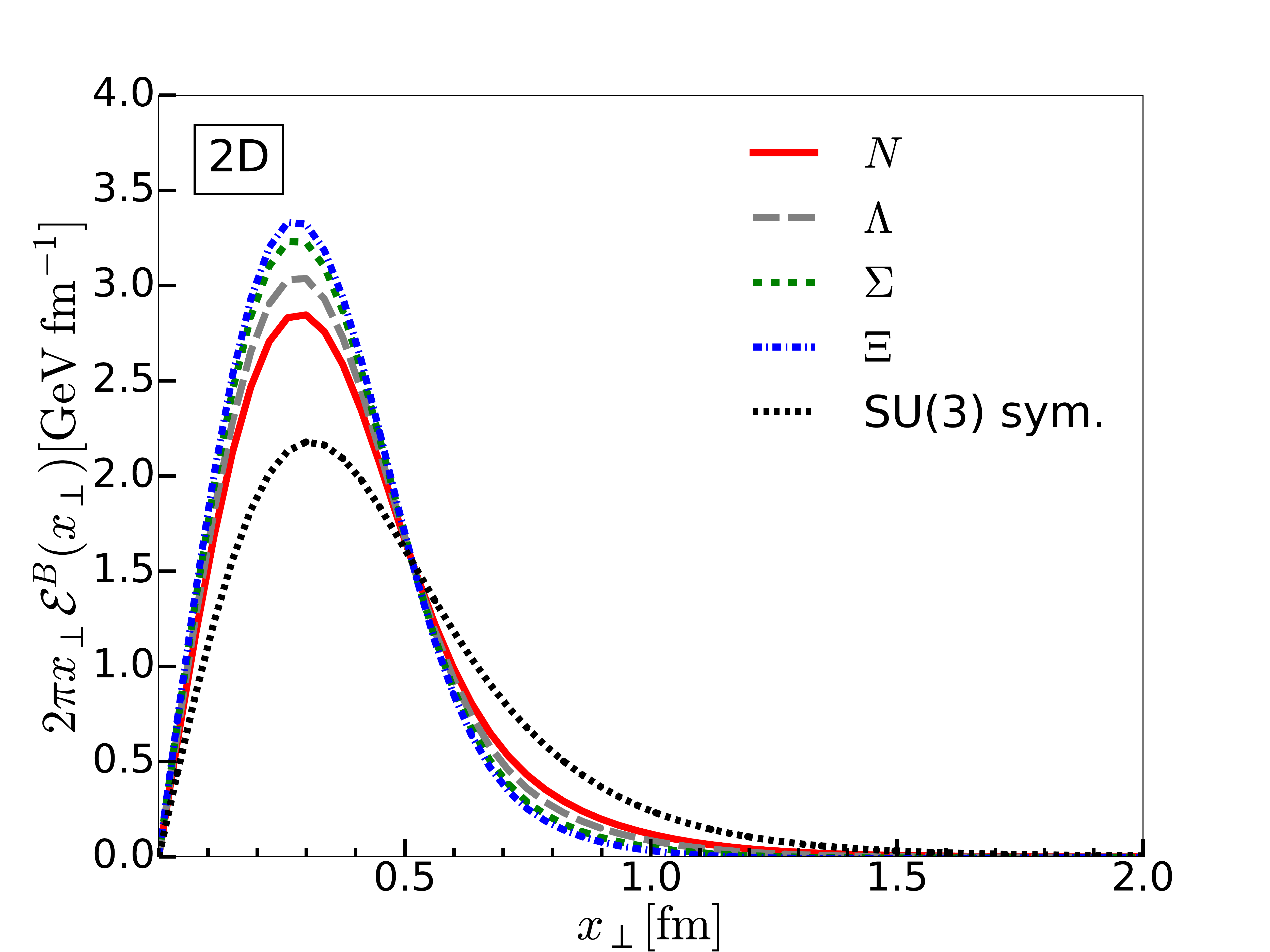}
  \includegraphics[scale=0.17]{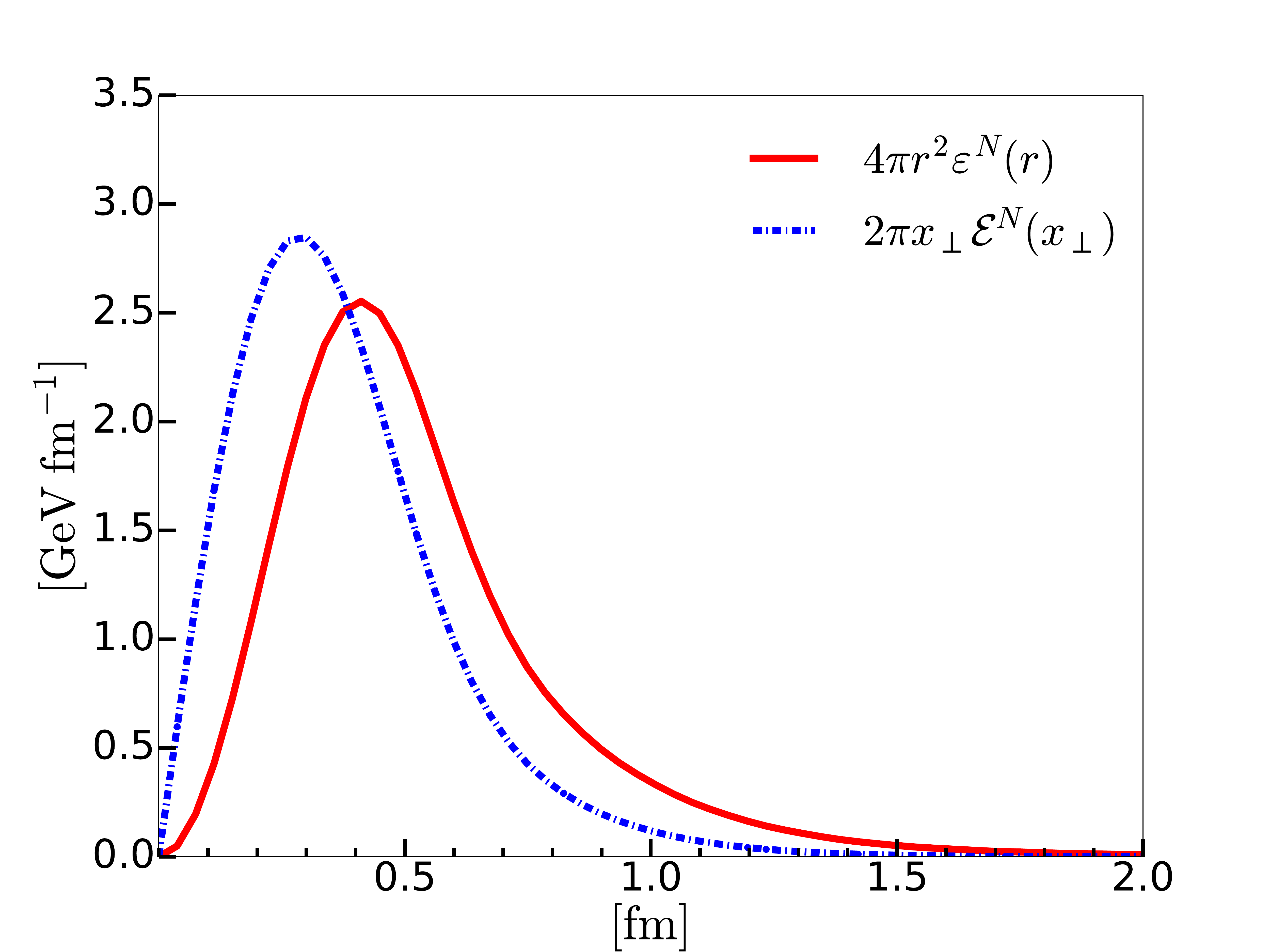}
\caption{3D BF and 2D IMF mass distributions (upper- and middle-left
  panels) and $r^{2}$-weighted ones (upper- and middle-left panels) of
  the baryon octet and the classical nucleon, and comparision (lower
  panel) between the 3D BF and 2D IMF ones of the nucleon. The
  solid~(red), long-dashed~(grey), short-dashed~(green),
  dashed-dotted~(blue), and dotted~(black) curves denote
  $\varepsilon^{B}(r)$ for the $N$, $\Lambda$, $\Sigma$, $\Xi$, and
  classical nucleon, respectively. } 
\label{fig:1}
\end{figure}
In the upper-left and -right panels of Fig.~\ref{fig:1}, we show
the 3D mass distributions and $r^{2}$-weighted ones of the baryon
octet. The magnitude of the mass distribution over $r$ becomes larger
as the strangeness increases, which indicates that the 
mass of the corresponding octet baryon also becomes larger with the
strangeness increased. Interestingly, as the mass of the 
baryon grows, the shape of the distribution gets more closely packed. 
This implies that the size of a heavier particle becomes more
compact. We will observe this fact soon. 
In the IMF, we draw the corresponding 2D distributions in the
middle-left and -right panels of Fig.~\ref{fig:1} by implementing the
IMF Abel transform~\eqref{eq:Abel_PS}. We find that the 2D mass
distributions get closer to the origin of the position space, compared
with the 3D ones. In the lower panel of Fig.~\ref{fig:1}, one can
clearly see a narrower 2D distribution for the nucleon as a
representative of the baryon octet. To quantify how strongly the
mass distributions stretch out over position space, we introduce both
the 2D and 3D mass radii in Eqs.~\eqref{eq:mass_rad}
and~\eqref{eq:radius}, respectively. They are found to be 
\begin{align}
\expval{r_{\varepsilon}^{2}}_{N}&=0.31~[\mathrm{fm}^{2}], \quad 
\expval{r_{\varepsilon}^{2}}_{\Lambda}=0.26~[\mathrm{fm}^{2}], \quad 
\expval{r_{\varepsilon}^{2}}_{\Sigma}=0.20~[\mathrm{fm}^{2}], \quad 
\expval{r_{\varepsilon}^{2}}_{\Xi}=0.17~[\mathrm{fm}^{2}],\cr
\expval{x_{\perp \mathcal{E}}^{2}}_{N}&=0.14~[\mathrm{fm}^{2}], \quad 
\expval{x_{\perp \mathcal{E}}^{2}}_{\Lambda}=0.11~[\mathrm{fm}^{2}], \quad 
\expval{x_{\perp \mathcal{E}}^{2}}_{\Sigma}=0.07~[\mathrm{fm}^{2}], \quad 
\expval{x_{\perp \mathcal{E}}^{2}}_{\Xi}=0.05~[\mathrm{fm}^{2}].
\end{align}
As mentioned above, the heavier hyperons are more compact than the
lighter ones: 
\begin{align}
&\expval{r_{\varepsilon}^{2}}_{N}>\expval{r_{\varepsilon}^{2}}_{\Lambda}>
\expval{r_{\varepsilon}^{2}}_{\Sigma} >
\expval{r_{\varepsilon}^{2}}_{\Xi} ,\cr
&\expval{x_{\perp \mathcal{E}}^{2}}_{N}> \expval{x_{\perp
    \mathcal{E}}^{2}}_{\Lambda}> \expval{x_{\perp
    \mathcal{E}}^{2}}_{\Sigma} > \expval{x_{\perp \mathcal{E}}^{2}}_{\Xi}.
\end{align}
Thus, we draw an important conclusion that the heavier octet baryon is
energetically  more compact in both the 3D BF and the 2D
IMF. Thus, this is not changed by the IMF Abel transformation. 
Interestingly, including the $m_{s}$ corrections 
to the mass distributions results in the energetically more compact
nucleon and yields a larger value at the center of the distribution,
compared to the classical nucleon (see Table~\ref{tab:2} and
~\ref{tab:3}).

\subsection{Angular momentum density}
Before we discuss the numerical results for the angular momentum
density, we want to remark on the total angular momentum in the
$\chi$QSM. In principle, the total angular momentum consists of the
orbital angular momentum and spin of the quarks and gluons. 
In the $\chi$QSM, however, the gluonic degrees of freedom are
absent or integrated out through the instanton vacuum. It
implies that the baryon spin arises only from the orbital angular 
momentum and spin of the quarks and antiquarks. The intrinsic 
quark spin contribution or the singlet axial charge ($g_{A}^{0}$) can
be obtained from the experimental data on the structure function
$g_{1}$ extracted from the experimental data on polarized deep
inelastic scattering (DIS). In a series of polarized 
DIS experiments, the quarks carry a small fraction of the nucleon
spin, i.e. $g^{0}_{A}\sim  0.33$~\cite{Aidala:2012mv}. 
It leads one to posit that the orbital angular momentum of the quarks
and gluons, and the spin of the gluon may have considerable
contributions to the nucleon spin. The future EIC project may shed
light on the spin structure of the nucleon. In the $\chi$QSM, we can
explicitly decompose the intrinsic spin and orbital angular momentum
of the quarks. The singlet axial charge was already studied in this
model and found to be $g^{0}_{A}\sim 0.44$~\cite{Christov:1995vm,
  Suh:2022atr}. In this subsection, we will demonstrate that the
missing part of the nucleon spin originates solely from the
relativistic orbital motion of the quarks with the effects of the
flavor SU(3) symmetry breakdown. 

The spin density $\rho^{B}_{J}(r)$ arises from the mixed component of
the EMT current $T^{0i}_{\mathrm{BF},B}$, and is normalized as the
spin of a octet baryon: 
\begin{align}
J^{B}(0)=\int d^{3}r \rho^{B}_{J}(r) = \int d^{3}r \,
  \left(-\frac{1}{2I_{1}}\mathcal{I}_{1}(\bm{r}) 
  +2M_{8}\expval{D_{83}}_{B}\left[
    \frac{K_{1}}{I_{1}}\mathcal{I}_{1}(\bm{r})-\mathcal{K}_{1}(\bm{r})
  \right] \right)= \frac{1}{2}
\end{align}
with
\begin{align}
\int d^{3}r  \, \mathcal{I}_{1}(\bm{r})=-I_1, \quad \int d^{3}r \,
  \mathcal{K}_{1}(\bm{r})=-K_1. 
\end{align}
If we integrate the angular momentum distributions of the octet
baryon over 3D space, then we obtain the corresponding spin $1/2$ (see
Table~\ref{tab:2}). As we explained already, one of the 
interesting results in the $\chi$QSM is that the total angular
momentum can be decomposed into the orbital
angular momentum and spin carried by quarks as follow:  
\begin{align}
  J^{B}(0)&=L^{B}+\frac{1}{2}g_{A}^{0,B}.
\label{eq:70}
\end{align}
The explicit proof is provided in Appendix~\ref{app:B}. Using
Eq.~\eqref{eq:70}, we can separately define the orbital angular
momentum and spin distributions.  
\begin{figure}[htp]
  \centering
  \includegraphics[scale=0.17]{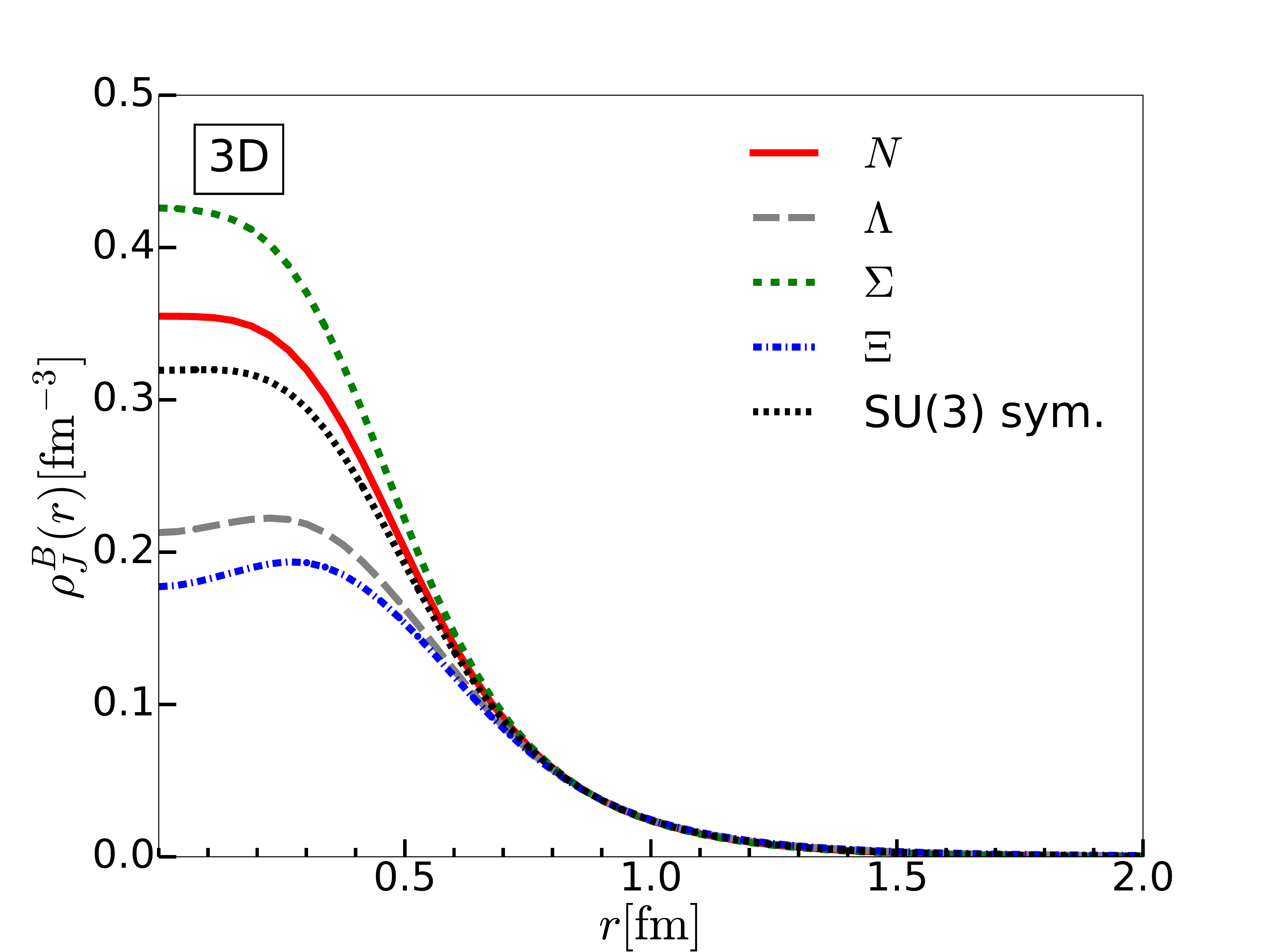}
  \includegraphics[scale=0.17]{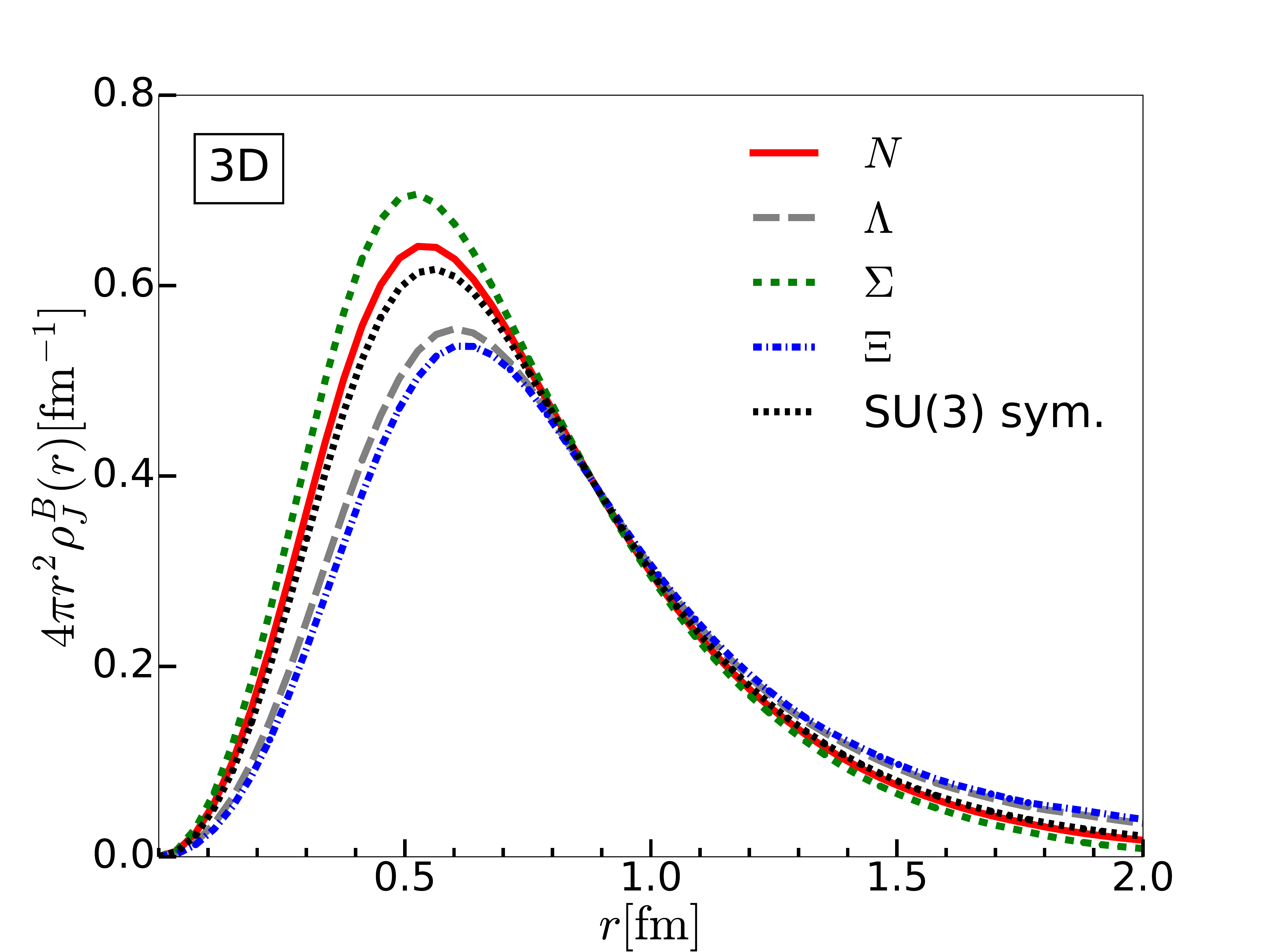}
  \includegraphics[scale=0.17]{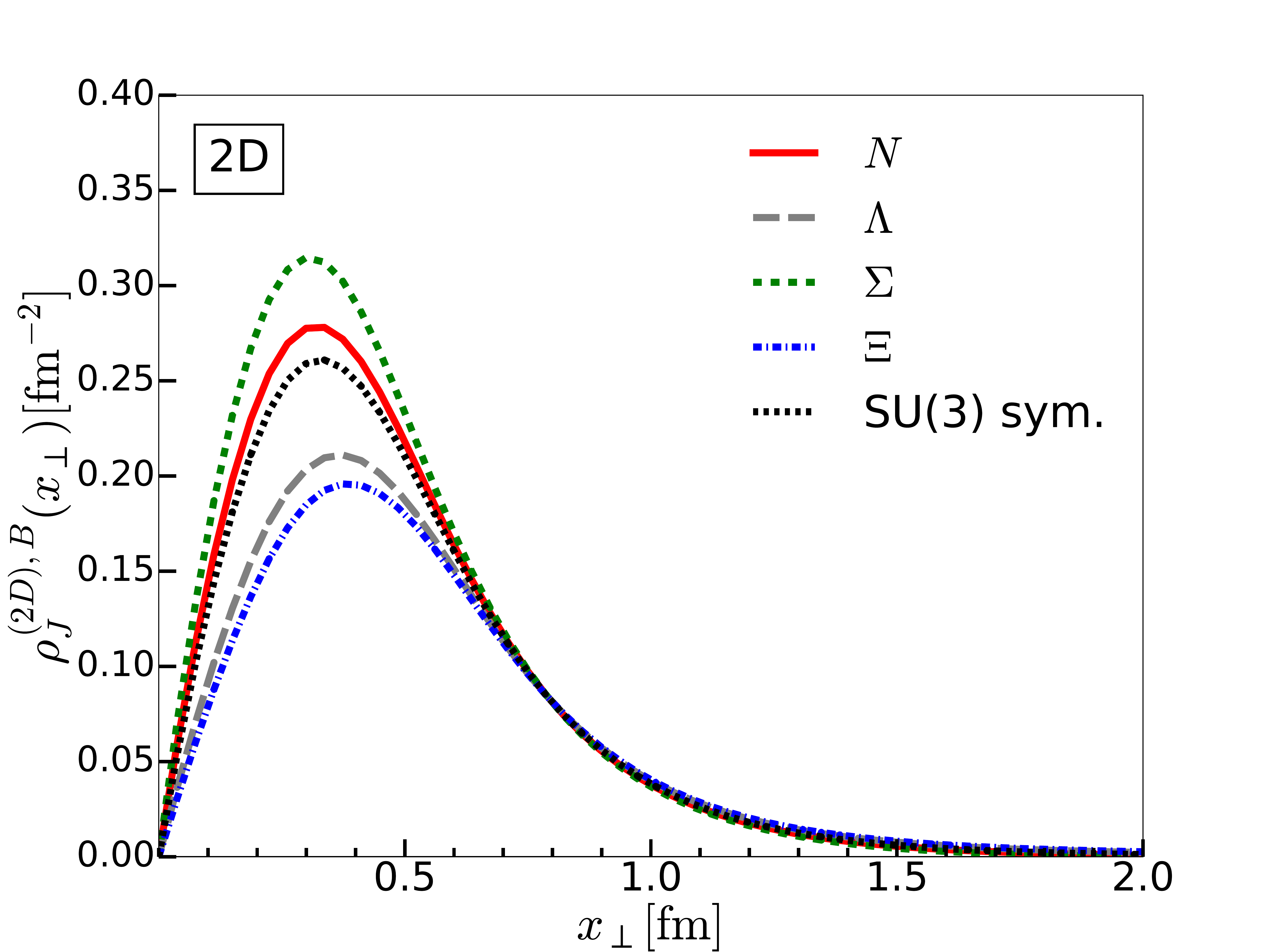}
  \includegraphics[scale=0.17]{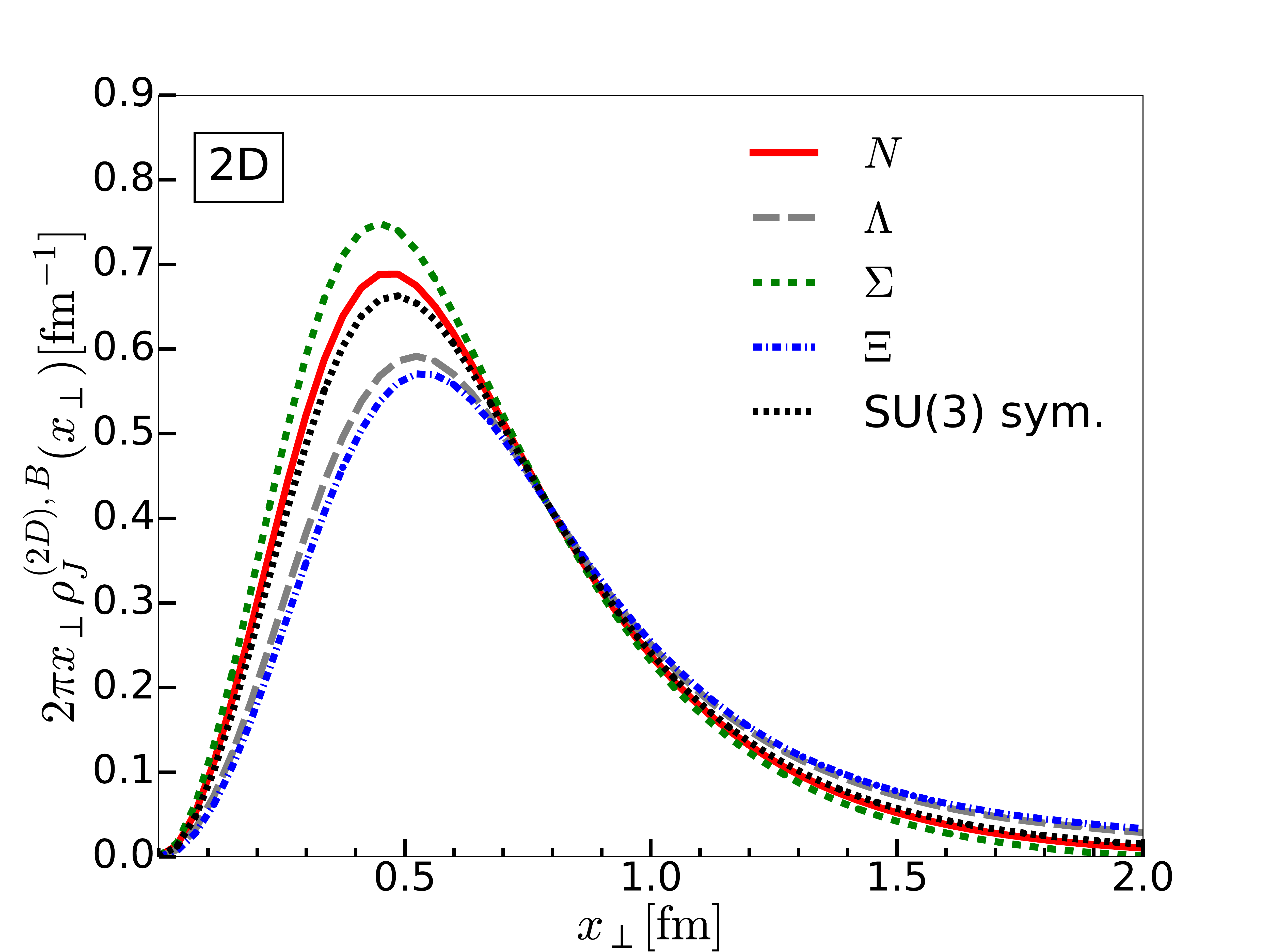}
  \includegraphics[scale=0.17]{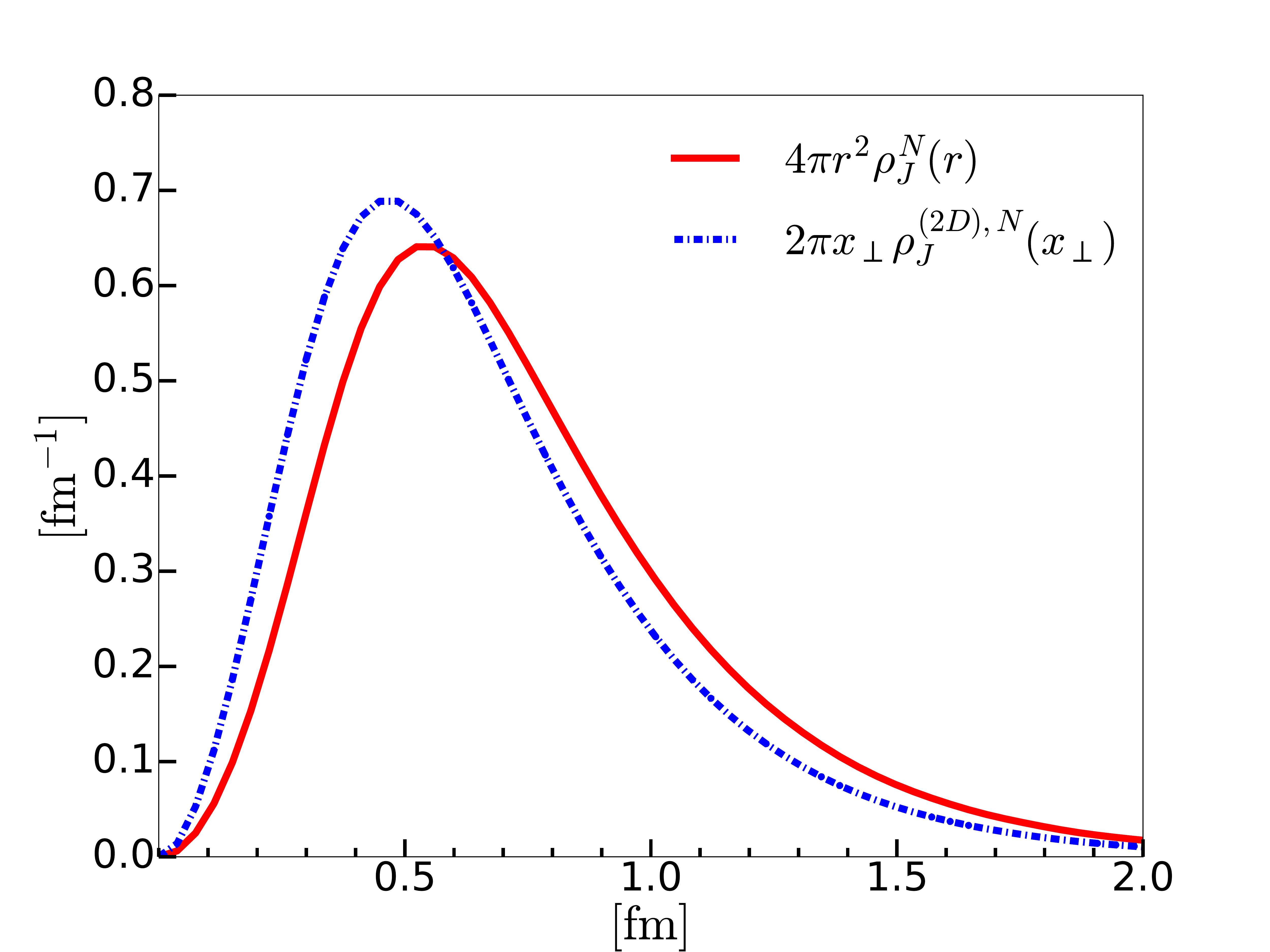}
\caption{3D BF and 2D IMF total angular momentum distributions (upper-
  and middle-left panels) and $r^{2}$-weighted ones (upper- and
  middle-right panels) of the baryon octet and the nucleon with the
  flavor SU(3) symmetry, and comparision (lower panel) between the 3D
  BF and 2D IMF ones of the nucleon. The notations are the same as in
  Fig.~\ref{fig:1}.} 
\label{fig:2}
\end{figure}
We first depict the total angular momentum distributions and
$r^{2}$-weighted ones in the 3D BF for the baryon octet in the
upper-left and -right panels of Fig.~\ref{fig:2}, respectively. The
distributions of the baryon octet are split up with respect to
that with the flavor SU(3) symmetry. While the strengths of the
distributions are enhanced for the nucleon and
$\Sigma$ baryon, those for the $\Xi$ and $\Lambda$ baryons are
diminished. However, the normalizations of the distributions are not
changed at all. In Table~\ref{tab:2}, we list the properly
normalized values of the $J^{B}(0)$ for the baryon octet. In the
middle-left, middle-right, and lower panels of Fig.~\ref{fig:2}, we
also present the same distributions in the 2D IMF. Similar to the 2D
mass distributions, the 2D IMF angular momentum distributions are
generically tilted inward, compared to the 3D BF distribution. 

\begin{figure}[htp]
  \centering
  \includegraphics[scale=0.17]{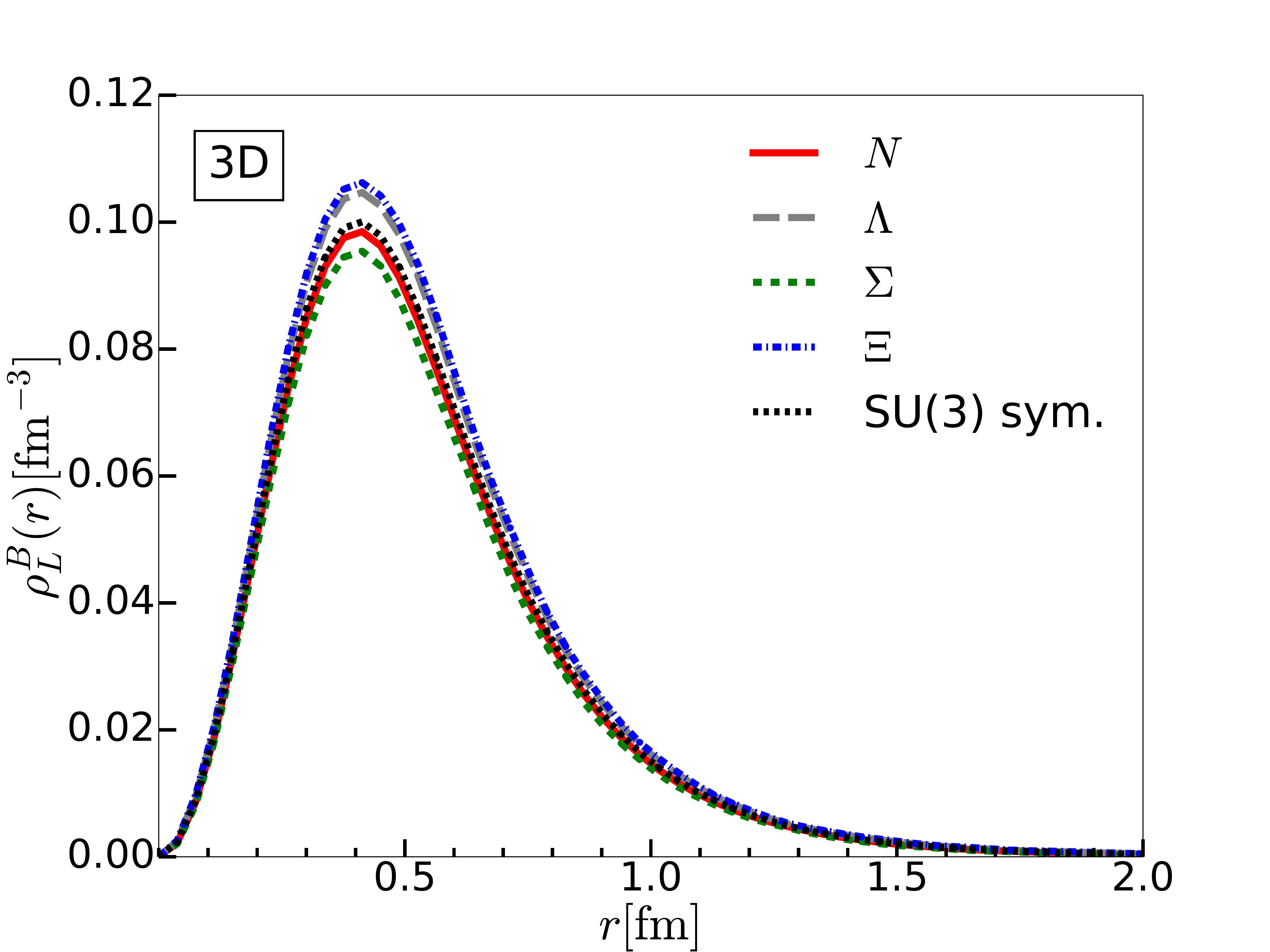}
  \includegraphics[scale=0.17]{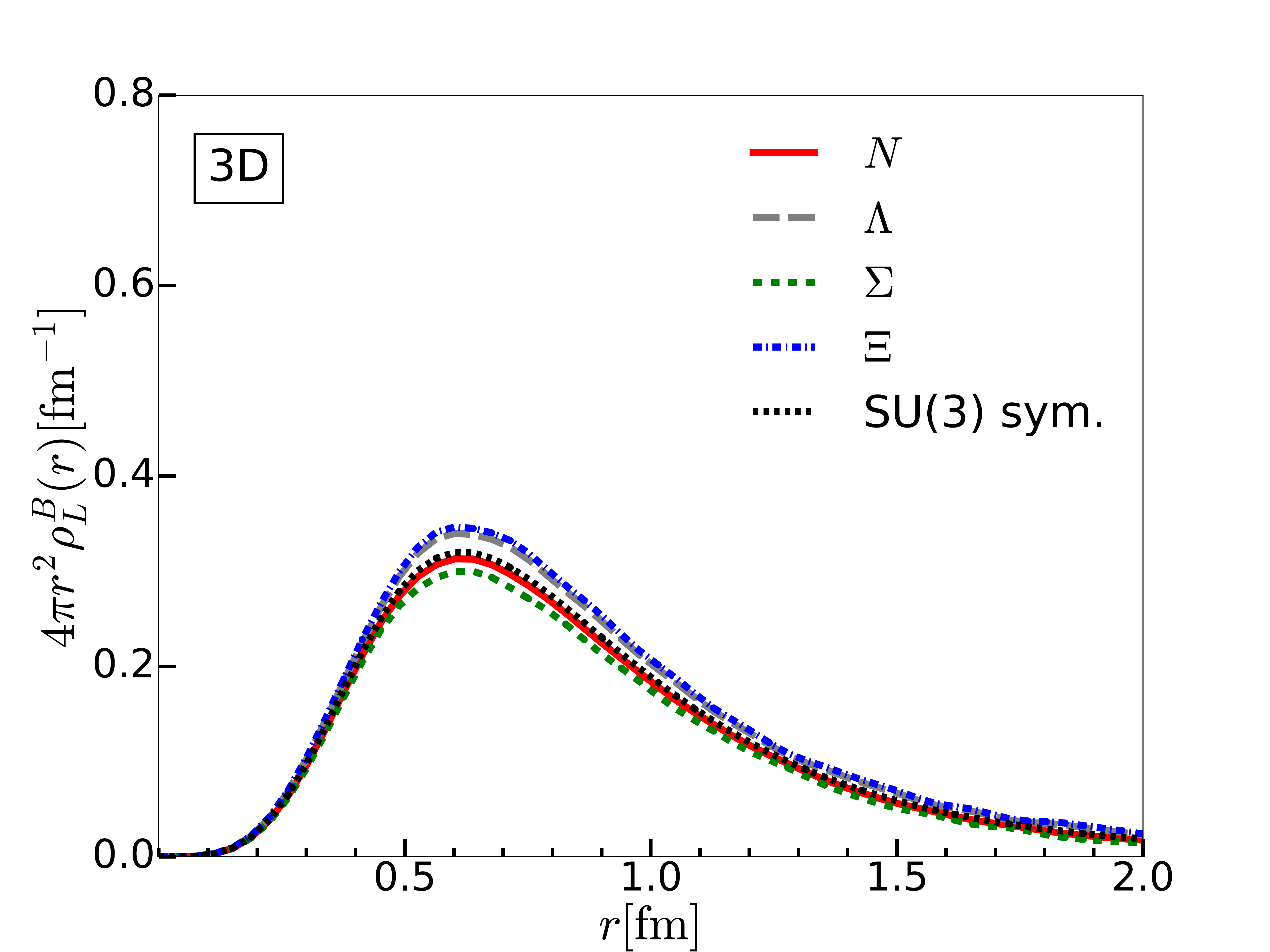}
  \includegraphics[scale=0.17]{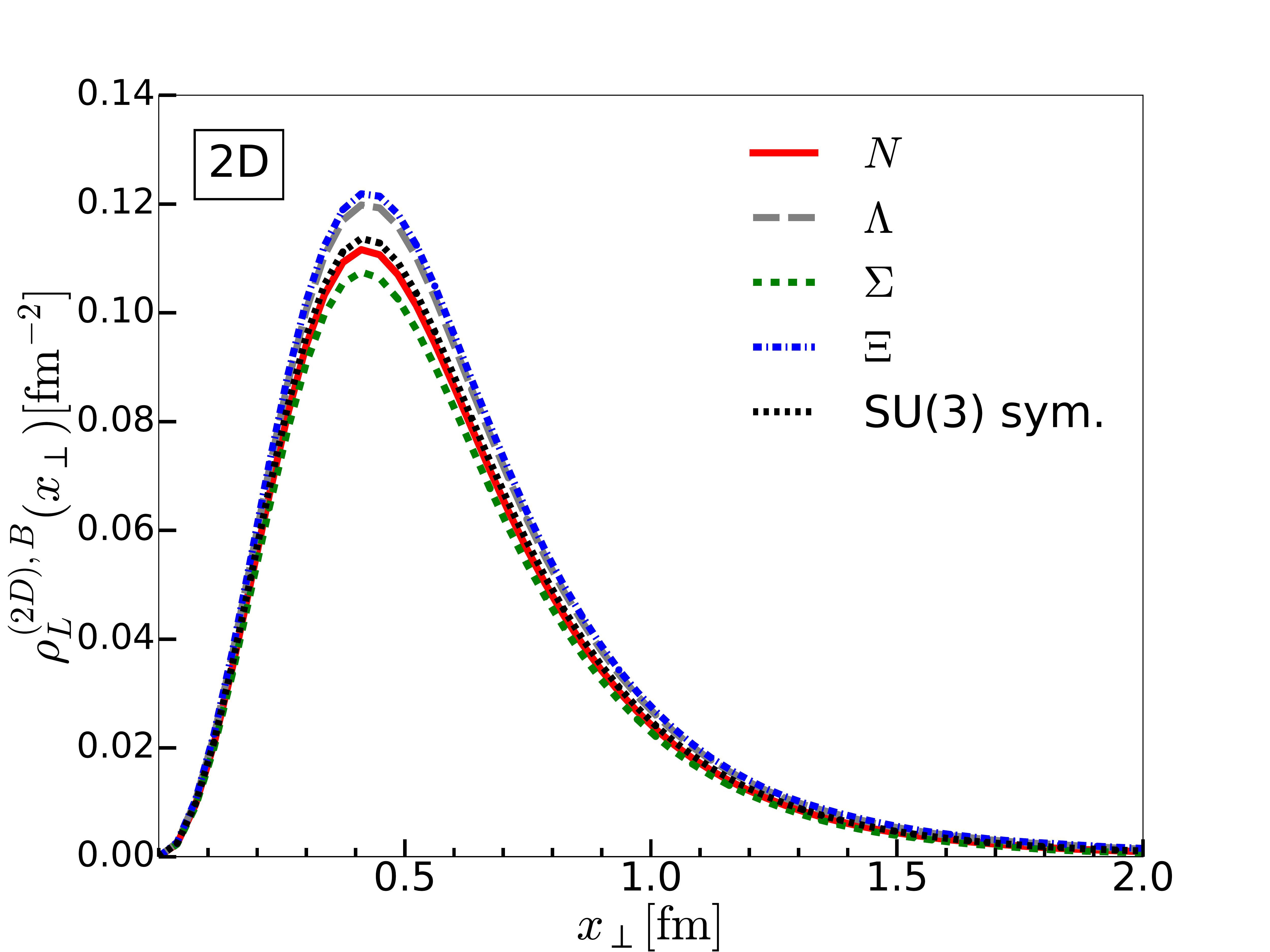}
  \includegraphics[scale=0.17]{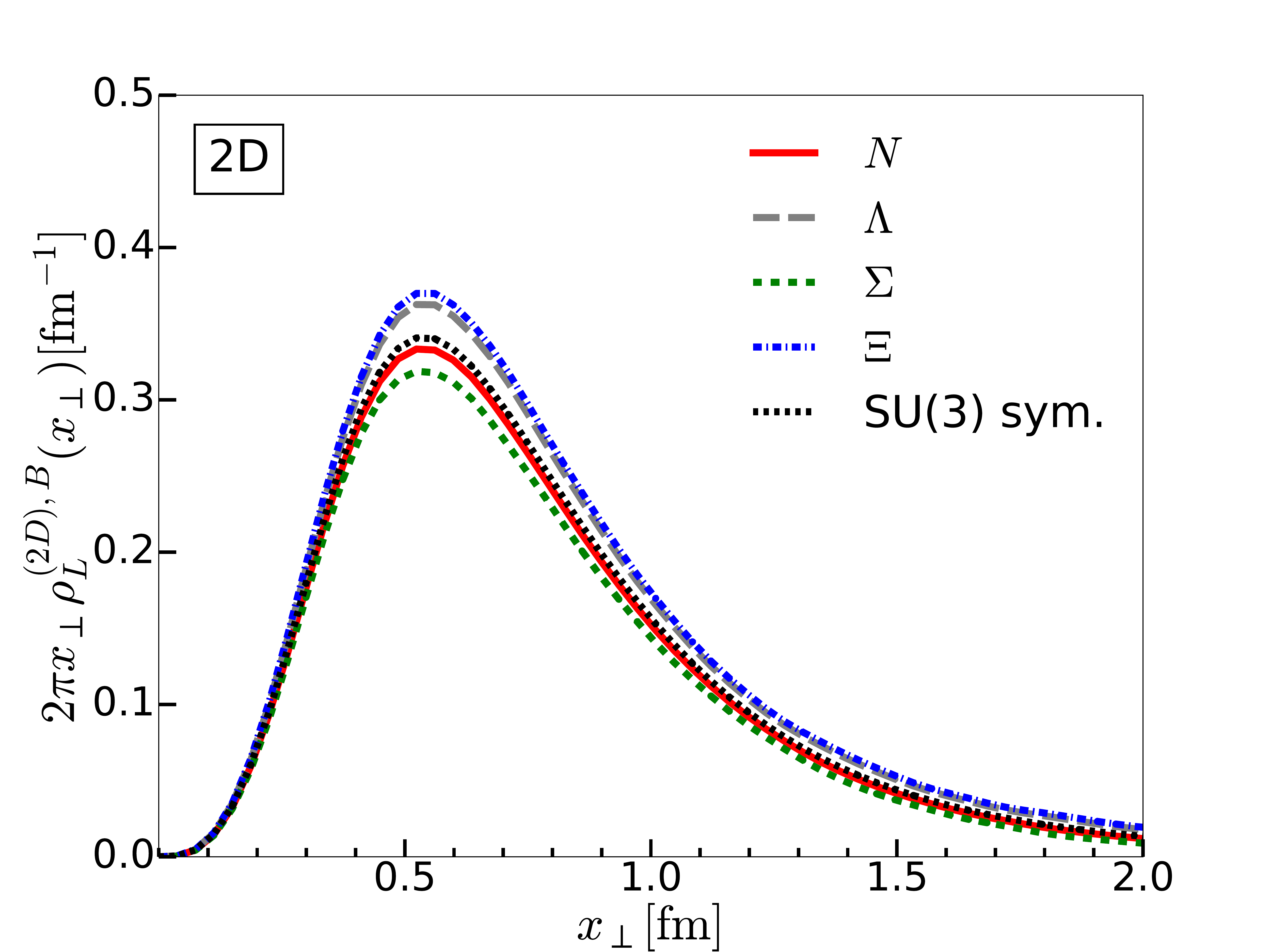}
  \includegraphics[scale=0.17]{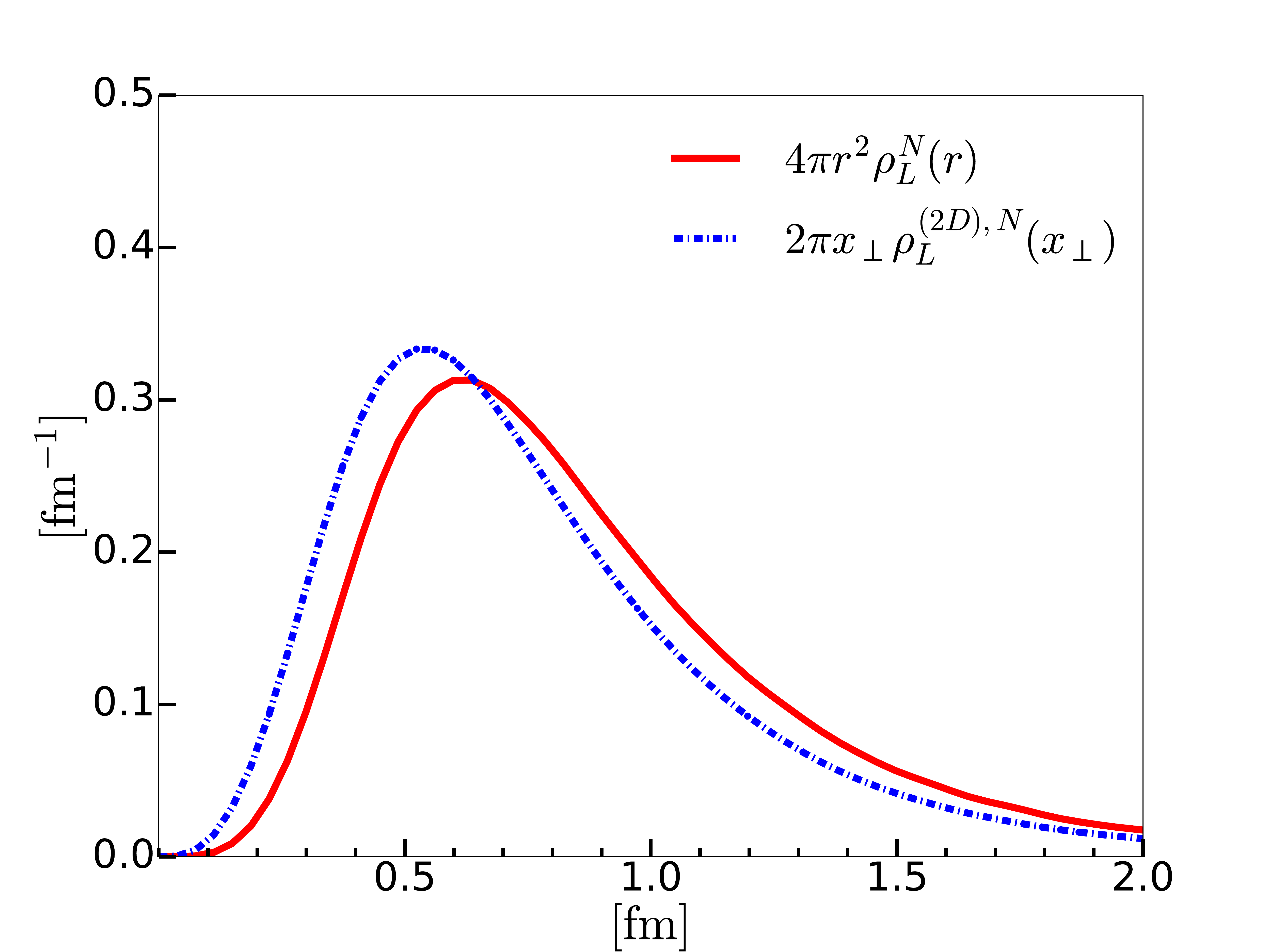}
\caption{3D BF and 2D IMF orbital angular momentum distributions
  (upper- and middle-left panels) and $r^{2}$-weighted ones (upper-
  and middle-right panels) of the baryon octet and the nucleon with
  the flavor SU(3) symmetry, and comparision (lower panel) between the
  3D BF and 2D IMF ones of the nucleon. The notations are the same as in
  Fig.~\ref{fig:1}.}  
\label{fig:3}
\end{figure}
\begin{figure}[htp]
  \includegraphics[scale=0.17]{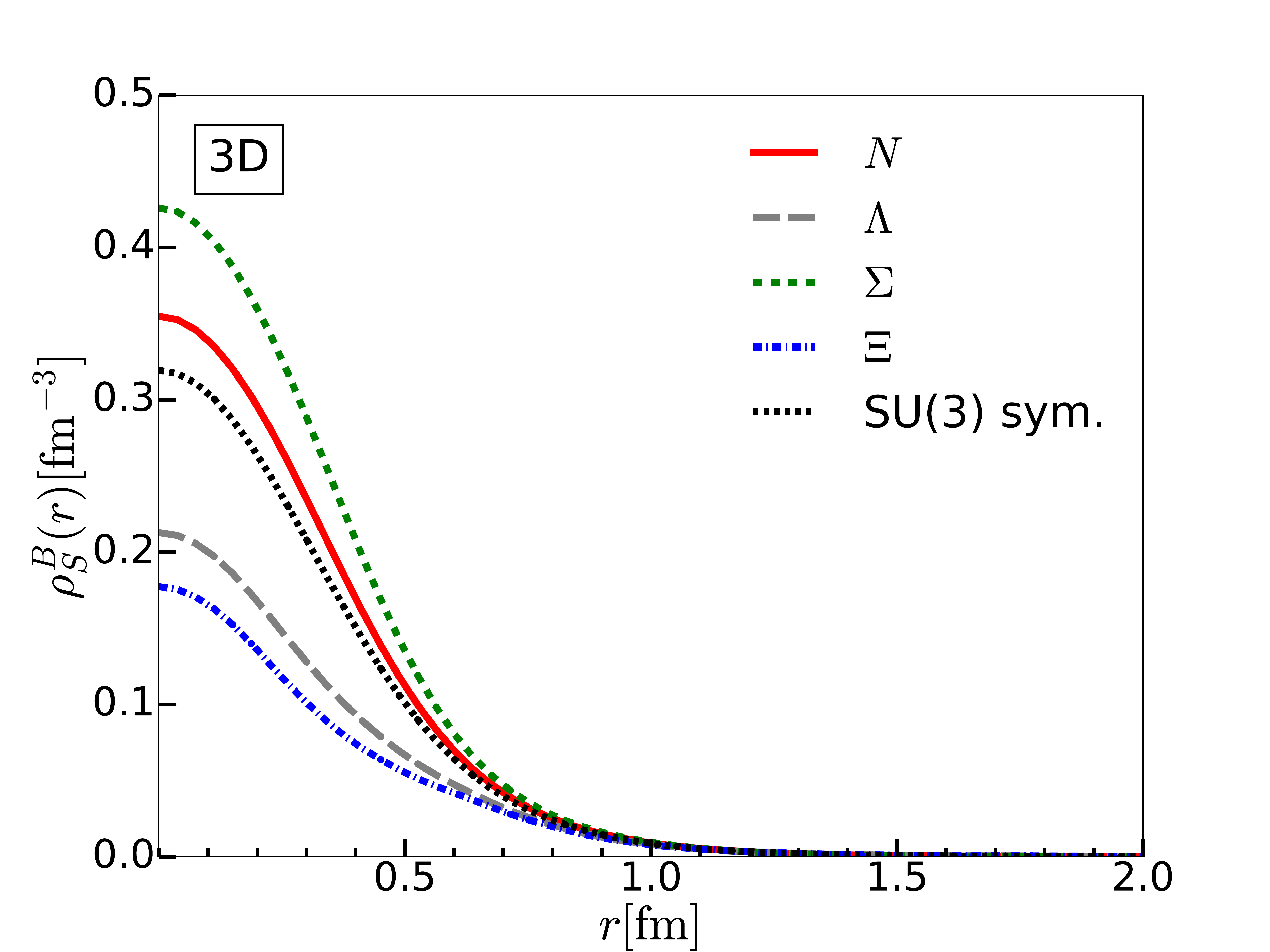}
  \includegraphics[scale=0.17]{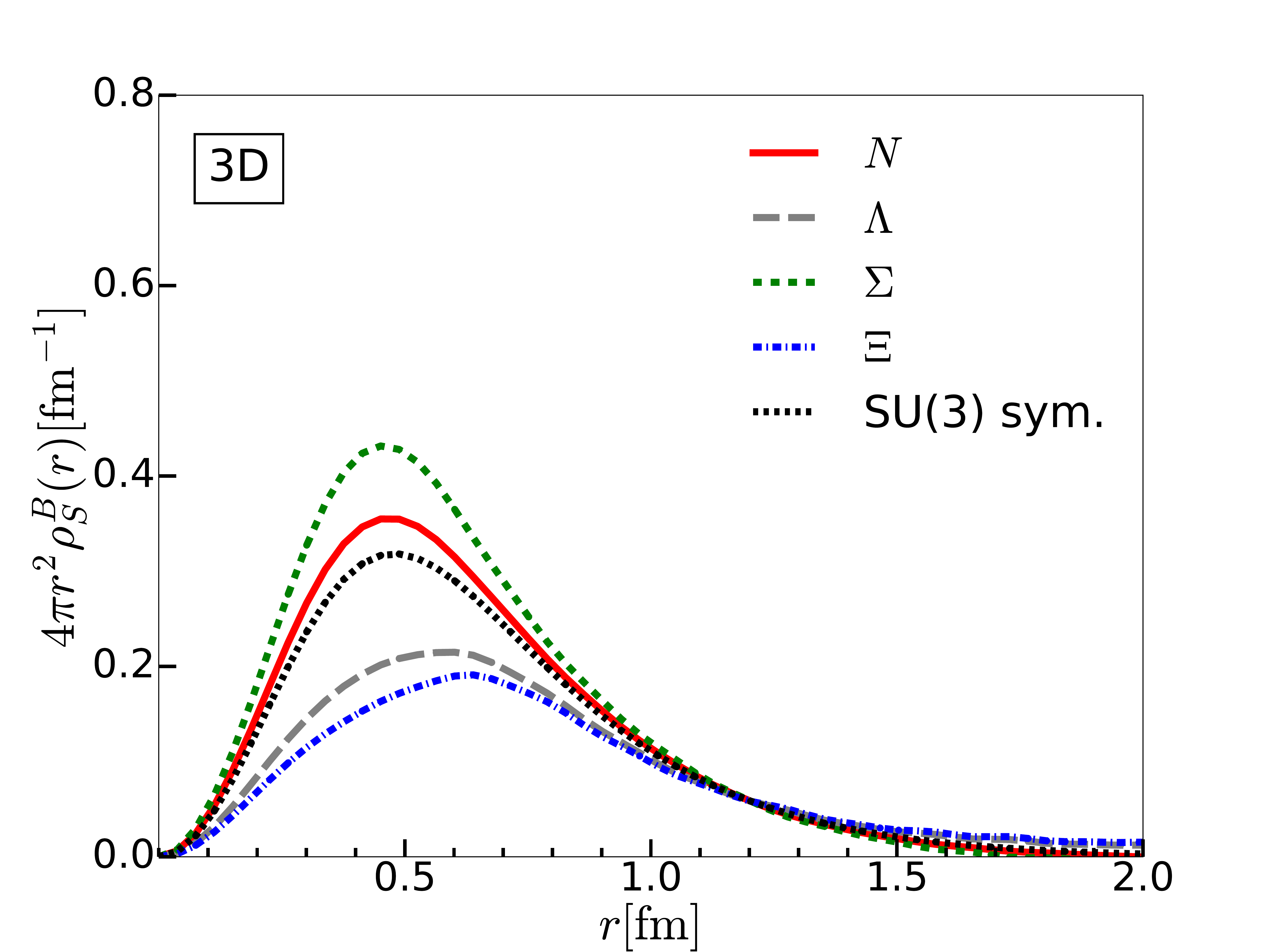}
  \includegraphics[scale=0.17]{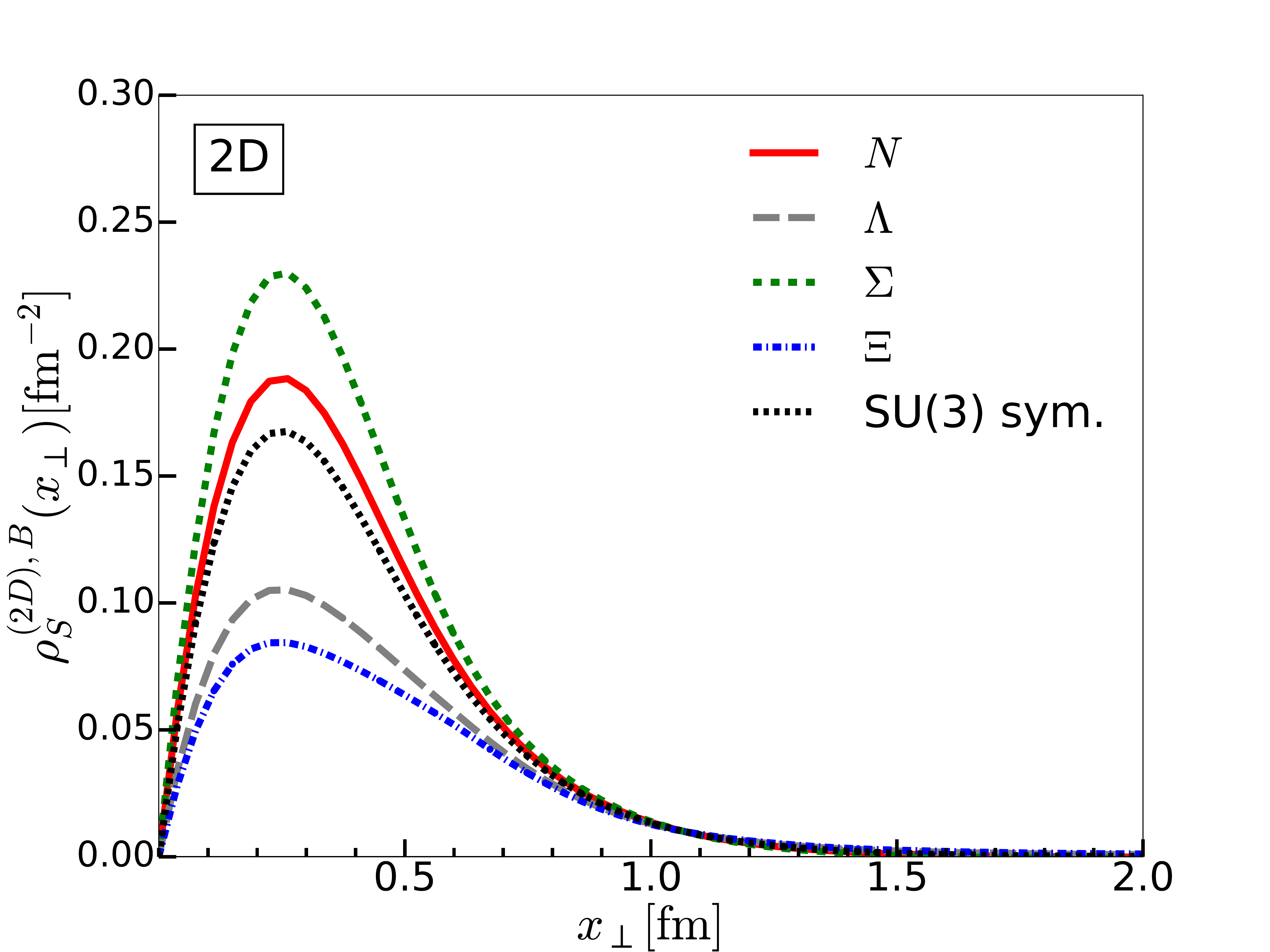}
  \includegraphics[scale=0.17]{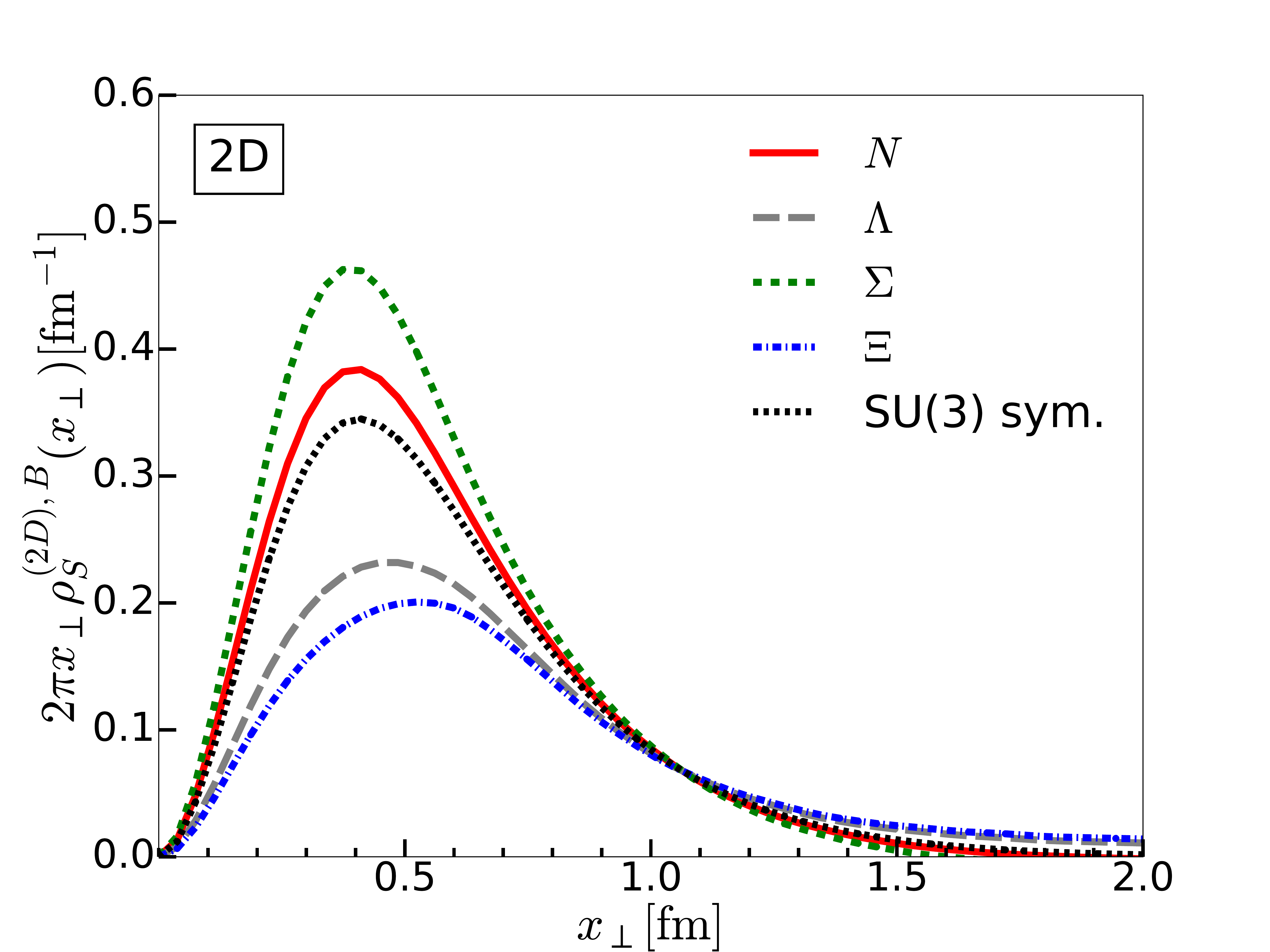}
  \includegraphics[scale=0.17]{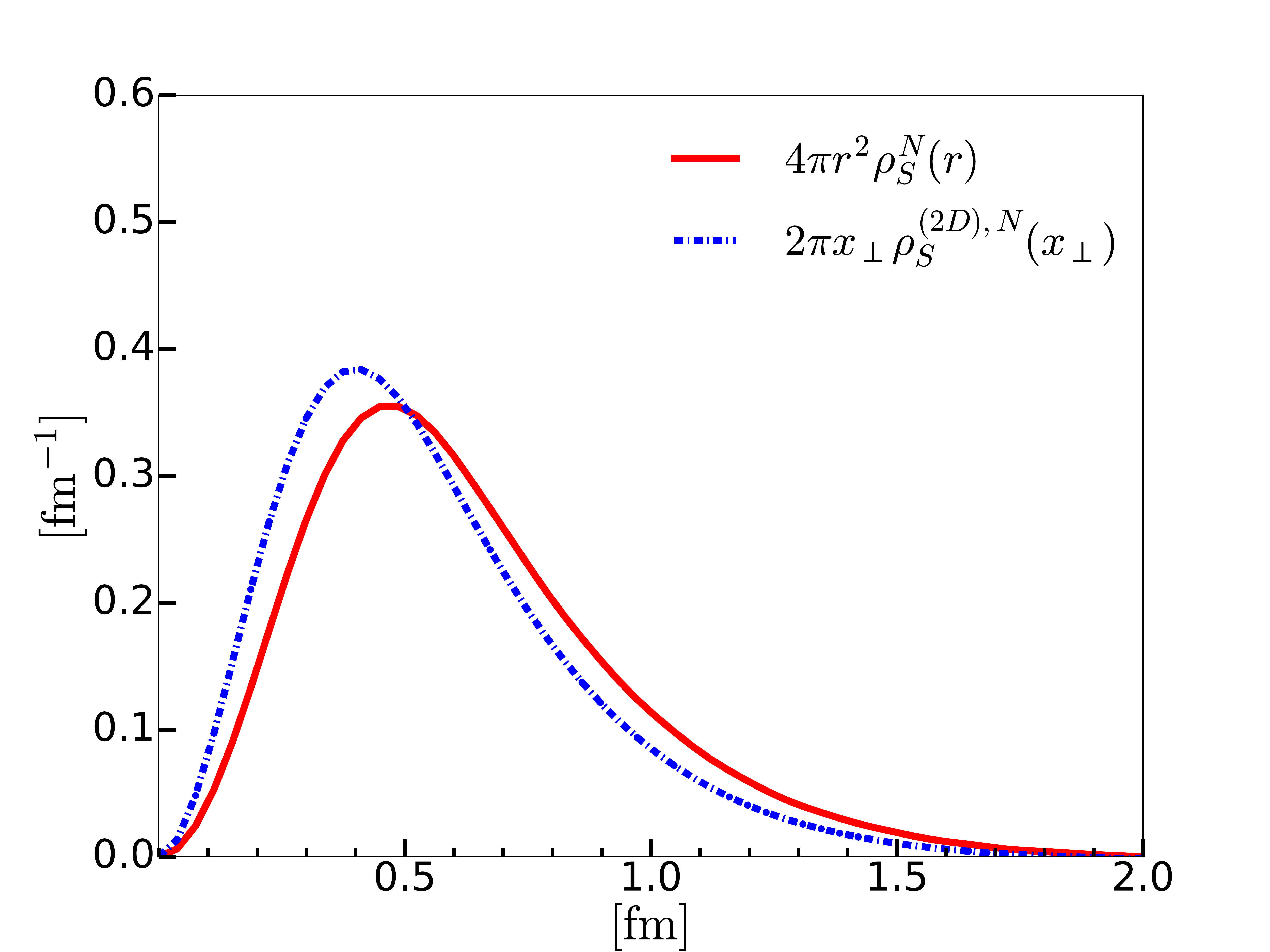}
\caption{3D BF and 2D IMF spin distributions (upper- and middle-left
  panels) and $r^{2}$-weighted ones (upper- and middle-right panels)
  of the baryon octet and the nucleon with the flavor SU(3) symmetry,
  and comparision (lower panel) between the 3D BF and 2D IMF ones of
  the nucleon. The notations are the same as in Fig.~\ref{fig:1}.
}
\label{fig:4}
\end{figure}
As defined in Appendix~\ref{app:B}, the total angular momentum
distribution is divided into the orbital angular momentum
$\rho^{B}_{L}$ and spin $\rho^{B}_{S}$ distributions, and we draw them
for both the 3D BF and the 2D IMF in Figs.~\ref{fig:3} and
~\ref{fig:4}, respectively. Integrating the distributions over 3D
space yields 
\begin{align}
\int d^{3} r \, \rho^{B}_{L}(r)= L^{B}, \quad \int d^{3} r \,
  \rho^{B}_{S}(r)= \frac{1}{2} g_{A}^{0,B}. 
\end{align}
The separate values of the $L^{B}$ and $g^{0,B}_{A}$ are listed in
Table~\ref{tab:2}. The fractions of the orbital angular momentum and
spin carried by quarks inside the baryon are estimated to be around
$\sim 50\%$ respectively and they are well balanced. Though the
relativistic effects on the baryon spin $J^{B}(0)$ for the
nucleon and $\Sigma$ baryon are slightly weaker than for the $\Lambda$
and $\Xi$ baryons, the orbital angular momentum is still very
important in understanding the missing contribution to the baryon
spin. The spin distributions are solely responsible for
the non-zero values at the center of the 
angular momentum distributions, and the orbital angular momentum
contributions dominate over the spin ones at the outer part. It means
that the orbital angular momentum governs the outer part of the
$\rho^{B}_{J}$. From those facts, one may expect that the ordering of
the values of the $\rho^{B}_{J}$ at the center is related to the
ordering of the axial charges of the baryon octet (see
Table~\ref{tab:2}). Note that the axial charges of the baryon octet
from the matrix element of the EMT current coincide with those from
the matrix element of the axial-vector current~\cite{Christov:1995vm,
  Ossmann:2005bbo, Suh:2022atr}.

\subsection{Mechanical properties and stability conditions}
We are now in a position to discuss the pressure and shear force
distributions. Before we present the numerical results for these
distributions, we want to mention how we acquire the stability
conditions in the $\chi$QSM. In the current work, we 
treat $1/N_{c}$ and $m_{s}$ corrections perturbatively. In the leading
$N_{c}$ approximation, the pressure $p(r)$ naturally satisfies the von
Laue condition that is equivalent to the equation of
motion~\cite{Goeke:2007fp}. However, once we introduce the
next-to-leading-order contribution (say, $m_{s}$ corrections) to
$p(r)$, it breaks the von Laue condition. To remedy this problem, two
different methods can be employed. The one is to minimize the baryon
mass after quantizing the soliton, the so-called ``variation after
quantization'' method. However, this method does not respect 
chiral symmetry in the large $r$ region~\cite{Cebulla:2007ei}. Thus,
instead of using this method, we introduce the ``quantization after
variation'' method. We first minimize the soliton mass and then
quantize the soliton. So, the rotational corrections are considered as
a small perturbation. Of course, we also have to pay the price in this
case: the von Laue stability condition will again be broken. However,
we can circumvent this problem by calculating the shear force
distribution to avoid the violation of the stability condition instead
of computing the pressure directly. We then reconstruct the pressure
distribution from the obtained shear force distribution by solving the
differential equation~\eqref{eq:diffEq}. Then, both the pressure and
shear-force distributions comply with the global stability
condition~\cite{Perevalova:2016dln, Kim:2020lrs}.   

Yet another ambiguous point appears at large $r$. The chiral
properties at large $r$ are significant in the description of the
GFFs, especially the $D$-term form factor. In the leading $N_{c}$
contribution, it was well studied in Refs.~\cite{Goeke:2007fp,
  Perevalova:2016dln, Polyakov:2018rew} and agrees analytically with
the results from chiral perturbation theory. However, in the
present work, this chiral property is numerically spoiled by the
finite box effects. We thus extrapolate the distribution at large $r$
by adopting the pion Yukawa tail~\cite{Perevalova:2016dln}. At the
same time, once we take into account the next-to-leading order~(NLO)
of $m_{s}$ or $1/N_{c}$ corrections, this chiral property is broken 
again. However, since we treat them perturbatively, the NLO
corrections weakly contribute to the distributions at any value of
$r$. Indeed we find that the $m_{s}$ correction to the shear force
distribution over $r$ is less than 50\% of the leading
contribution. At the large $r$, this correction is saturated to 20\%
of the leading contribution. Thus, we are able to safely approximate
shear force distribution at large $r$ by using the pion tail used in
the leading $N_{c}$ contribution.  

\begin{figure}[htp]
  \centering
  \includegraphics[scale=0.17]{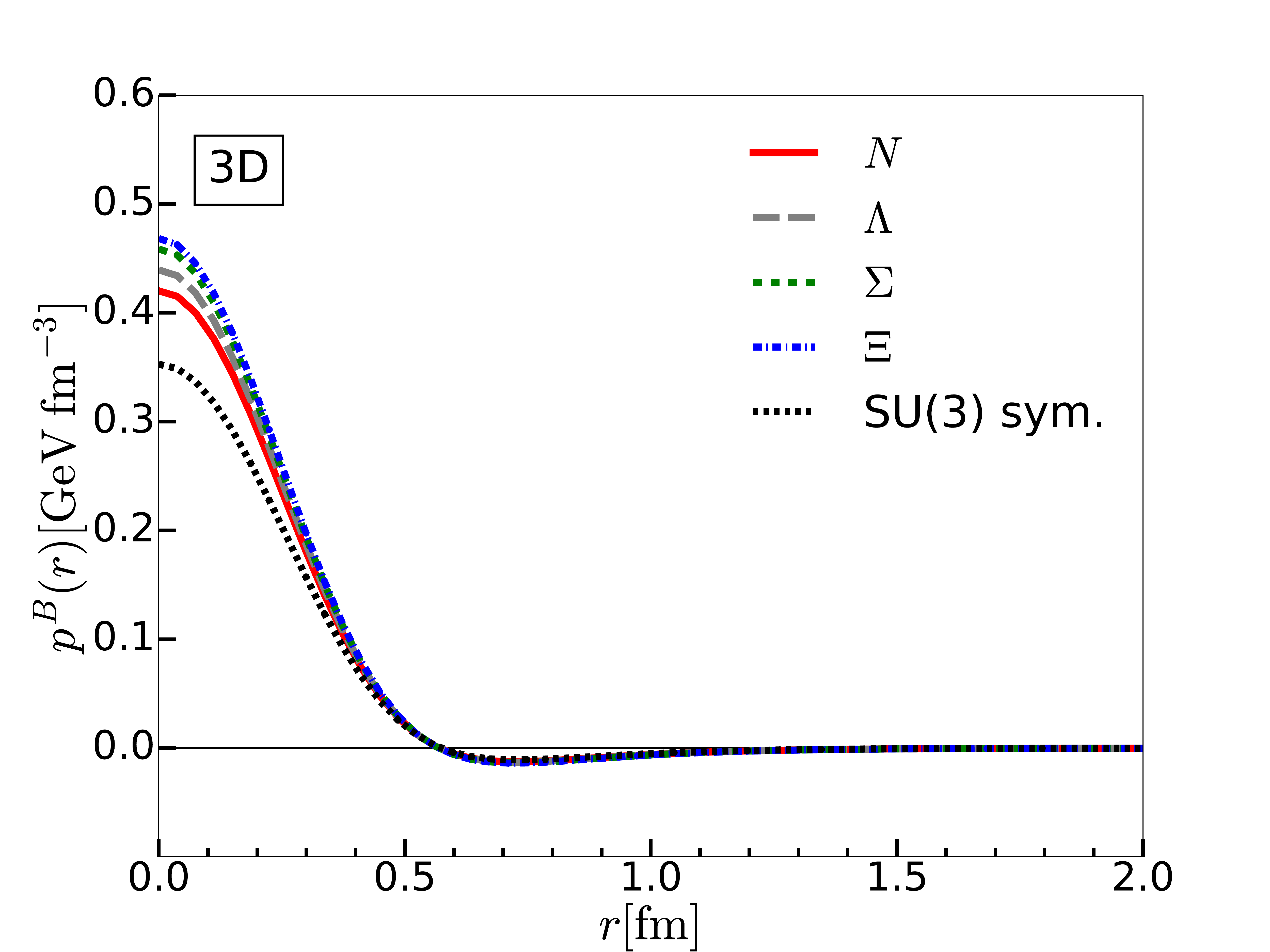}
  \includegraphics[scale=0.17]{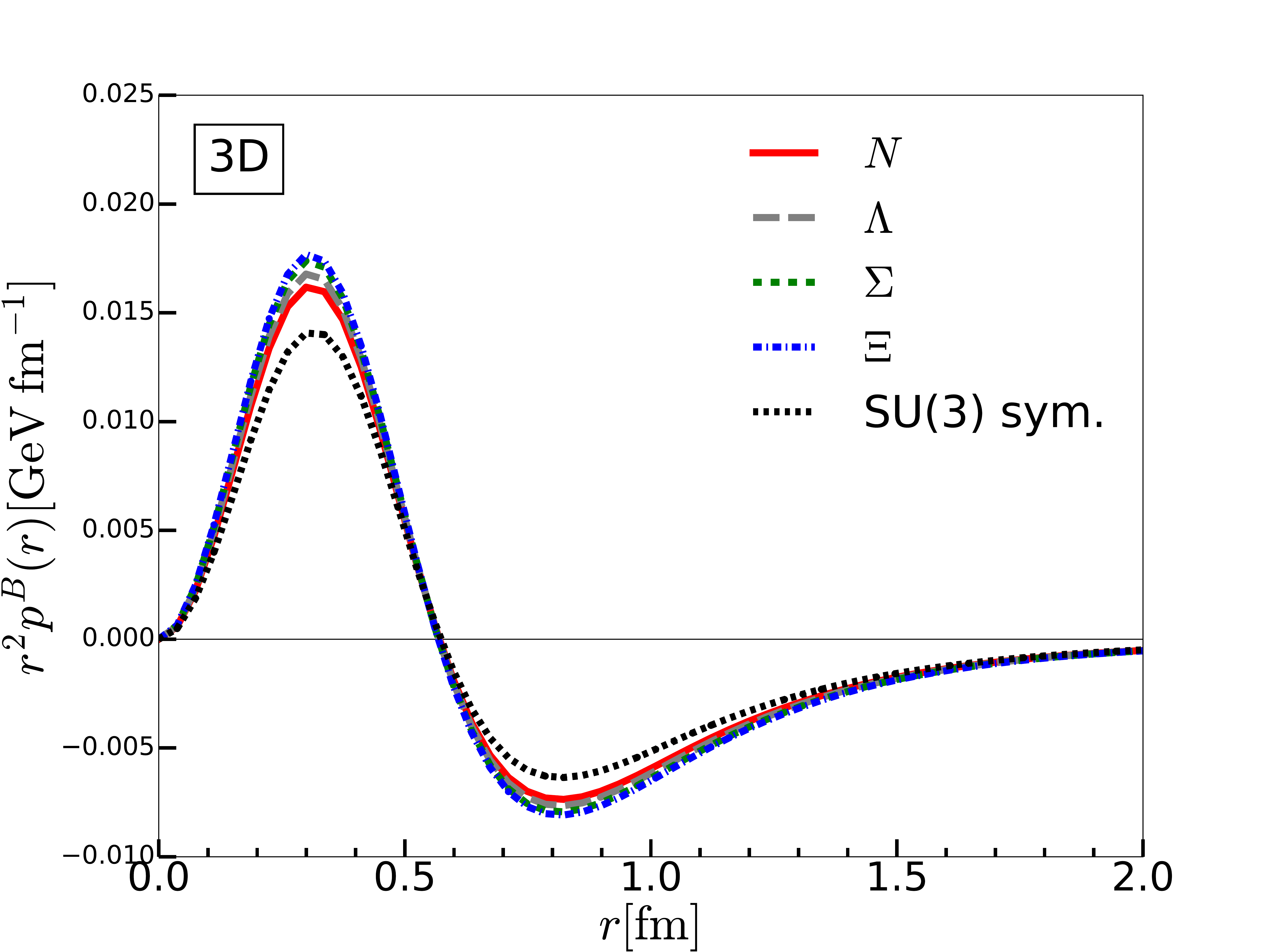}
  \includegraphics[scale=0.17]{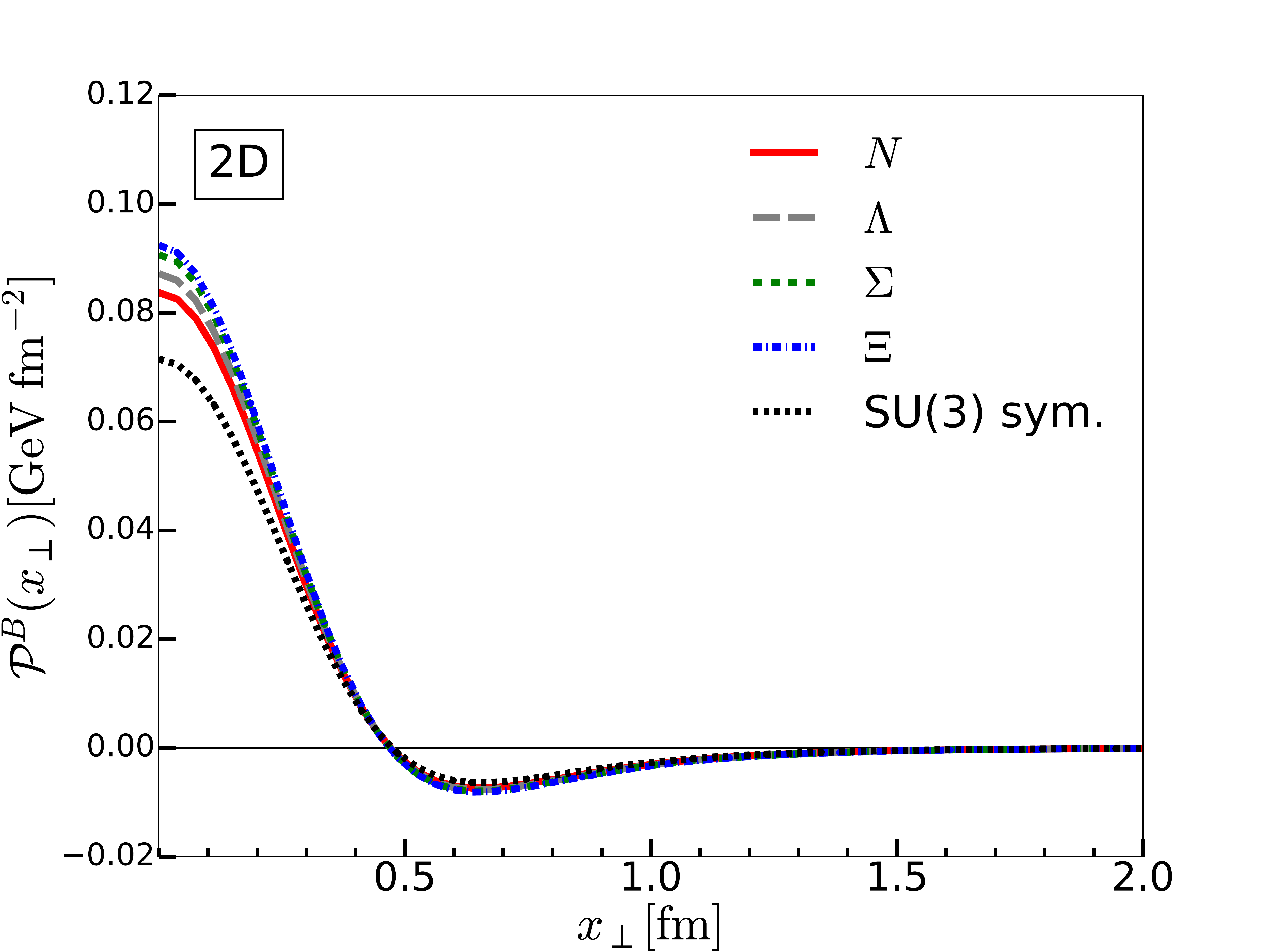}
  \includegraphics[scale=0.17]{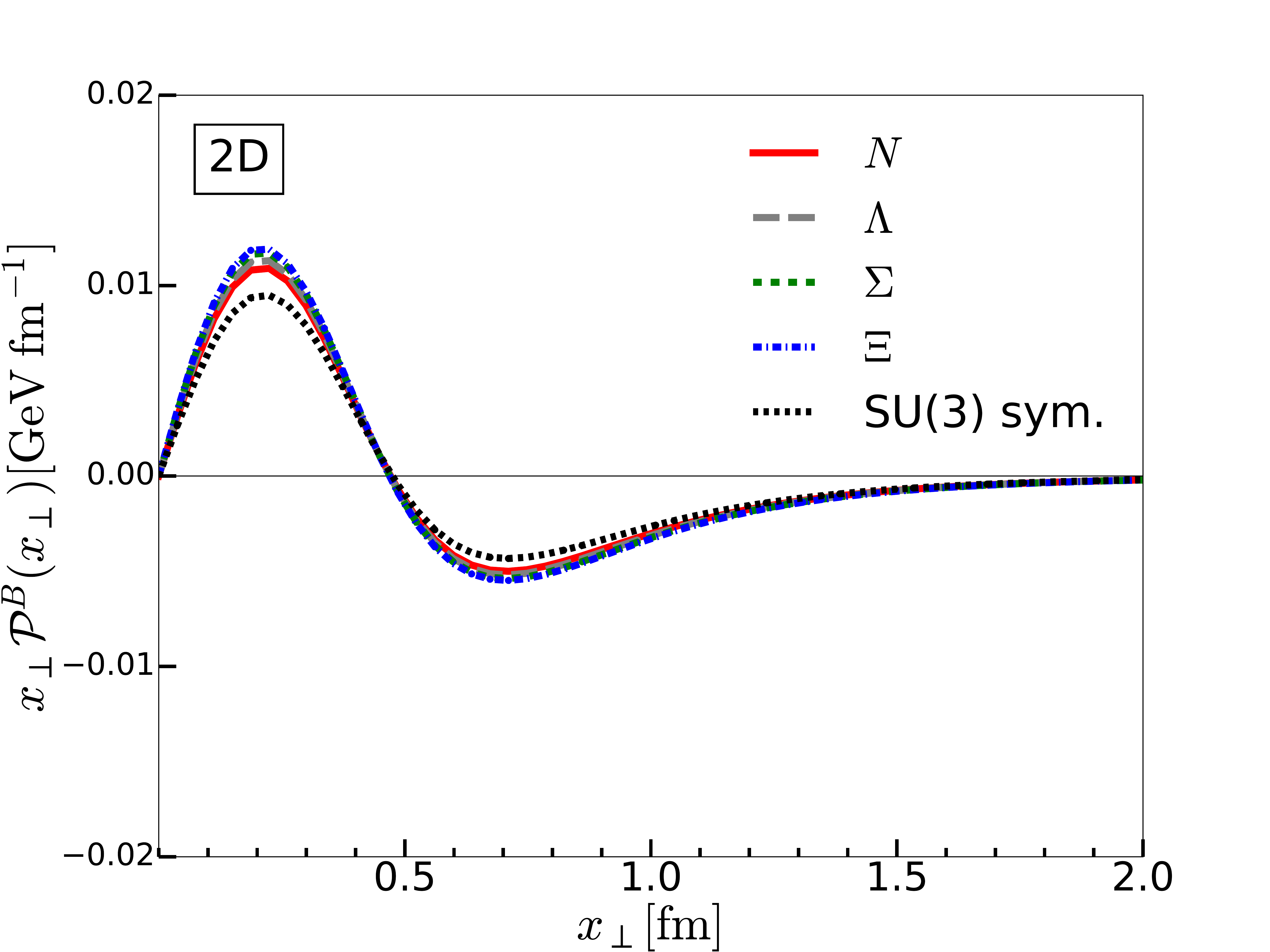}
  \includegraphics[scale=0.17]{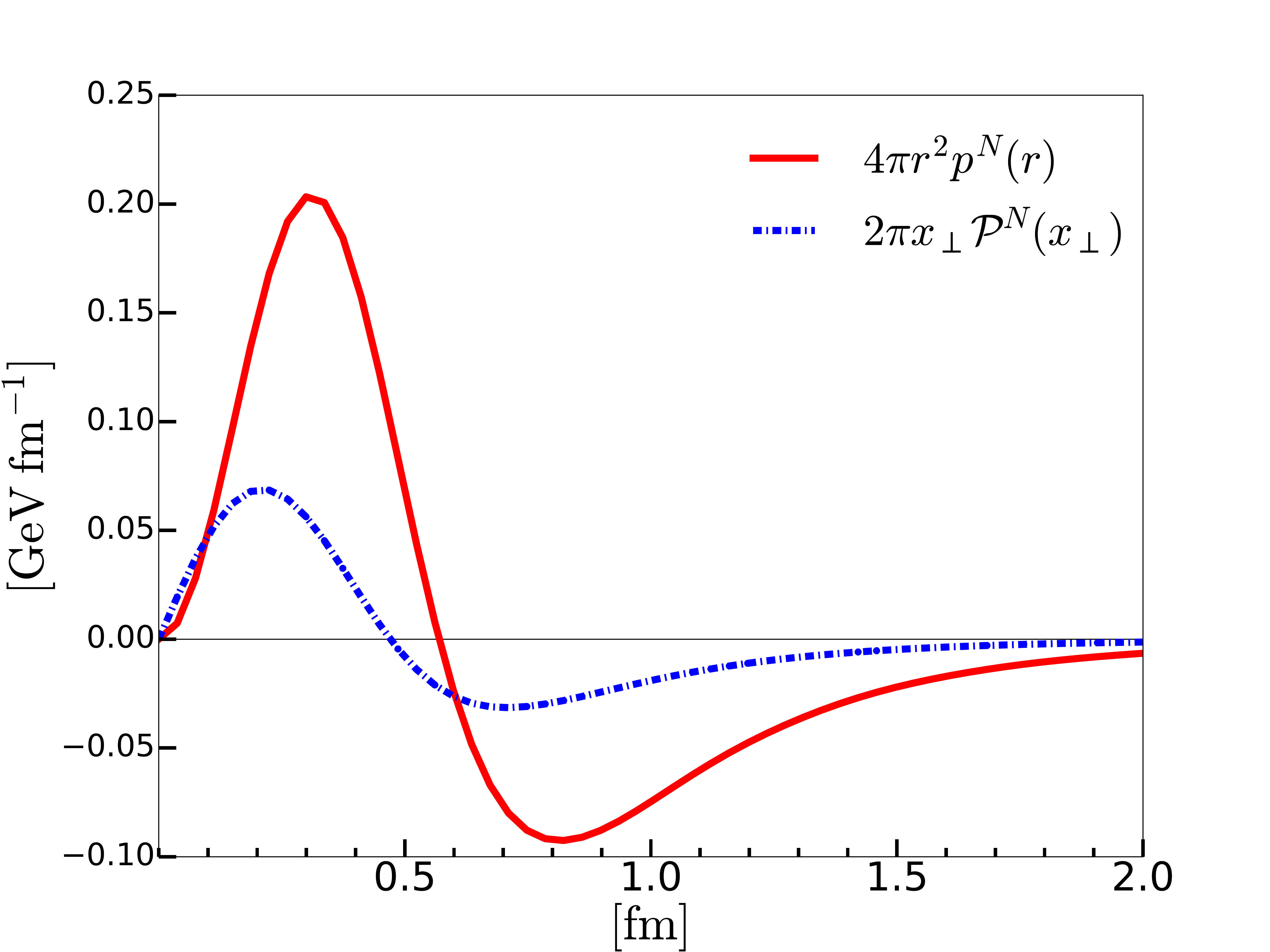}
\caption{3D BF and 2D IMF pressure distributions (upper- and
  middle-left panels) and $r^{2}$-weighted ones (upper- and
  middle-right panels) of the baryon octet and the nucleon with the
  flavor SU(3) symmetry, and comparision (lower panel) between the 3D
  BF and 2D IMF ones of the nucleon. The notations are the same as in
  Fig.~\ref{fig:1}.
} 
\label{fig:5}
\end{figure}
The reconstructed pressure distribution obviously satisfies the von
Laue stability condition: 
\begin{align}
\int dr \, r^{2} p(r)=0.
\end{align}
In Fig.~\ref{fig:5}, we present the pressure distributions of the
baryon octet in both the 3D BF and the 2D IMF, which are reconstructed
from the shear force distributions. The comparison tells us that the
size of the heavier octet baryon is mechanically more compact than
that of the lighter octet baryon, as in the case of the mass
distributions. It can be clearly seen by introducing the $(r_{0})_{B}$
and $(x_{\perp 0} )_{B}$ at which the pressure distribution vanishes for
3D BF and 2D IMF ones, respectively. Note that, to comply with the von
Laue condition, this nodal point is necessary. As shown in the right
panel of Fig.~\ref{fig:5}, the inner and outer parts are explicitly
canceled out, so that the von Laue condition is satisfied. We find the 
following ordering for both the 3D BF and the 2D IMF: 
\begin{align}
(r_{0})_{N}<(r_{0})_{\Lambda}< (r_{0})_{\Sigma}< (r_{0})_{\Xi}, \quad
  (x_{\perp 0})_{N}<(x_{\perp 0})_{\Lambda}< (x_{\perp 0})_{\Sigma}<
  (x_{\perp 0})_{\Xi}. 
\end{align}
Indeed, the heavier octet baryon is a more compact object than the
lighter one. We also find that for the heavier octet baryon the
pressures for both the 3D BF and the 2D IMF at the core part are
larger than the lighter ones~(see Tables~\ref{tab:2} and ~\ref{tab:3}).  

\begin{figure}[htp]
  \centering
  \includegraphics[scale=0.17]{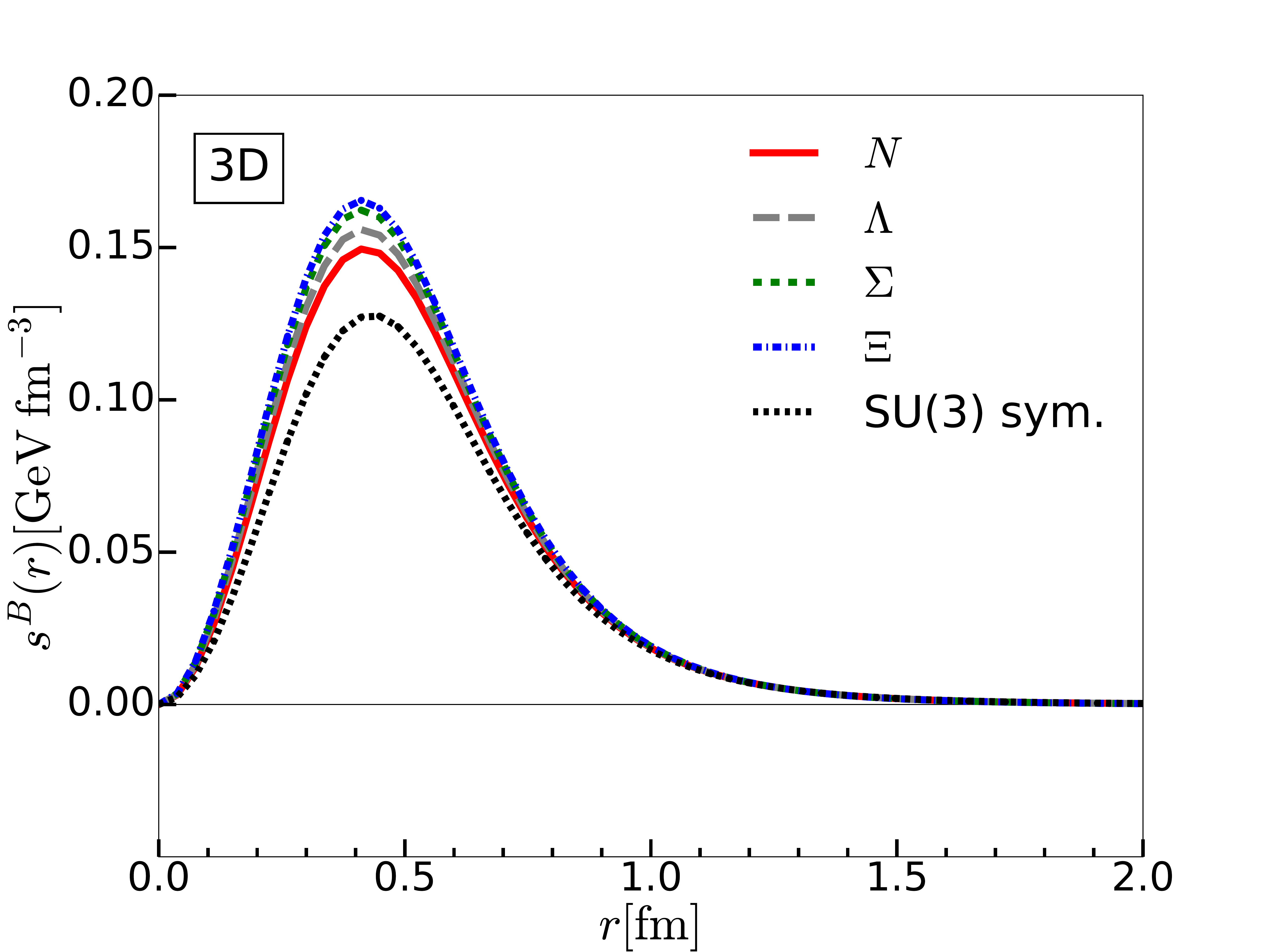}
  \includegraphics[scale=0.17]{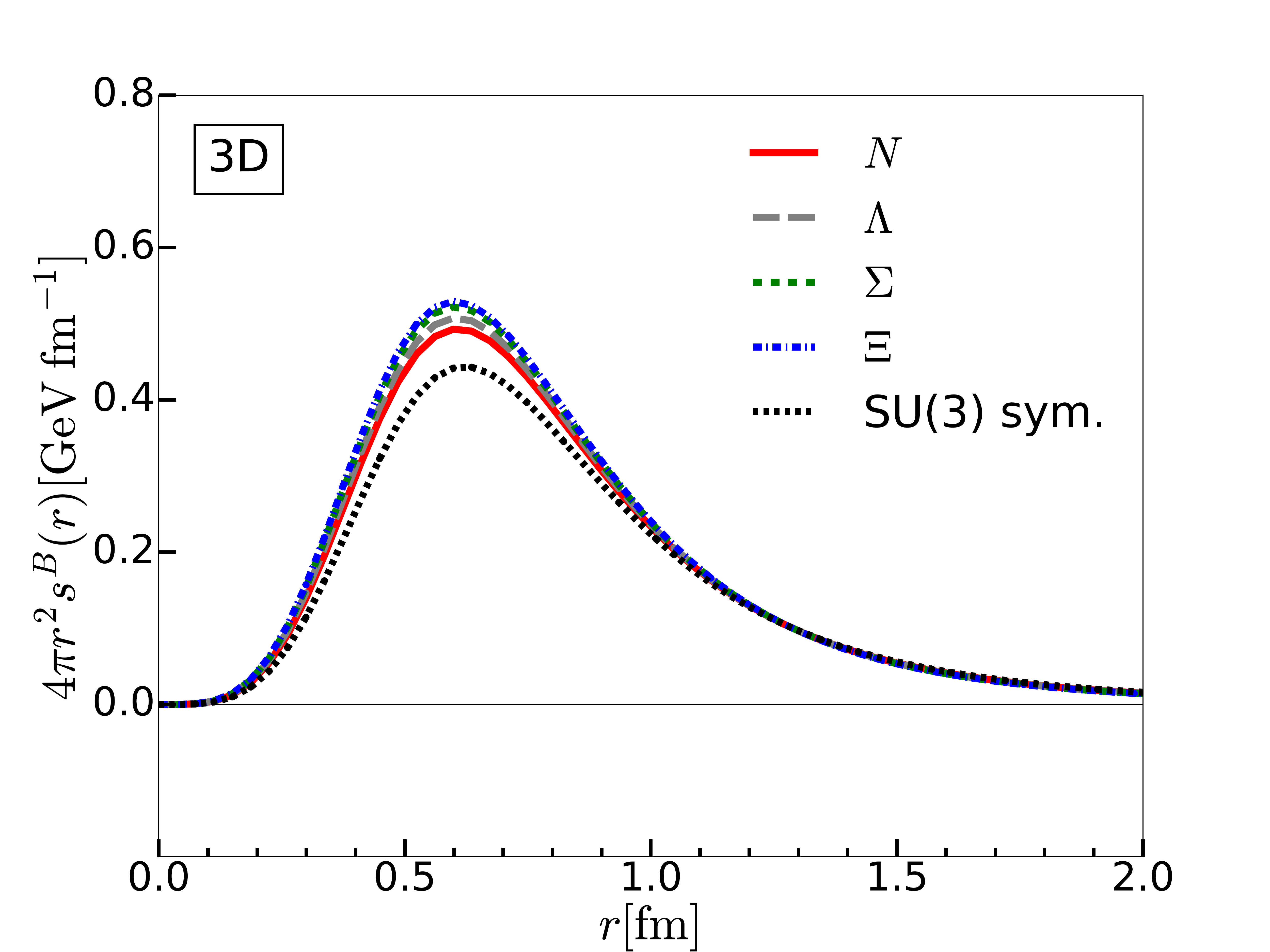}
  \includegraphics[scale=0.17]{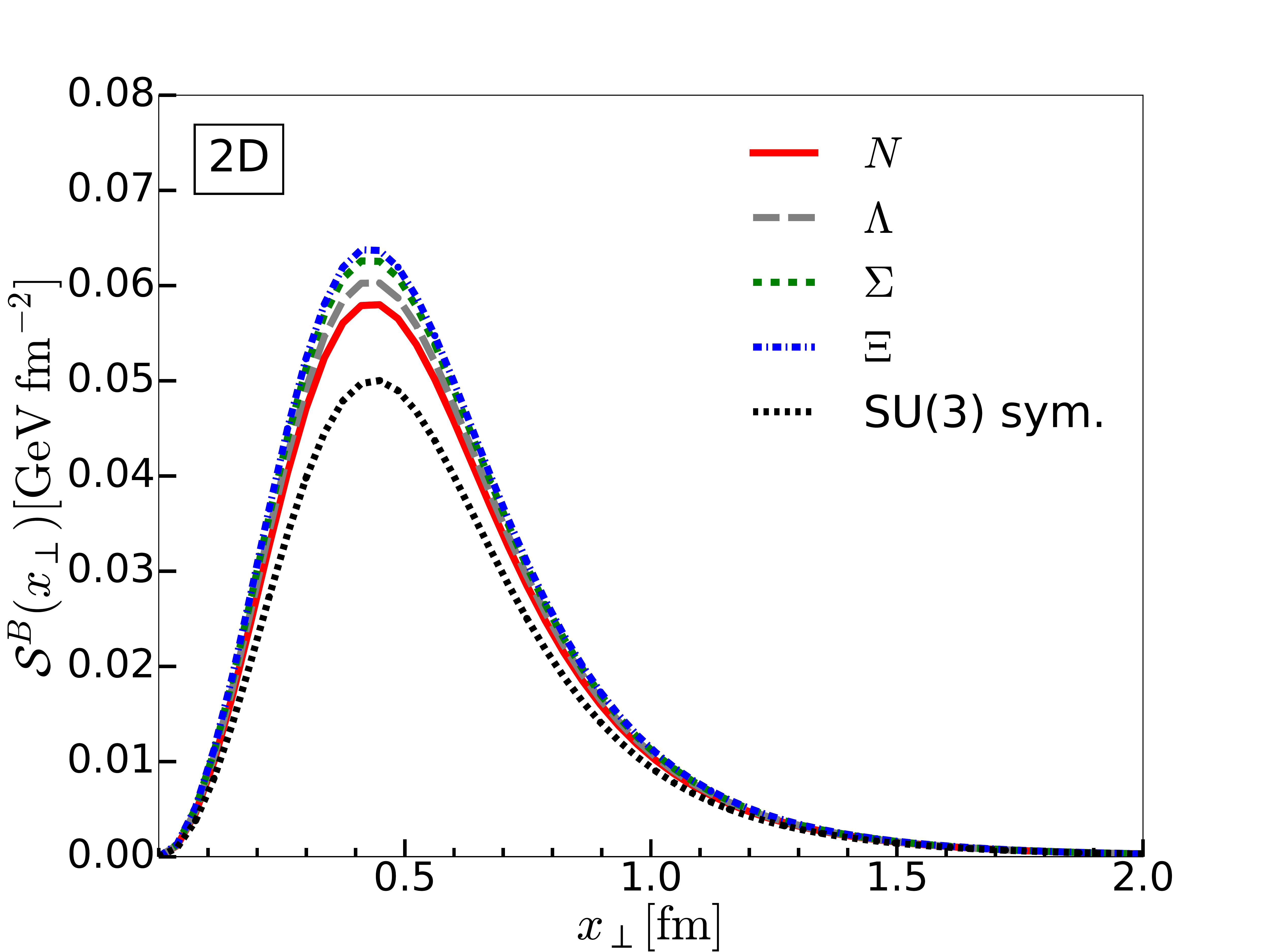}
  \includegraphics[scale=0.17]{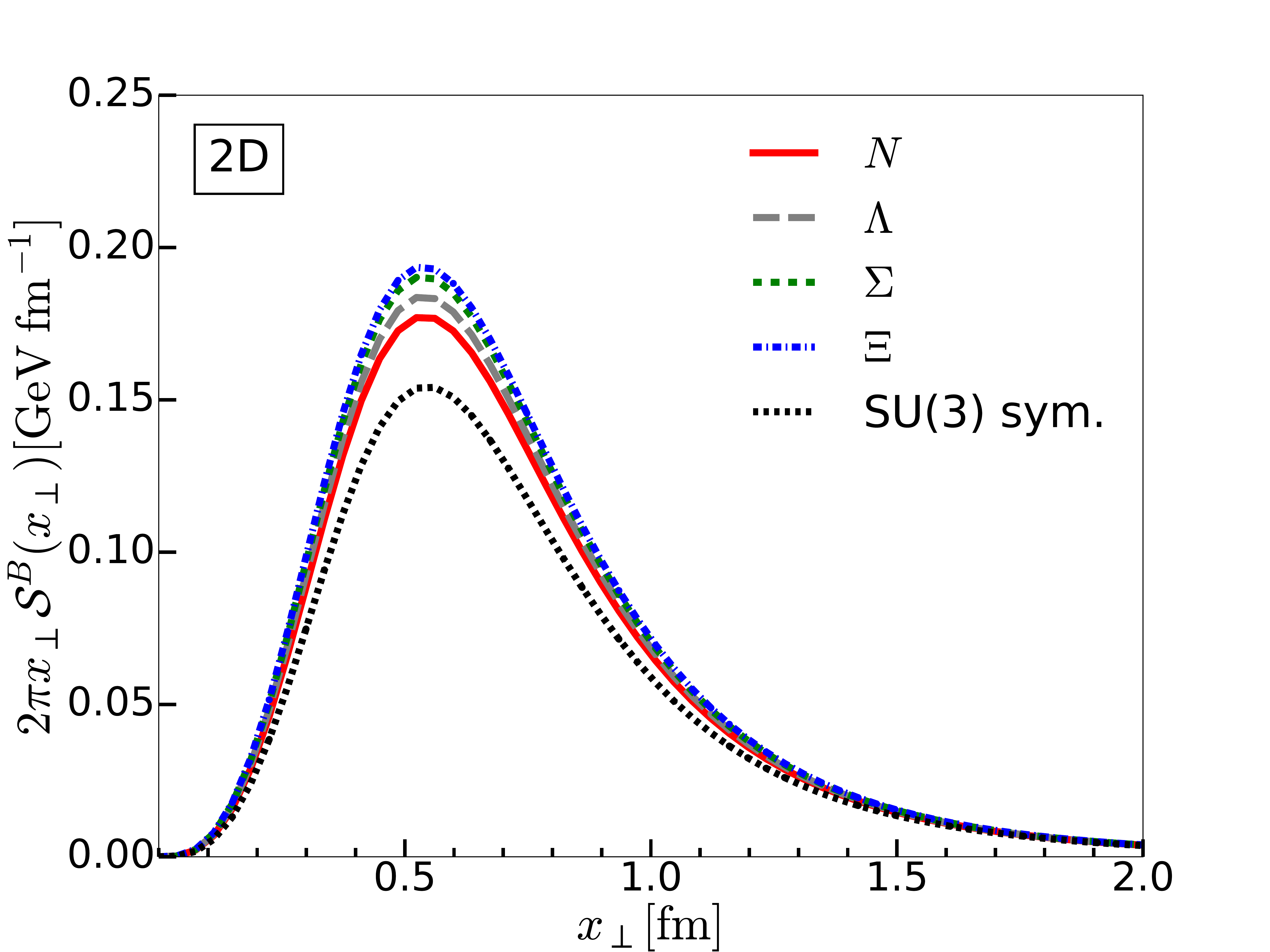}
  \includegraphics[scale=0.17]{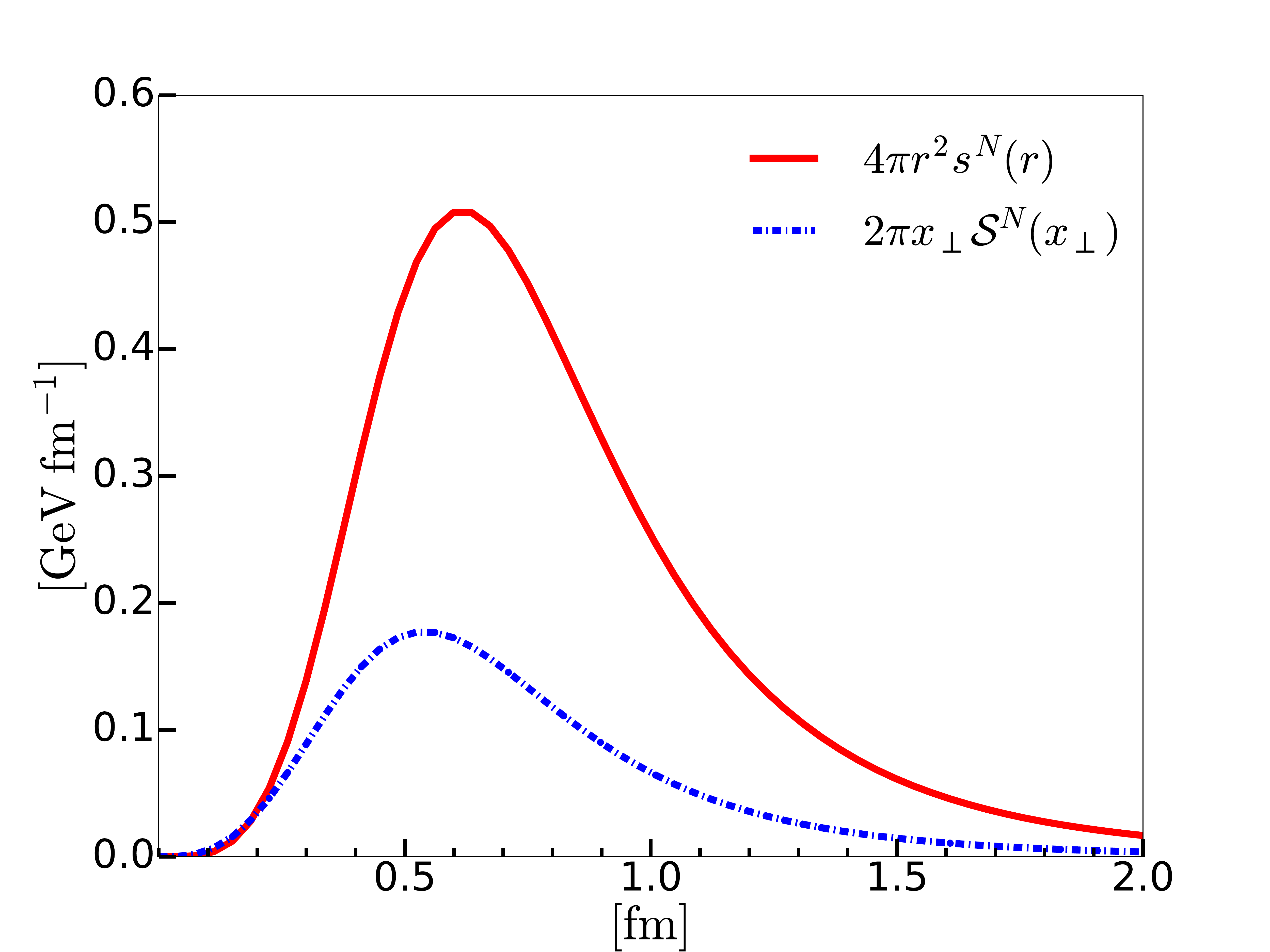}
\caption{3D BF and 2D IMF shear-force distributions (upper- and
  middle-left panels) and $r^{2}$-weighted ones (upper- and
  middle-right panels) of the baryon octet and the nucleon with the
  flavor SU(3) symmetry, and comparision (lower panel) between the 3D
  BF and 2D IMF ones of the nucleon. 
The notations are the same as in Fig.~\ref{fig:1}.
}
\label{fig:6}
\end{figure}
\begin{figure}[htp]
  \centering
  \includegraphics[scale=0.15]{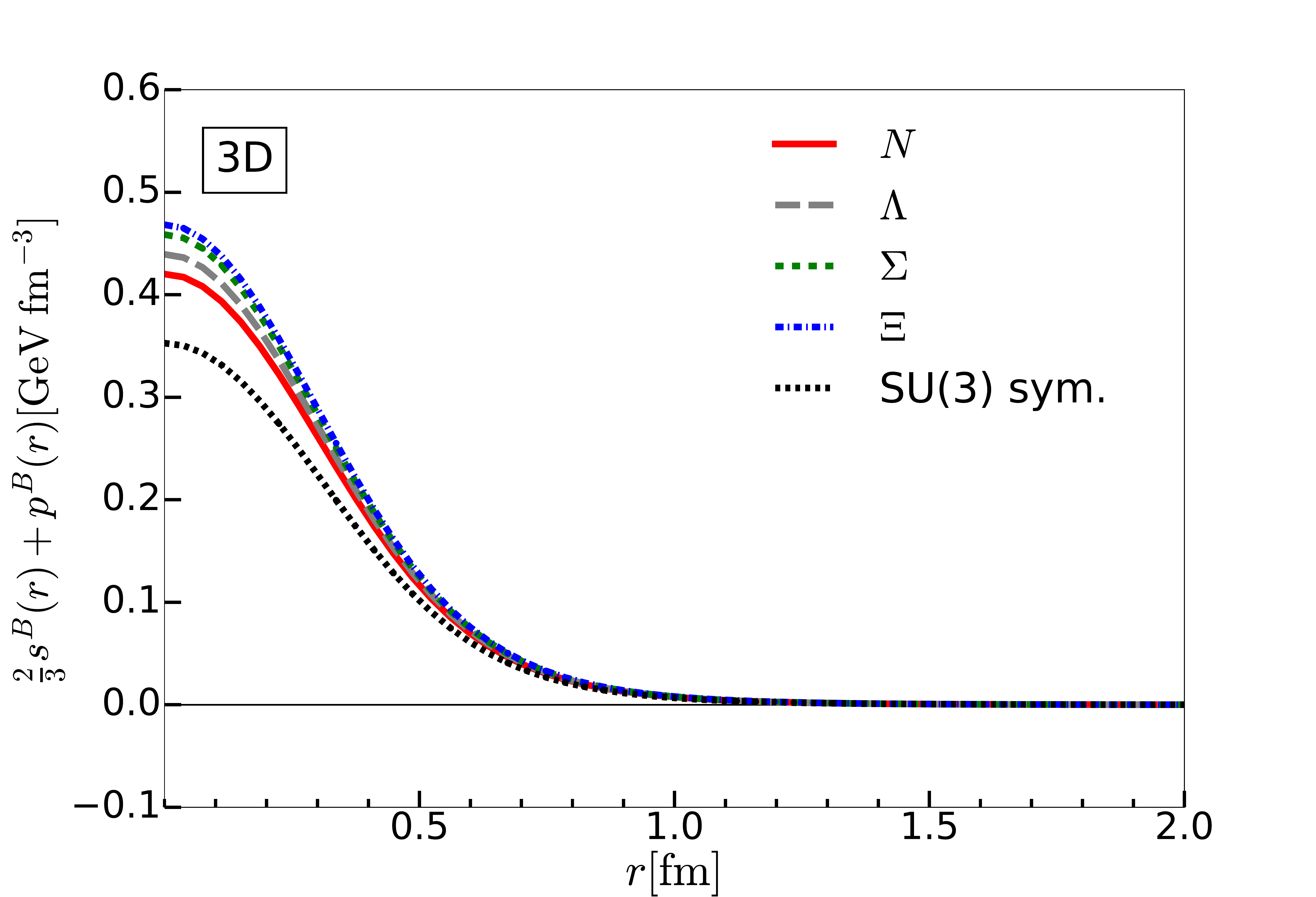}
  \includegraphics[scale=0.15]{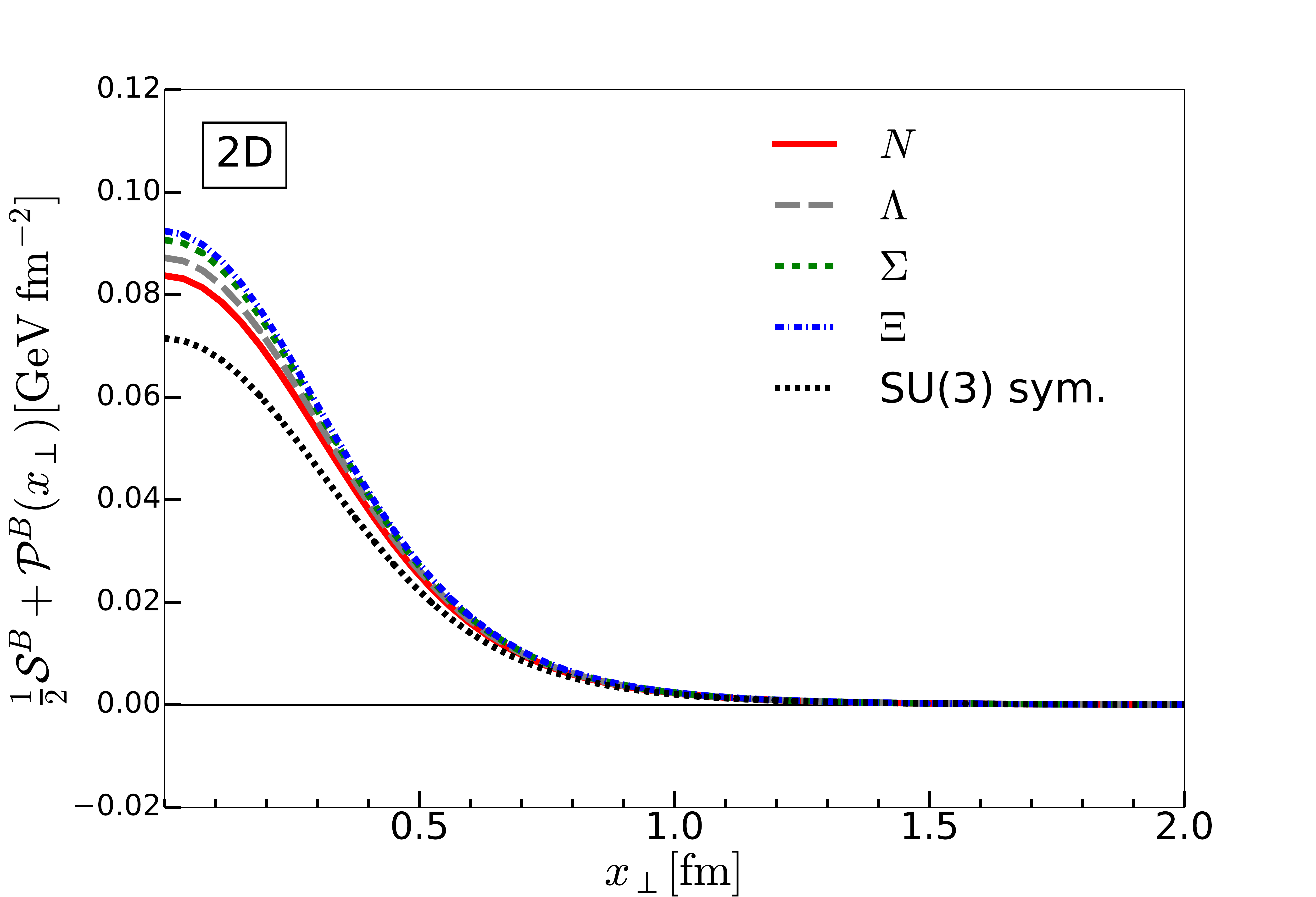}
\caption{3D BF and 2D IMF normal force distributions of the baryon
  octet and the nucleon with the flavor SU(3) symmetry. 
The notations are the same as in Fig.~\ref{fig:1}.
}
\label{fig:7}
\end{figure}
In Fig.~\ref{fig:6}, we draw the shear-force distributions of the
baryon octet in the 3D BF and the 2D IMF. We find that they are
always positive over $r$. We can deduce from the fact that
the 3D normal force is also positive for all values of $r$ (see
Eq.~\eqref{eq:diffEq}): 
\begin{align}
\frac{2}{3}s^{B}(r)+p^{B}(r)>0, \quad s^{B}(r) > 0.
\end{align}
It implies that the 2D one should also be positive over $r$:
\begin{align}
\frac{1}{2}\mathcal{S}^{B}(x_{\perp})+ \mathcal{P}^{B}(x_{\perp}) >0,
  \quad \mathcal{S}^{B}(x_{\perp}) > 0. 
\end{align}
This positive shear-force distribution, $s^{B}(r)>0$ or
$\mathcal{S}^{B}(x_{\perp})>0$, is the signature of the negative
$D$-term form factor at the zero momentum transfer. In
Fig.~\ref{fig:7}, we also present the 3D BF and 2D IMF normal force
distributions as a function of $r$ for the baryon octet. They indeed
satisfy the positivity over $r$.  
By integrating either $r^{2}$-weighted pressure or shear-force
distributions, one can obtain the $D$-terms. The numerical results for
the $D$-terms of the baryon octet are derived as  
\begin{align}
D^{N}(0)=-3.08, \quad 
D^{\Lambda}(0)=-3.22, \quad \quad 
D^{\Sigma}(0)=-3.37, \quad \quad 
D^{\Xi}(0)=-3.45.
\label{eq:Dterm}
\end{align}
As expected, we obtain the negative $D$-terms for the baryon
octet. Interestingly, we find that the heavier octet baryon has a
larger value of the $D$-term. To quantify the mechanical size of the
octet baryon, we estimate the 3D BF and 2D IMF mechanical radii of the
baryon octet as follows:  
\begin{align}
\expval{r_{\mathrm{mech}}^{2}}_{N}&=0.53~[\mathrm{fm}^{2}], \quad 
\expval{r_{\mathrm{mech}}^{2}}_{\Lambda}=0.53~[\mathrm{fm}^{2}], \quad 
\expval{r_{\mathrm{mech}}^{2}}_{\Sigma}=0.52~[\mathrm{fm}^{2}], \quad 
\expval{r_{\mathrm{mech}}^{2}}_{\Xi}=0.52~[\mathrm{fm}^{2}],\cr
\expval{x_{\perp \mathrm{mech}}^{2}}_{N}&=0.36~[\mathrm{fm}^{2}], \quad 
\expval{x_{\perp \mathrm{mech}}^{2}}_{\Lambda}=0.35~[\mathrm{fm}^{2}], \quad 
\expval{x_{\perp \mathrm{mech}}^{2}}_{\Sigma}=0.35~[\mathrm{fm}^{2}], \quad 
\expval{x_{\perp \mathrm{mech}}^{2}}_{\Xi}=0.35~[\mathrm{fm}^{2}].
\end{align}
This indicates that the heavier octet baryon is mechanically a more
compact object than the lighter one. Since, however, the $m_{s}$
corrections to the mechanical radii are negligible, the mechanical
sizes of octet baryons are rather comparable. All the relevant
physical observables are listed in Tables~\ref{tab:2} and
~\ref{tab:3}. Last but not least, it is of great interest to see the
ordering of the magnitude of the nucleon radii with the flavor SU(3)
symmetry breaking. We observe the following ordering:
\begin{align}
\langle r^{2}_{\varepsilon} \rangle_{N}  < \langle
  r^{2}_{\mathrm{mech}} \rangle_{N} < \langle r^{2}_{\mathrm{charge}}
  \rangle_{N}\cramped[]{} ,
\end{align}
where $\langle r^{2}_{\mathrm{charge}} \rangle_{N}$ is the charge
radius of the nucleon taken from Ref.~\cite{Christov:1995vm}. We
find that the ordering of the radii are kept to be the same as the
results with flavor SU(3) symmetry for both the 3D BF and the 2D IMF. 

\begin{table}[htb] 
\caption{Various observables for the baryon octet and the nucleon with
  the SU(3) symmetry in the 3D BF:  energy densities at the center
  $\varepsilon^{B}(0)$;  mean square radii
  $\expval{r_{\varepsilon}^{2}}_{B}$ and
  $\expval{r_{\mathrm{mech}}^{2}}_{B}$; normalized total angular
  momentum $2J^{B}(0)$; orbital angular momentum $2L^{B}$; iso-singlet
  axial charge $g_{A}^{0,B}$; pressure densities $p^{B}(0)$ at the
  center of each baryon; nodal point of the pressure distribution
  $(r_{0})_B$; $D$-term $D(0)$.}  
\begin{center}
  \renewcommand{\arraystretch}{1.5}
\scalebox{1.05}{%
\begin{tabular}{cccccccccc} 
  \hline
  \hline
  $B$ & $\varepsilon^{B}(0)$    & $\expval{r_{\varepsilon}^{2}}_{B}$
    & $2J^{B}(0)$     & $g_{A}^{0,B}$    & $2L^{B}$    & $p^{B}(0)$ 
    & $(r_{0})_{B}$  & $D^{B}(0)$ & $\expval{r_{\mathrm{mech}}^{2}}_{B}$  \\
    & $\mathrm{GeV}/\mathrm{fm}^{3}$      & $\mathrm{fm}^{2}$
    &  & & & $\mathrm{GeV}/\mathrm{fm}^{3}$      & $\mathrm{fm}$   
    &  & $\mathrm{fm}^{2}$  \\
  \hline
  $N$         &  $2.85$   &  $0.31$
              &  $1.00$   &  $0.48$   &  $0.52$   
              &  $0.42$   & $0.57$   
              & -$3.08$   &  $0.53$   \\
  $\Lambda$   &  $3.12$   &  $0.26$
              &  $1.00$   &  $0.40$   &  $0.60$   
              &  $0.44$   & $0.57$   
              & -$3.22$   &  $0.53$   \\      
  $\Sigma$    &  $3.40$   &  $0.20$
              &  $1.00$   &  $0.53$   &  $0.47$   
              &  $0.46$   & $0.57$   
              & -$3.37$   &  $0.52$   \\     
  $\Xi$       &  $3.53$   &  $0.17$
              &  $1.00$   &  $0.38$   &  $0.62$   
              &  $0.47$   & $0.57$   
              & -$3.45$   &  $0.52$   \\      
  \hline
  SU(3) sym.  &  $1.89$ &  $0.54$
              &  $1.00$  &  $0.46$  &  $0.54$  
              &  $0.35$  &  $0.57$  
              & -$2.60$  & $0.55$  \\
  \hline
  \hline
\end{tabular}}
\end{center}
\label{tab:2}
\end{table}

\begin{table}[htb] 
\caption{Various observables for the baryon octet and the nucleon with
  the SU(3) symmetry in the 2D IMF:  energy densities at the center
  $\mathcal{E}^{B}(0)$; mean square radii $\expval{x_{\perp
      \mathcal{E}}^{2}}_{B}$ and $\expval{x_{\perp
      \mathrm{mech}}^{2}}_{B}$; pressure densities
  $\mathcal{P}^{B}(0)$ at the center of each baryon; nodal point of
  the pressure distribution $(x_{\perp 0})_B$.}  
\begin{center}
  \renewcommand{\arraystretch}{1.5}
\scalebox{1.05}{%
\begin{tabular}{cccccc} 
  \hline
  \hline
  $B$ & $\mathcal{E}^{B}(0)$                     
      & $\expval{x_{\perp \mathcal{E}}^{2}}_{B}$             
      & $\mathcal{P}^{B}(0)$  
      & $(x_{\perp 0})_{B}$            
      & $\expval{x_{\perp \mathrm{mech}}^{2}}_{B}$    \\
      & $\mathrm{GeV}/\mathrm{fm}^{3}$      & $\mathrm{fm}^{2}$           
                      & $\mathrm{GeV}/\mathrm{fm}^{3}$      & $\mathrm{fm}$   
      & $\mathrm{fm}^{2}$  \\
  \hline
  $N$         &  $2.56$   &  $0.14$   
              &  $0.084$   & $0.47$   
              &  $0.36$   \\
  $\Lambda$   &  $2.76$   &  $0.11$   
              &  $0.087$   & $0.47$   
              &  $0.35$   \\      
  $\Sigma$    &  $2.96$   &  $0.07$   
              &  $0.091$   & $0.47$   
              &  $0.35$   \\     
  $\Xi$       &  $3.01$   &  $0.05$   
              &  $0.092$   & $0.47$   
              &  $0.35$   \\      
  \hline
  SU(3) sym.  &  $1.87$   &  $0.30$
              &  $0.071$   &  $0.47$  
              & $0.37$    \\
  \hline
  \hline
\end{tabular}}
\end{center}
\label{tab:3}
\end{table}

\subsection{Results for the gravitational form factors}
\begin{figure}[htb]
  \centering
  \includegraphics[scale=0.17]{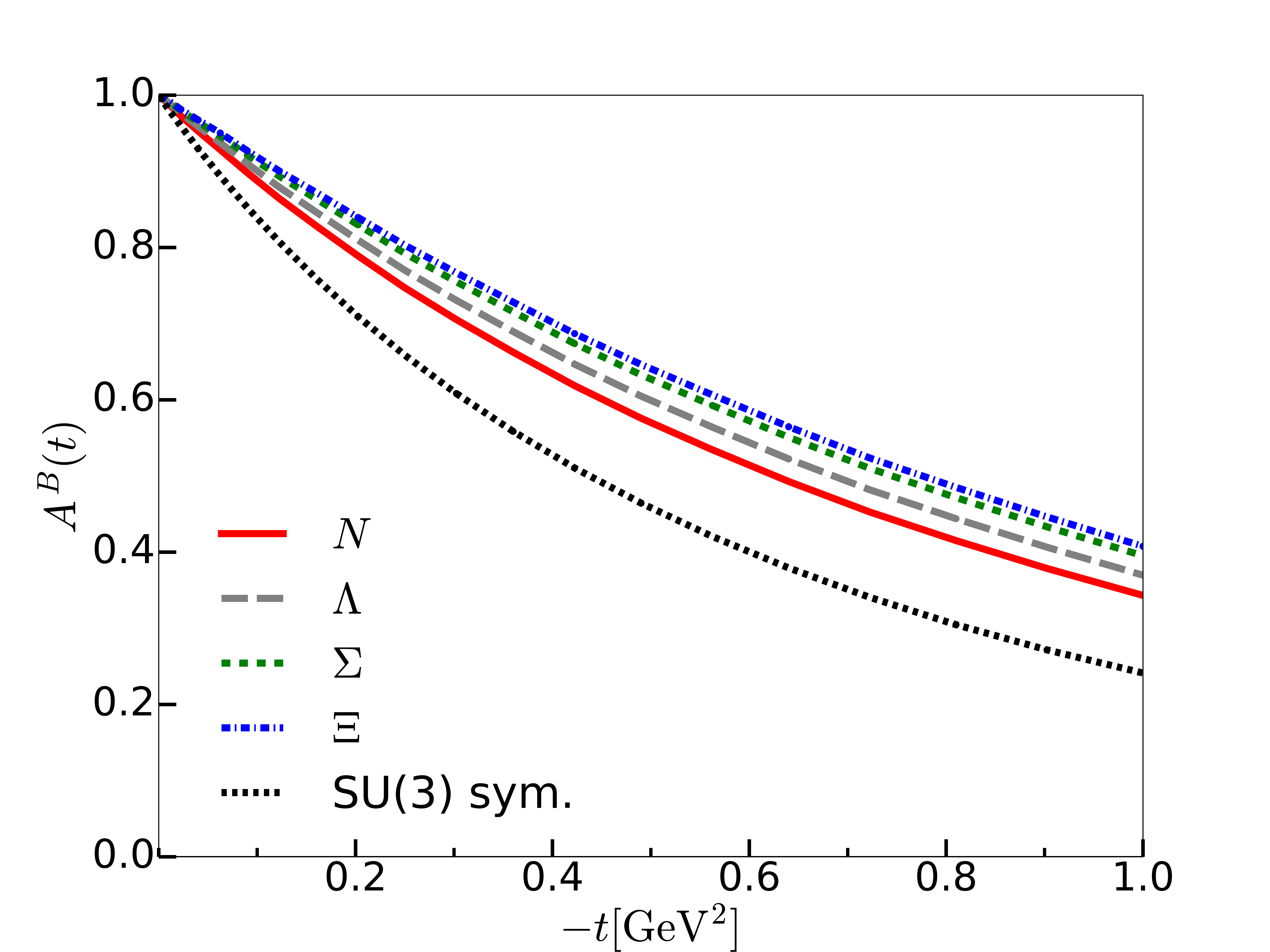}
  \includegraphics[scale=0.17]{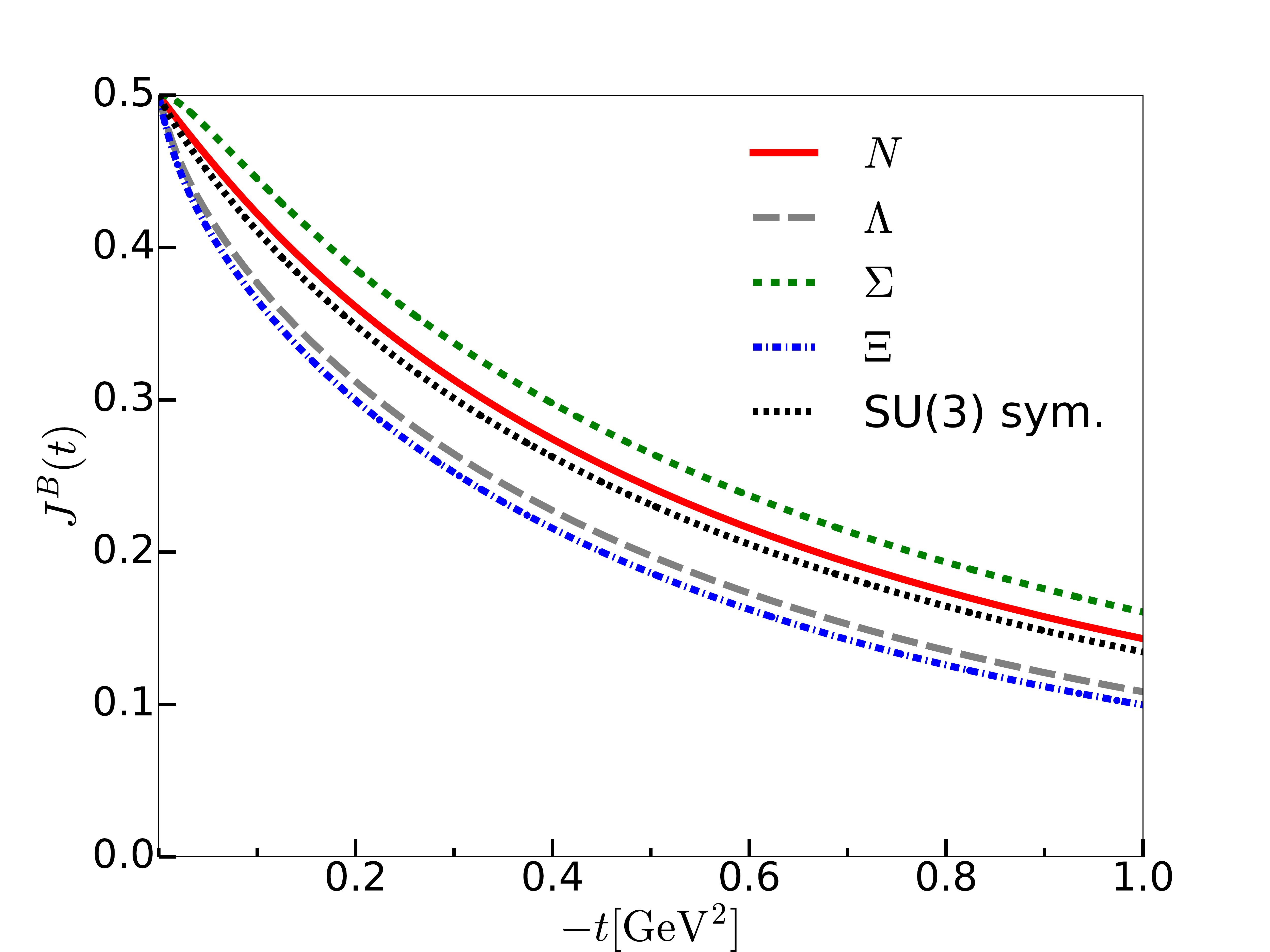}
  \includegraphics[scale=0.17]{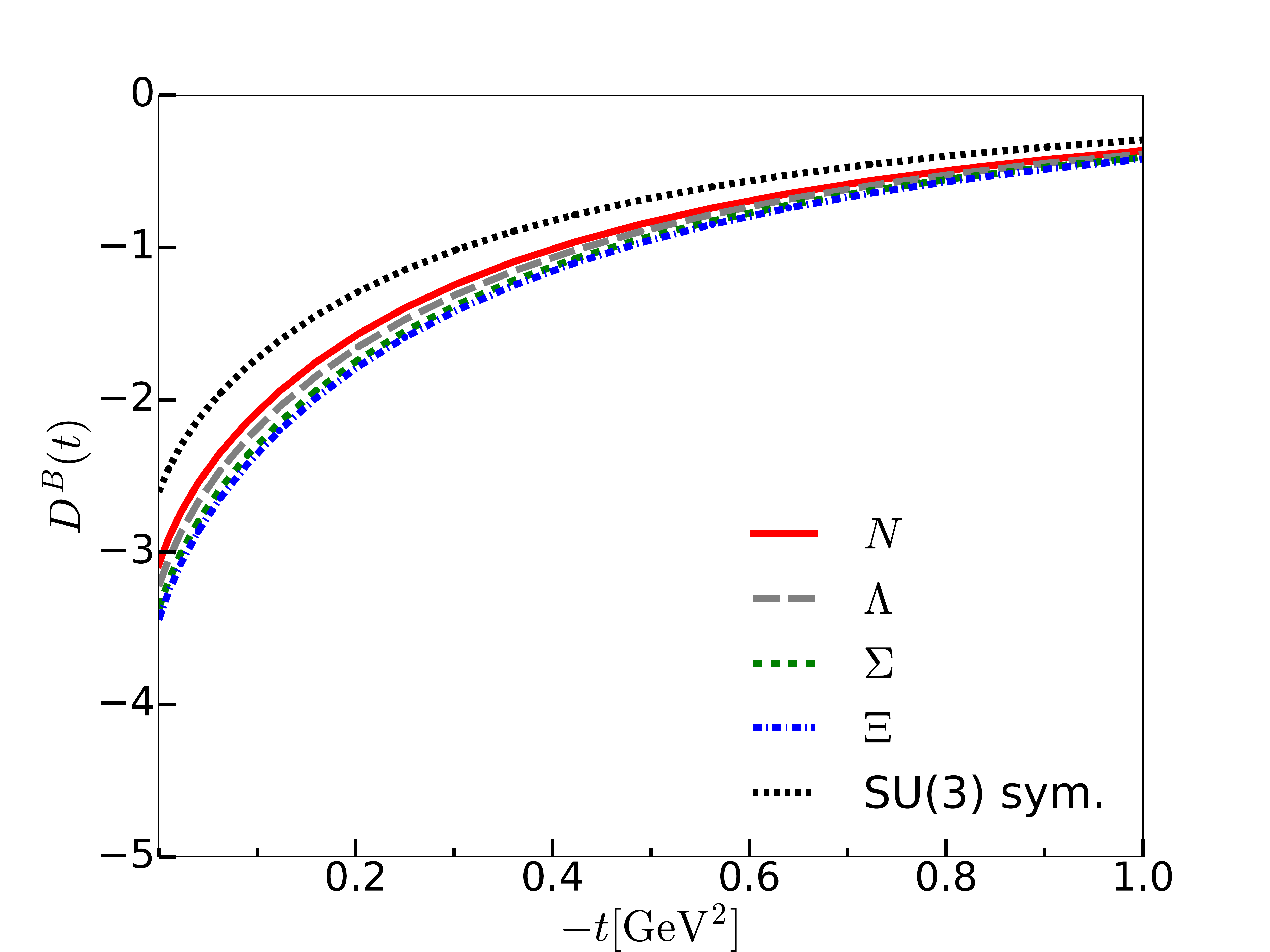}
\caption{Results for the EMT form factors $A^{B}(t)$, $D^{B}(t)$, and
  $J^{B}(t)$ of the baryon octet and the nucleon with the flavor SU(3)
  symmetry. The solid~(red), long-dashed~(grey), short-dashed~(green),
  dashed-dotted~(blue), and dotted~(black) curves denote $s^{B}(r)$
  for the $N$, $\Lambda$, $\Sigma$, $\Xi$, and nucleon with the flavor
  SU(3) symmetry, respectively. The notations are the same as in
  Fig.~\ref{fig:1}.}  
\label{fig:8} 
\end{figure}

The GFFs are obtained by the Fourier transform of the corresponding
EMT distributions. In Fig.~\ref{fig:8}, we present the numerical
results for the GFFs of the baryon octet as functions of the momentum 
transfer $t$. In the upper left panel of Fig.~\ref{fig:8}, the results
of the $A^{B}(t)$ show that the form factor of the heavier octet
baryon falls off slowly in comparison with that of the lighter one. It
reflects the fact that the heavier octet baryon is energetically more
compact than the lighter one. A similar feature was found in the case
of mass distribution of the heavy baryon~\cite{Kim:2020nug}. In the
upper right panel of Fig.~\ref{fig:8}, the results of the $J^{B}(t)$
show somewhat different features as observed in the angular momentum
distribution in the previous subsection. The form factor $J^{B}(t)$
for the nucleon and $\Sigma$ baryon falls off slowly in comparison
with those for the $\Xi$ and $\Lambda$ baryons. Lastly, in the lower
panel of Fig.~\ref{fig:8}, the $D$-term form factors are drawn. The
negativity of the $D$-term for the octet baryon can be also deduced
from the positive shear-force distributions over $r$. Thus, the
negative $D$-term is connected to the positivity of the normal force
distribution. We find that the heavier octet baryon possesses the
larger absolute value of the $D$-term. The mechanical radius can be
also obtained from the $D$-term form factor, but unlike a typical form
factor, the slope of the $D$-term form factor does not give the
mechanical radius of a baryon. Interestingly, the mechanical radii
have opposite behavior compared to the $D$-term. The heavier octet
baryon possesses a smaller size of mechanical radius than the lighter
one. 

\section{Summary and outlook \label{sec:6}}
In the present work, we aimed at investigating the gravitational form
factors of the baryon octet and the corresponding energy-momentum
tensor distributions within the SU(3) chiral quark-soliton model,
considering the effects of flavor SU(3) symmetry
breaking. Starting from the matrix element of the energy-momentum
tensor current for the baryon octet, we were able to compute the four
different densities: mass, angular momentum, pressure, and shear force
in both the 3D Breit and 2D infinite momentum frames. Integrating the
energy density over position space yielded the masses of the baryon
octet, so that the mass form factor was properly normalized to be
$A^{B}(0)=1$. We also found that, with the flavor SU(3) symmetry
breaking, the mass radius of the heavier octet baryon is larger than
that of the lighter one. It implies that the heavier octet baryon is
energetically a more compact object, compared with the lighter
one. We then examined the angular momentum densities of the octet 
baryons. Integrating the angular momentum densities over position
space gave the spins of the baryon octet even in the flavor SU(3)
symmetry broken case, so that the spin form factors were properly
normalized to be $J^{B}(0)=1/2$. Interestingly, the total angular
momentum was decomposed into the flavor-singlet axial charge and
orbital angular momentum $J^{B}= L^{B}+ g^{0,B}_{A}/2$. The quark spin
contributions to the angular momentum of the octet baryon were
estimated at around 50~\%. Since there are no gluonic degrees of
freedom in the chiral quark-soliton model, the missing contributions
were solely explained by the orbital motion of the quarks. While the
$m_{s}$ corrections differently contribute to the angular momentum
distributions for the octet baryons, their effects were rather
mild. So, both the spin and the orbital angular momentum contributions
to the angular momentum distributions for the octet baryons are well
balanced overall. Lastly, we computed the shear-force distribution
from the model calculation and then reconstructed the pressure
distribution from the shear-force distribution by using the
equilibrium differential equation. In addition, we extrapolated the
shear-force distribution at large $r$ to remove the numerical
redundancy by tagging the pion Yukawa tail. So, the pressure obviously
complied with the von Laue condition. One of the remarkable results
observed in this work is that the shear force is always positive for
any values of the $r$. It indicates the positive normal force over $r$
and the negative $D$-term. It means that the local stability condition
is still preserved even if we take into account the effects of the
flavor SU(3) symmetry breaking. We also estimated the $D$-terms for
the octet baryons and found that the heavier octet baryon has a larger
absolute value of the $D$-term than the lighter one. On the other
hand, the mechanical radius of the heavier octet baryon was smaller
than that of the lighter one. It implies that the heavier octet baryon
is mechanically a more compact object than the lighter one. We
presented the numerical results for the gravitational form factors of
the baryon octet as functions of the momentum transfer $t$ by the
Fourier transform of the given energy-momentum tensor
distributions. The mass and angular momentum form factor $A^{B}(t)$
and $J^{B}(t)$ were properly normalized to one and 1/2, respectively,
their slopes reflect the values of the distribution radius. When it
comes to the $D$-term form factors, as expected, their negative values
were obtained. In addition, $m_{s}$ corrections to the $D$-terms were
found to be rather small.  

\section*{Acknowledgments}
The work was supported by the Basic Science Research Program through
the National Research Foundation of Korea funded by the Korean
government (Ministry of Education, Science and Technology, MEST),
Grant-No. 2021R1A2C2093368 and 2018R1A5A1025563.

\clearpage
\appendix

\section{Densities and regularization functions \label{app:A}}
In this section, we collect the explicit expressions for the EMT
distributions. The mass distribution is written as 
\begin{align}
\frac{1}{N_{c}}\mathcal{E}(\bm{r})&=E_{v}
\psi_{v}^{\dagger}(\bm{r})\psi_{v}(\bm{r})
+\sum_{n=\mathrm{all}}\psi_{n}^{\dagger}(\bm{r})\psi_{n}(\bm{r})R_{0}(E_{n}),\cr
\frac{1}{N_{c}}\mathcal{S}(\bm{r})&=\psi_{v}^{\dagger}(\bm{r})
                                    \gamma^{0} \psi_{v}(\bm{r}) 
+\sum_{n=\mathrm{all}}\psi_{n}^{\dagger}(\bm{r})\gamma^{0}
                                    \psi_{n}(\bm{r}) R_{1}(E_{n}),\cr
\frac{1}{N_{c}}\mathcal{C}(\bm{r})&=\frac{1}{2}\sum_{n\neq v}
\frac{E_{n}+E_{v}}{E_{n}-E_{v}}\bra{n}\gamma^{0}\ket{v}
\psi_{v}^{\dagger}(\bm{r})\psi_{n}(\bm{r})
+\frac{1}{4}\sum_{\substack{n=\mathrm{all} \\ m=\mathrm{all}}}(E_{n}+E_{m})
\bra{n}\gamma^{0}\ket{m}\psi_{m}^{\dagger}(\bm{r}) 
\psi_{n}(\bm{r})R_{5}(E_{n},E_{m}),
\end{align}
and the angular momentum distribution is given by
\begin{align}
\frac{1}{N_{c}}\mathcal{I}_{1}(\bm{r})&=
\sum_{n\neq v}\frac{1}{E_{n}-E_{v}}
\mel{n}{\tau_{3}}{v}\psi_{v}^{\dagger}(\bm{r})\hat{J}_{3}\psi_{n}(\bm{r})
+\frac{1}{2}\sum_{\substack{n=\mathrm{all} \\ m=\mathrm{all}}}
\mel{n}{\tau_{3}}{m}\psi_{m}^{\dagger}(\bm{r})\hat{J}_{3}
\psi_{n}(\bm{r})R_{3}(E_{n},E_{m}),\cr
\frac{1}{N_{c}}\mathcal{K}_{1}(\bm{r})&=
\sum_{n\neq v}\frac{1}{E_{n}-E_{v}}
\mel{n}{\gamma^{0}\tau_{3}}{v}\psi_{v}^{\dagger}(\bm{r})
\hat{J}_{3}\psi_{n}(\bm{r})
+\frac{1}{2}\sum_{\substack{n=\mathrm{all} \\ m=\mathrm{all}}}
\mel{n}{\gamma^{0}\tau_{3}}{m}\psi_{m}^{\dagger}(\bm{r})
\hat{J}_{3}\psi_{n}(\bm{r})R_{5}(E_{n},E_{m}).
\end{align}
The shear-force distributions are expressed as 
\begin{align}
\frac{1}{N_{c}}\mathcal{N}_{1}(\bm{r})&=
\frac{3}{2}\left[\psi_{v}^{\dagger}(\bm{r})
\left(\gamma^{0}\left(\bm{\hat{n}}\cdot\bm{\gamma}\right)
\left(\bm{\hat{n}}\cdot\bm{p}\right)-\frac{1}{3}
\gamma^{0}(\bm{\gamma}\cdot\bm{p})\right)\psi_{v}(\bm{r})\right.\cr
&+\left.\sum_{n=\mathrm{all}}\psi_{n}^{\dagger}(\bm{r})
\left(\gamma^{0}\left(\bm{\hat{n}}\cdot\bm{\gamma}\right)
\left(\bm{\hat{n}}\cdot\bm{p}\right)-\frac{1}{3}
\gamma^{0}(\bm{\gamma}\cdot\bm{p})\right)\psi_{n}(\bm{r})
R_{1}(E_{n})\right],\cr
\frac{1}{N_{c}}\mathcal{N}_{2}(\bm{r})&=
\frac{3}{2}\left[\sum_{n\neq v}\frac{1}{E_{n}-E_{v}}
\mel{n}{\gamma^{0}}{v}
\psi_{v}^{\dagger}(\bm{r})\left(\gamma^{0}\left(\bm{\hat{n}}\cdot
\bm{\gamma}\right)
\left(\bm{\hat{n}}\cdot\bm{p}\right)-\frac{1}{3}
\gamma^{0}(\bm{\gamma}\cdot\bm{p})\right)
\psi_{n}(\bm{r})\right.\cr
&+\frac{1}{2}\left.\sum_{\substack{n=\mathrm{all} \\ m=\mathrm{all}}}
\mel{n}{\gamma^{0}}{m}
\psi_{m}^{\dagger}(\bm{r})\left(\gamma^{0}\left(\bm{\hat{n}}
\cdot\bm{\gamma}\right)
\left(\bm{\hat{n}}\cdot\bm{p}\right)-\frac{1}{3}
\gamma^{0}(\bm{\gamma}\cdot\bm{p})\right)
\psi_{n}(\bm{r})R_{2}(E_{n},E_{m})\right],
\end{align}
where the regularization functions are defined by
\begin{align}
  R_{0}(E_{n})&=\frac{1}{4\sqrt{\pi}}
\int\frac{du}{u^{3/2}}\phi(u,\Lambda)e^{-uE_{n}^{2}},\cr
  R_{1}(E_{n})&=-\frac{E_{n}}{2\sqrt{\pi}}
\int\frac{du}{\sqrt{u}}\phi(u,\Lambda)e^{-uE_{n}^{2}},\cr
  R_{2}(E_{n},E_{m})&=\frac{1}{2\sqrt{\pi}}
\int\frac{du}{\sqrt{u}}\phi(u,\Lambda)
  \frac{E_{n}e^{-uE_{n}^{2}}-E_{m}e^{-uE_{m}^{2}}}{E_{n}-E_{m}},\cr
  R_{3}(E_{n},E_{m})&=\frac{1}{2\sqrt{\pi}}
\int\frac{du}{\sqrt{u}}\phi(u,\Lambda)
\bigg[\frac{1}{u}\frac{e^{-uE_{n}^{2}}-e^{-uE_{m}^{2}}}{E_{m}^{2}-E_{n}^{2}}
  -\frac{E_{n}e^{-uE_{n}^{2}}+E_{m}e^{-uE_{m}^{2}}}{E_{n}+E_{m}}\bigg],\cr
  R_{5}(E_{n},E_{m})&=\frac{1}{2}
\frac{\mathrm{sign}(E_{n})-\mathrm{sign}(E_{m})}{E_{n}-E_{m}}.
\end{align}
with $\psi_{v}(\bm{r}):=\langle \bm{r}| v \rangle$,
$\psi_{n}(\bm{r}):=\langle \bm{r}| n \rangle$ and
$\hat{J}_{3}=\hat{L}_{3}+\hat{S}_{3}$. $| v \rangle$ and $| n \rangle$
denote the states of the valence and sea quarks with the corresponding
eigenenergies $E_{v}$ and $E_{n}$ of the single-quark Hamiltonian
$h(U)$, respectively. In addition, the dynamical parameters are
defined as follows: 
\begin{align}
 &\frac{1}{N_{c}}I_{1}
  =\frac{1}{2}\Bigg[\sum_{n\neq v}\frac{1}{E_{n}-E_{v}}
  \mel{n}{\tau_{3}}{v}
  \mel{v}{\tau_{3}}{n}
  +\frac{1}{2}\sum_{\substack{n=\mathrm{all} \\ m=\mathrm{all}}}
  \mel{n}{\tau_{3}}{m}
  \mel{m}{\tau_{3}}{n}
  R_{3}(E_{n},E_{m})\Bigg],\cr
&\frac{1}{N_{c}}K_{1}= \frac{1}{2}\Bigg[\sum_{n\neq v}\frac{1}{E_{n}-E_{v}}
  \mel{n}{\gamma^{0}\tau_{3}}{v}
  \mel{v}{\tau_{3}}{n}
  +\frac{1}{2}\sum_{\substack{n=\mathrm{all} \\ m=\mathrm{all}}}
  \mel{n}{\gamma^{0}\tau_{3}}{m}
  \mel{m}{\tau_{3}}{n}
  R_{5}(E_{n},E_{m})\Bigg].
  \label{eq:dynamical_para}
\end{align}

\section{Angular-momentum decomposition \label{app:B}}
We show how the total angular momentum can be
decomposed into the spin and orbital angular momentum 
contributions. It was derived in the SU(2)
$\chi$QSM~\cite{Ossmann:2004bp, Goeke:2007fp}, and we generalize it 
in SU(3). The angular momentum distribution is obtained
to be 
\begin{align}
  \rho^{B}_{J}(\bm{r})&=
  -\frac{1}{2I_{1}}\mathcal{I}_{1}(\bm{r})
  +2M_{8}\expval{D_{83}}_{B}\left(
    \frac{K_{1}}{I_{1}}\mathcal{I}_{1}(\bm{r})-\mathcal{K}_{1}(\bm{r})
  \right),
\end{align}
where each density is given by
\begin{align}
  \mathcal{I}_{1}(\bm{r})&=
  \frac{N_{c}}{4}\Bigg[\sum_{n\neq v}\frac{1}{E_{n}-E_{v}}
  \mel{n}{\tau_{3}}{v}
  \psi_{v}^{\dagger}(\bm{r})\left(2\hat{L}_{3}+(E_{n}+E_{m})
\gamma_{5}(\bm{\hat{r}}\times\sigma)_{3}\right)\psi_{n}(\bm{r})\cr
  &+\frac{1}{2}\sum_{\substack{n=\mathrm{all} \\ m=\mathrm{all}}}
  \mel{n}{\tau_{3}}{m}
  \psi_{m}^{\dagger}(\bm{r})\left(2\hat{L}_{3}+(E_{n}+E_{m})
\gamma_{5}(\bm{\hat{r}}\times\sigma)_{3}\right)\psi_{n}(\bm{r})
  R_{3}(E_{n},E_{m})\Bigg],\cr
  \mathcal{K}_{1}(\bm{r})&=
  \frac{N_{c}}{4}\Bigg[\sum_{n\neq v}\frac{1}{E_{n}-E_{v}}
  \mel{n}{\gamma^{0}\tau_{3}}{v}
  \psi_{v}^{\dagger}(\bm{r})\left(2\hat{L}_{3}+(E_{n}+E_{m})
\gamma_{5}(\bm{\hat{r}}\times\sigma)_{3}\right)\psi_{n}(\bm{r})\cr
  &+\frac{1}{2}\sum_{\substack{n=\mathrm{all} \\ m=\mathrm{all}}}
  \mel{n}{\gamma^{0}\tau_{3}}{m}
  \psi_{m}^{\dagger}(\bm{r})\left(2\hat{L}_{3}+(E_{n}+E_{m})
\gamma_{5}(\bm{\hat{r}}\times\sigma)_{3}\right)\psi_{n}(\bm{r})
  R_{5}(E_{n},E_{m})\Bigg].
\end{align}
To avoid numerical error discussed in Ref.~\cite{Ossmann:2005bbo}, we
manipulate the given densities. The second terms of each density can
be easily converted into the spin and orbital angular momentum
operators as follows: 
\begin{align}
\psi_{m}^{\dagger}(\bm{r})
\left((E_{n}+E_{m})\gamma_{5}(\bm{\hat{r}}\times\sigma)_{3}\right)
\psi_{n}(\bm{r})&=\varepsilon_{3jk}\psi_{m}^{\dagger}(\bm{r})
\left\{H,\gamma_{5}\bm{\hat{r}}^{j}\sigma^{k}\right\}
\psi_{n}(\bm{r})=\psi_{m}^{\dagger}(\bm{r})
\left(2\hat{L}_{3}+2\sigma_{3}\right)\psi_{n}(\bm{r}),
\end{align}
After that we are able to rewrite the densities as follows:
\begin{align}
  \frac{1}{N_{c}}\mathcal{I}_{1}(\bm{r})
  &=\sum_{n\neq v}\frac{1}{E_{n}-E_{v}}
  \mel{n}{\tau_{3}}{v}
  \psi_{v}^{\dagger}(\bm{r})\hat{J}_{3}
\psi_{n}(\bm{r})+\frac{1}{2}\sum_{\substack{n=\mathrm{all} \\ m=\mathrm{all}}}
  \mel{n}{\tau_{3}}{m}
  \psi_{m}^{\dagger}(\bm{r})\hat{J}_{3}\psi_{n}(\bm{r})
  R_{3}(E_{n},E_{m}),\cr
  \frac{1}{N_{c}}\mathcal{K}_{1}(\bm{r})
  &=\sum_{n\neq v}\frac{1}{E_{n}-E_{v}}
  \mel{n}{\gamma^{0}\tau_{3}}{v}
  \psi_{v}^{\dagger}(\bm{r})\hat{J}_{3}\psi_{n}(\bm{r})+\frac{1}{2}
\sum_{\substack{n=\mathrm{all} \\ m=\mathrm{all}}}
  \mel{n}{\gamma^{0}\tau_{3}}{m}
  \psi_{m}^{\dagger}(\bm{r})\hat{J}_{3}\psi_{n}(\bm{r})
  R_{5}(E_{n},E_{m}),
  \label{eq:app1}
\end{align}
where $\hat{J}_{3}=\hat{L}_{3}+\hat{S}_{3}$ and the spin operator
$\hat{S}$ are defined as $\hat{S}_{3}=\frac{1}{2}\sigma_{3}$. Thus, we
can define the orbital and spin densities by replacing the total
angular momentum operator $\hat{J}_{3}$ by either $\hat{L}_{3}$ or
$\hat{S}_{3}$: 
\begin{align}
\rho_{J}^{B}(r)=\rho_{L}^{B}(r)+\rho_{S}^{B}(r).
\end{align}
Here we are able to simplify Eq.~\eqref{eq:app1}
by using the fact that the quark states are eigenstates of the grand
spin operator~($\hat{G}_{3}=\hat{J}_{3}+\hat{T}_{3}$): 
\begin{align}
 \frac{1}{N_{c}} \int d^{3}r \mathcal{I}_{1}(\bm{r})
  &=\sum_{n\neq v}\frac{1}{E_{n}-E_{v}}
  \mel{n}{\tau_{3}}{v}
  \mel{v}{\left(\hat{G}_{3}-\hat{T}_{3}\right)}{n}
+\frac{1}{2}\sum_{\substack{n=\mathrm{all} \\ m=\mathrm{all}}}
  \mel{n}{\tau_{3}}{m}
  \mel{m}{\left(\hat{G}_{3}-\hat{T}_{3}\right)}{n}
  R_{3}(E_{n},E_{m})\cr
  \frac{1}{N_{c}}\int d^{3}r \mathcal{K}_{1}(\bm{r})
  &=\sum_{n\neq v}\frac{1}{E_{n}-E_{v}}
  \mel{n}{\gamma^{0}\tau_{3}}{v}
  \mel{v}{\left(\hat{G}_{3}-\hat{T}_{3}\right)}{n}
+\frac{1}{2}\sum_{\substack{n=\mathrm{all} \\ m=\mathrm{all}}}
  \mel{n}{\gamma^{0}\tau_{3}}{m}
  \mel{m}{\left(\hat{G}_{3}-\hat{T}_{3}\right)}{n}
  R_{5}(E_{n},E_{m}).
\end{align}
Note that the matrix elements
$\mel{m}{\hat{G}_{3}}{n}=G_{3}\delta_{mn}$ vanish for both 
densities. By integrating both sides over the 3D space, the densities 
$\mathcal{I}_{1}$ and $\mathcal{K}_{1}$ becomes the dynamical
parameter $I_{1}$ and $K_{1}$ defined in
Eq.~\eqref{eq:dynamical_para}: 
\begin{align}
  \int d^{3}r \mathcal{I}_{1}(\bm{r})=-I_{1}, \quad 
\int d^{3}r \mathcal{K}_{1}(\bm{r})=-K_{1}.
\end{align}
Therefore, the integration of the angular momentum density over the 3D
space always gives the spin normalization 
\begin{align}
  \int d^{3}r\rho_{J}^{B}(r)&=\frac{1}{2}.
\end{align}

\bibliography{OctetEMT}
\bibliographystyle{apsrev4-2}

\end{document}